\documentclass[preprint,aps,amsmath,nofootinbib]{revtex4}
\usepackage{graphicx}
\usepackage{amssymb}
\usepackage{color}
\definecolor{joerg}{rgb}{1.0,0.0,0.0}

\newcommand{\eq}{\begin{equation}}
\newcommand{\feq}{\end{equation}}

\newcommand{\bl}{\left}
\newcommand{\br}{\right}

\allowdisplaybreaks[1]

\def\eg{{\it e.g.}}
\def\ie{{\it i.e.}}

\def\calo{\mathcal{O}}

\def\theories{\mathcal{Q}}

\def\crit{{\cal C}}

\newcommand{\be}{\begin{equation}}
\newcommand{\ee}{\end{equation}}

\makeatother
\begin{document}
\title{Investigating the Ultraviolet Properties of Gravity with a Wilsonian Renormalization Group
Equation}
\author{Alessandro Codello}
\email{a.codello@gmail.com}
\affiliation{Institut f\"ur Physik, Johannes-Gutenberg-Universit\"at, Staudingerweg 7, D-59099 Mainz, Germany}
\author{Roberto Percacci}
\email{percacci@sissa.it}
\affiliation{Institute for Theoretical Physics, Utrecht University, Leuvenlaan 4, NL-3584, The Netherlands}
\affiliation{SISSA, via Beirut 4, I-34014 Trieste, Italy, and INFN, Sezione di Trieste, Italy}
\author{Christoph Rahmede}
\email{rahmede@sissa.it}
\affiliation{SISSA, via Beirut 4, I-34014 Trieste, Italy, and INFN, Sezione di Trieste, Italy}
%\pacs{04.60.-m, 11.10.Hi}
%\vskip 1cm
\begin{abstract}
We review and extend in several directions recent results on the
``asymptotic safety'' approach to quantum gravity.
The central issue in this approach is the search of a Fixed Point
having suitable properties, and the tool that is used is
a type of Wilsonian renormalization group equation.
We begin by discussing various cutoff schemes, \ie\ ways of implementing the
Wilsonian cutoff procedure.
We compare the beta functions of the gravitational couplings
obtained with different schemes, studying first the contribution
of matter fields and then the so--called Einstein--Hilbert truncation,
where only the cosmological constant and Newton's constant are retained.
In this context we make connection with old results, in particular
we reproduce the results of the epsilon expansion and the perturbative
one loop divergences.
We then apply the Renormalization Group to higher derivative gravity.
In the case of a general action quadratic in curvature we recover,
within certain approximations, the known asymptotic freedom
of the four--derivative terms, while Newton's constant
and of the cosmological constant have a nontrivial fixed point.
In the case of actions that are polynomials in the scalar curvature
of degree up to eight we find that the theory has a fixed
point with three UV--attractive directions, so that the requirement
of having a continuum limit constrains the couplings to lie in a
three--dimensional subspace, whose equation is explicitly given.
We emphasize throughout the difference between scheme--dependent and
scheme--independent results,
and provide several examples of the fact that only dimensionless
couplings can have ``universal'' behavior.
\end{abstract}
%\vfil
%\eject
\maketitle

\section{Introduction}

It is well known that general relativity
can be treated as an effective quantum field theory
\cite{Burgess,Donoghue,Bjerrum}.
This means that it is possible to compute quantum effects
due to graviton loops, as long as the momenta of the particles in
the loops are cut off at some scale.
For example, in this way it has been possible to unambiguously compute
quantum corrections to the Newtonian potential \cite{Donoghue}.
The results are independent of the structure of
any ``ultraviolet completion'', and therefore constitute genuine low
energy predictions of any quantum theory of gravity.
When one tries to push this effective field theory to energy scales
comparable to the Planck scale, or beyond, well-known
difficulties appear. It is convenient to distinguish two orders of
problems. The first is that the strength of the gravitational
coupling grows without bound. For a particle with energy $p$ the
effective strength of the gravitational coupling is measured by the
dimensionless number $\sqrt{\tilde G}$, with $\tilde G=G p^2$.
This is because the gravitational couplings involve derivatives of the metric.
The consequence of this is that if we let $p\to\infty$, also $\tilde G$
grows without bound. The second problem is the need of introducing
new counterterms at each order of perturbation theory. Since each
counterterm has to be fixed by an experiment, the ability of the
theory to predict the outcome of experiments is severely limited.

Could a Quantum Field Theory (QFT) get around these obstacles?
Recall that Newton's constant, as any coupling constant in a QFT,
must be subject to Renormalization Group (RG) flow. It is
conceivable that when $p\to\infty$, $G(p)\sim p^{-2}$, in which
case $\tilde G$ would cease to grow and would reach a finite limit,
thereby avoiding the first problem. If this is the case,
we say that Newton's constant has an UV Fixed Point (FP).
More generally, if we allow the action to contain several couplings
$g_i$ with canonical mass dimension $d_i$, we say that the theory
has a FP if all the dimensionless parameters
\begin{equation}
\label{couplings} \tilde g_i=g_i k^{-d_i}
\end{equation}
tend to finite values in the UV limit
\footnote{Strictly speaking only the essential couplings, {\it i.e.}
those that cannot be eliminated by field redefinitions, need to reach
a FP. See \cite{Perini3} for a related discussion in a gravitational context.}.
This particular RG behavior
would therefore solve the first of the two problems mentioned above
and would guarantee that the theory has a sensible UV limit.

In order to address the second problem we have to investigate the
set of RG trajectories that have this good behavior. We want to use
the condition of having a good UV limit as a criterion for selecting
a QFT of gravity. If all trajectories were attracted to the FP in
the UV limit, we would encounter a variant of the second
problem: the initial conditions for the RG flow would be arbitrary,
so determining the RG trajectory of the real world would require in
principle an infinite number of experiments and the theory would
lose predictivity. At the other extreme, the theory would have
maximal predictive power if there was a single trajectory ending at
the FP in the UV. However, this may be too much to ask.
An acceptable intermediate situation occurs when the trajectories
ending at the FP in the UV are parametrized by a finite number of
parameters. A theory with these properties was called
``asymptotically safe'' in \cite{Weinberg}.

To better understand this property, imagine, in the spirit of
effective field theories, a general QFT with all possible terms in
the action which are allowed by the symmetries. We can
parametrize the (generally infinite dimensional) ``space of all
theories'', $\theories$, by the dimensionless couplings $\tilde
g_i$. We assume that redundancies in the description of physics due
to the freedom to perform field redefinitions have been eliminated,
\ie\ all couplings are ``essential'' (such couplings can be defined
\eg\ in terms of cross sections in scattering experiments). We then
consider the Renormalization Group (RG) flow in this space; it is
given by the beta functions
\begin{equation}
\beta_i=k\frac{d \tilde g_i}{d k}\ .
\end{equation}
If there is a FP, {\it i.e.} a point with coordinates $\tilde
g_{i*}$ such that all $\beta_i(\tilde g_*)=0$, we call $\crit$ its
``critical surface'', defined as the locus of
points that are attracted towards the FP when $k\to\infty$
\footnote{RG transformations lead towards lower energies,
and the trajectories lying in $\crit$ are repelled by the FP
under these transformations. For this reason, $\crit$ is also
called the ``unstable manifold''. Since we are interested in
studying the UV limit, it is more convenient to study the
flow for increasing $k$.}.
One can determine the tangent space to the critical surface at the FP by
studying the linearized flow
\begin{equation}
k\frac{d(\tilde g_i-\tilde g_{i*})}{dk}=M_{ij}(\tilde g_j-\tilde g_{j*})\ ,
\end{equation}
where
\begin{equation}
M_{ij}=\frac{\partial\beta_i}{\partial\tilde g_j}\Bigr|_{*} .
\end{equation}
The attractivity properties of a FP are determined by the signs of
the critical exponents $\vartheta_i$, defined to be minus the
eigenvalues of $M$. The couplings corresponding to negative
eigenvalues (positive critical exponent) are called relevant and
parametrize the UV critical surface; they are attracted towards the
FP for $k\to\infty$ and can have arbitrary values. The ones that correspond
to positive eigenvalues (negative critical exponents) are called
irrelevant; they are repelled by the FP and must be set to zero.
%All this is familiar from the theory of critical phenomena.

A free theory (zero couplings) has vanishing beta functions, so the
origin in $\theories$ is a FP, called the Gau\ss ian FP. In the
neighborhood of the Gau\ss ian FP one can apply perturbation theory,
and one can show that the critical exponents are then equal to the
canonical dimensions ($\vartheta_i=d_i$), so the relevant couplings
are the ones that are power--counting renormalizable
\footnote{The behavior of dimensionless or marginal couplings may require a more
sophisticated analysis.}. In a local theory they are usually finite
in number. Thus, a QFT is perturbatively renormalizable and
asymptotically free if and only if the critical surface of the Gau\ss ian FP is
finite dimensional. Points outside $\crit$ flow to infinity, or to other FP's.
A theory with these properties makes sense to arbitrarily high
energies, because the couplings do not diverge in the UV, and is
predictive, because all but a finite number of parameters are fixed
by the condition of lying on $\crit$.
%This is a special example of what was called asymptotic safety.

If the Gau\ss ian FP is replaced by a more general, nontrivial FP,
we are led to a form of nonperturbative renormalizability.
It is this type of behavior that was called "asymptotic safety".
It is completely equivalent, in the language of critical phenomena,
to the assumption that the world is described by a ``renormalized trajectory''
emanating from some generic FP \cite{Wilson}.
%An asymptotically safe theory
%would have the same good properties of a renormalizable and
%asymptotically free one: the couplings would have a finite UV limit
%and the condition of lying on $\crit$ would leave only a finite
%number of parameters to be determined by experiment.
In general, studying the properties of such theories requires
the use of nonperturbative tools. If the nontrivial FP
is sufficiently close to the Gau\ss ian one, its properties can
also be studied in perturbation theory,
but unlike in asymptotically free theories, the results of perturbation
theory do not become better and better at higher energies.

In order to establish whether gravity has this type of behavior,
several authors \cite{Weinberg,Kawai:1989yh,Gastmans}
applied the $\epsilon$
expansion around two dimensions, which is the critical dimension
where Newton's constant is dimensionless. The beta function of
Newton's constant then has the form
\begin{equation}
\label{betaepsilon}
\beta_{\tilde G}=\epsilon \tilde G+B_1\tilde G^2\ ,
\end{equation}
where $\tilde G=G k^{-\epsilon}$ and $B_1<0$ \cite{Weinberg,Kawai:1989yh},
so there is a FP at $\tilde G=-\epsilon/B_1>0$.
Unfortunately this result is only reliable for small
$\epsilon$ and it is not clear whether it will extend to four dimensions.

In the effective field theory approach (in $d=4$),
Bjerrum-Bohr, Donoghue and Holstein have proposed
interpreting a class of one loop diagrams as giving the scale
dependence of Newton's constant \cite{Bjerrum}. They calculate
$$
G(r)=G_0\left[1-\frac{167}{30\pi}\frac{G_0}{r^2}\right]\ ,
$$
where $r$ is the distance between two gravitating point particles.
If we identify $k=1/ar$, with $a$ a constant of order one,
this would correspond to a beta function
\begin{equation}
\label{betabjerrum}
\beta_{\tilde G}=2\tilde G-a^2\frac{167}{15\pi}\tilde G^2\ .
\end{equation}
This beta function has the same form as (\ref{betaepsilon})
in four dimensions,
and, most important, the second term is again negative. This means
that the dimensionful Newton constant $G$ {\it decreases} towards
lower distances or higher energies, \ie\ gravity is {\it
antiscreening}. This is the behavior that is necessary for a FP to exist,
and indeed this beta function predicts a FP for
$\tilde G=\frac{30\pi}{167 a^2}$.
This calculation was based on perturbative methods and
since the FP occurs at a not very small value of $\tilde G$,
it is not clear that one can trust the result.
What we can say with confidence is that the onset of the running of $G$
has the right sign.
Clearly in order to make progress on this issue we need different tools.

In this paper we will discuss the application of Wilsonian
renormalization group methods to the UV behavior of gravity.
In section II we will introduce a particularly convenient tool,
called the ``Exact Renormalization Group Equation'' (ERGE) which
can be used to calculate the ``beta functional'' of a QFT.
Renormalizability is not necessary and the theory may have infinitely many couplings.
In section III we illustrate the use of the ERGE by calculating
the contribution of minimally coupled matter fields to the gravitational beta functions.
In this simple setting,
we will review the techniques that are used to extract from the beta
functional the beta functions of individual couplings,
emphasizing those results that are ``scheme independent''
in the sense that they are the same irrespective of technical
details of the calculation.
In section IV we apply the same techniques to the calculation of the beta
functions for the cosmological constant and Newton's constant in
Einstein's theory in arbitrary dimensions,
extending in various ways the results of earlier studies
\cite{Reuter,Souma,Lauscher,Fischer,reuterweyer1}.
We also show that the FP that is found in four--dimensional gravity is
indeed the continuation for $\epsilon\to 2$ of the FP that is found
in the $2+\epsilon$ expansion.
We compare various ways of defining the Wilsonian cutoff
and find the results to be qualitatively stable.
In sections V and VI we make connection with old results from perturbation theory.
In section V we rederive the 't Hooft--Veltman one loop divergence from the ERGE
and we show it to be scheme--independent. We also discuss why the
Goroff--Sagnotti two loop divergence cannot be seen with this method
and we discuss the significance of this fact.
In sections VI and VII we consider higher derivative gravity.
In section VI we derive the existence of the FP in the most general
truncation involving four derivatives at one loop, and we highlight the differences
between the Wilsonian procedure \cite{Codello} and earlier calculations.
In section VII we consider higher powers of curvature, restricting
ourselves to polynomials in the scalar curvature.
We give more details of our recent calculations \cite{CPR} and extend them to
polynomials of order eight.
In section VIII we assess the present status of this
approach to quantum gravity and discuss various open problems.

%%%%%%%%%%%%%%%%%%%%%%%%%%%%%%%%%%%%%%%%%%%%%%%%%%%%%%%%%%%%%%%%%%%%%%%%%%%%%%%%%%%%%%%%%

\section{The ERGE and its approximations}

The central lesson of Wilson's analysis of QFT is that the
``effective'' (as in ``effective field theory'') action describing
physical phenomena at a momentum scale $k$ can be thought of as the
result of having integrated out all fluctuations of the field with
momenta larger than $k$ \cite{Wilson}.
At this general level of discussion, it is not necessary to specify
the physical meaning of $k$: for each
application of the theory one will have to identify the physically
relevant variable acting as $k$\ \ \footnote{In scattering
experiments $k$ is usually identified with some external momentum.
See \cite{Hewett} for a discussion of this choice in concrete
applications to gravity.}.
Since $k$ can be regarded as the lower limit of some
functional integration, we will usually refer to it as the infrared
cutoff. The dependence of the ``effective'' action on $k$ is the
Wilsonian RG flow.

There are several ways of implementing this idea in practice,
resulting in several forms of the RG equation. In the specific
implementation that we shall use, instead of introducing a sharp
cutoff in the functional integral, we suppress the contribution of the field
modes with momenta lower than $k$. This is
obtained by modifying the low momentum end of the propagator, and
leaving all the interactions unaffected. We describe here this
procedure for a scalar field.
We start from a bare action $S[\phi]$, and we add to it a suppression
term $\Delta S_{k}[\phi]$ that is quadratic in the field.
In flat space this term can be written simply in momentum
space. In order to have a procedure that works in an arbitrary
curved spacetime we choose a suitable differential operator $\calo$
whose eigenfunctions $\varphi_n$, defined by
$\calo\varphi_n=\lambda_n\varphi_n$, can be taken as a basis in the
functional space we integrate over:
\begin{equation*}
\phi(x)=\sum_n \tilde\phi_n\varphi_n(x)\ ,
\end{equation*}
where $\tilde\phi_n$ are generalized Fourier components of the
field. (We will use a notation that is suitable for an operator with
a discrete spectrum.) Then, the additional term can be written in
either of the following forms:
\begin{equation}
\label{cutoffterm}
\Delta S_{k}[\phi]=
\frac{1}{2}\int dx\,\phi(x)\, R_{k}(\calo)\,\phi(x)
=\frac{1}{2}\sum_n \tilde\phi_n^2 R_k(\lambda_n)\,.
\end{equation}
The kernel $R_{k}(\calo)$ will also be called ``the cutoff''. It is
arbitrary, except for the general requirements that $R_k(z)$ should be
a monotonically decreasing function both in $z$ and $k$,
that $R_k(z)\to 0$ for $z \gg k$ and $R_k(z)\not= 0$ for $z \ll k$.
These conditions are enough to guarantee that the contribution to
the functional integral of field modes $\tilde\phi_n$
corresponding to eigenvalues $\lambda_n\ll k^2$ are
suppressed, while the contribution of field modes corresponding to
eigenvalues $\lambda_n\gg k^2$ are unaffected.
We will further fix $R_k(z)\to k^2$ for $k\to 0$.
We define a $k$-dependent generating functional of connected Green functions by
\begin{equation*}
e^{-W_{k}\left[J\right]}= \int D\phi\exp\left\{ -S[\phi]-\Delta
S_{k}[\phi]-\int dx\, J\phi\right\}
\end{equation*}
and a modified $k$-dependent Legendre transform
\begin{equation*}
\Gamma_{k}[\phi]=W_{k}\left[J\right]-\int dx\, J\phi-\Delta
S_{k}[\phi]\,,
\end{equation*}
where $\Delta S_{k}[\phi]$ has been subtracted. The functional
$\Gamma_k$ is sometimes called the ``effective  average action'',
because it is closely related to the effective action for fields
that have been averaged over volumes of order $k^{-d}$ ($d$ being
the dimension of spacetime) \cite{average}. The ``classical fields'' $\delta
W_{k}/\delta J$ are denoted again $\phi$ for notational simplicity.
In the limit $k\to 0$ this functional tends to
the usual effective action $\Gamma[\phi]$, the generating functional
of one-particle irreducible Green functions.
It is similar in spirit to the Wilsonian effective action,
but differs from it in the details of the implementation.

The average effective action $\Gamma_{k}[\phi]$, used at tree level,
gives an accurate description of processes occurring at momentum
scales of order $k$. In the spirit of effective field theories, we
shall assume that $\Gamma_k$ exists and is quasi--local in the sense
that it admits a derivative expansion of the form
\begin{equation}
\label{Gammak}
\Gamma_{k}(\phi,g_{i})=
\sum_{n=0}^\infty\sum_{i}g_{i}^{(n)}(k)\calo_{i}^{(n)}\left(\phi\right)\ ,
\end{equation}
where $g_{i}^{(n)}(k)$ are coupling constants and $\calo_{i}^{(n)}$
are all possible operators constructed with the field $\phi$ and
$n$ derivatives, which are compatible with the
symmetries of the theory. The index $i$ is used here to label
different operators with the same number of derivatives.

From the definition given above, it is easy to show that the
functional $\Gamma_k$ satisfies the following ``Exact Renormalization Group Equation''
(or ERGE) \cite{Wetterich,Morris}
\begin{equation}
\label{ERGE}
k\frac{d\Gamma_k}{dk}=
\frac{1}{2}\mathrm{Tr}\left[
\Gamma_k^{(2)}+R_{k}\right]^{-1}k\frac{dR_k}{dk}\ ,
\end{equation}
where the trace in the r.h.s. is a sum over the eigenvalues of the
operator $\calo$ (in flat space it would correspond to a momentum
integration) and we have introduced the notation
$\Gamma_k^{(2)}=\frac{\delta^{2}\Gamma_{k}}{\delta\phi\delta\phi}$
for the inverse propagator of the field $\phi$
defined by the functional $\Gamma_k$.
The r.h.s. of (\ref{ERGE}) can be regarded as the
``beta functional'' of the theory, giving the $k$--dependence of all
the couplings of the theory. In fact, taking the derivative of
(\ref{Gammak}) one gets
\begin{equation}
\label{dtGamma}
k\frac{d\Gamma_k}{dk}=\sum_{n=0}^\infty\sum_{i}\beta_{i}^{(n)}\calo_{i}^{(n)}\ .
\end{equation}
where
\begin{equation}
\label{betafunctions}
\beta_{i}^{(n)}(g_{j},k)=k\frac{dg_{i}^{(n)}}{dk}=\frac{dg_{i}^{(n)}}{dt}
\end{equation}
are the beta functions of the (generally dimensionful) couplings.
Here we have introduced $t=\log(k/k_{0})$, $k_{0}$ being an arbitrary initial value.
If we expand the trace on the r.h.s. of (\ref{ERGE}) in operators
$\calo_i^{(n)}$ and compare with (\ref{dtGamma}), we can in principle
read off the beta functions of the individual couplings.

This formalism can be easily generalized to the case of
multicomponent fields $\Phi^A$, where $A$ may denote both internal
and spacetime (Lorentz) indices. In this case the cutoff term will
have the form $\Phi^A R_{k AB} \Phi^B$,
and the trace in the r.h.s. of the ERGE will also involve a finite
trace over the indices $A,B$.
In the case of gauge theories there are further complications due to
the fact that the cutoff interferes with gauge invariance.
One way of dealing with this issue, which we shall use here,
is to use the background field method \cite{Reuter,Pawlowski}
(for another approach see \cite{rosten}).
One defines a functional of two fields, the
background field and the classical field, which is invariant under
background gauge transformations.
In the end the two fields are identified and one obtains a gauge
invariant functional of one gauge field only.

The ERGE can be seen formally as a RG--improved one loop equation.
To see this, recall that given a bare action $S$ (for a bosonic field),
the one loop effective action $\Gamma^{(1)}$  is
\begin{equation}
\Gamma^{(1)}=S+\frac{1}{2}\textrm{Tr}\log\left[\frac{\delta^{2}S}{\delta\phi\delta\phi}\right]\ .
\end{equation}
Let us add to $S$ the cutoff term (\ref{cutoffterm}); the functional
\begin{equation}
\label{oneloopaction}
\Gamma^{(1)}_{k}=S+\frac{1}{2}\textrm{Tr}\log\left[\frac{\delta^{2}S}{\delta\phi\delta\phi}+R_{k}\right]\,,
\end{equation}
may be called the ``one loop effective average action''.
It satisfies the equation
\begin{equation}
\label{oneloopERGE}
k\frac{d\Gamma_k^{(1)}}{dk}=\frac{1}{2}\mathrm{Tr}\left[\frac{\delta^{2}S}
{\delta\phi\delta\phi}+R_{k}\right]^{-1}k\frac{dR_k}{dk}\ ,
\end{equation}
which is formally identical to (\ref{ERGE}) except that in the
r.h.s. the renormalized running couplings $g_{i}(k)$ are replaced everywhere by
the ``bare'' couplings $g_{i}$, appearing in $S$.
Thus the ``RG improvement'' in the ERGE consists in replacing the bare couplings
by the running renormalized couplings.
In this connection, note that in general the cutoff function $R_{k}$ may contain the
couplings $g_i$ and therefore the term $k\frac{d}{dk}R_{k}$ in the
r.h.s. of (\ref{ERGE}) will itself contain the beta functions.
Thus, extracting the beta functions from the ERGE generally implies
solving an algebraic equation where the beta functions appear on
both sides \footnote{For example, in a scalar theory with action
$\int d^4x\,\left[Z(\partial\phi)^2+m^2\phi^2+\lambda\phi^4\right]$
it is natural to choose the cutoff of the form $R_k(z)=Z_k k^2
r(z/k^2)$. Then, the r.h.s. of (\ref{ERGE}) will contain $\partial_t Z$.}.
This complication can be avoided by choosing the cutoff in such a way that
it does not contain any coupling.
Then, the entire content of the ERGE is in the (RG--improved)
one loop beta functions.
The result is still ``exact'' insofar as one is able to keep track of
{\it all} possible couplings of the theory.

The formal derivation of the ERGE from a path integral makes use of
a bare action $S$.
To explore the relation between $S$ and the (renormalized) average effective action
$\Gamma_k$, for any $k$, requires that an ultraviolet regularization be defined
(in addition to the infrared regularization provided by $R_k$).
We will not need to discuss this point,
since the bare action is an unphysical quantity and all the
physics is encoded in the running renormalized action $\Gamma_k$.
In this connection note that the trace in the r.h.s. of (\ref{ERGE}), which includes an
integration over momenta, is perfectly ultraviolet convergent
and does not require any UV regulator.
This is because the function $k\frac{d}{dk} R_k$ in the r.h.s. goes
to zero for momenta greater than $k$ and makes the integration convergent.
So, we can regard the derivation given above as merely formal manipulations
that motivate the form of the ERGE, but then the ERGE itself is perfectly well defined,
without the need of introducing an UV regulator.
If we assume that at a given scale $k$ physics is described by a renormalized
action $\Gamma_k$, the ERGE gives us a way of studying the dependence of this
functional on $k$, and
the behavior of the theory at high energy can be studied by taking the limit
of $\Gamma_k$ for $k\to\infty$ (which need not coincide with the bare action $S$).

In most cases it is impossible to follow the flow of
infinitely many couplings and a common procedure is to consider a
truncation of the theory, namely to retain only a finite subset of
terms in the effective action $\Gamma_{k}$. For example one could
consider the derivative expansion (\ref{Gammak}) and retain all
terms up to some given order $n$. Whatever the choice, one
calculates the coefficients of the retained operators in the r.h.s.
of (\ref{ERGE}) and in this way the corresponding beta functions are
computed. In general the set of couplings that one chooses in this
way will not be closed under RG evolution, so one is neglecting the
potential effect of the excluded couplings on the ones that are
retained. Still, in this way one can obtain genuine nonperturbative
information, and this procedure has been applied to a variety of
physical problems, sometimes with good quantitative results. For
reviews, see \cite{Bagnuls,Berges,morrisrev}.

If we truncate the effective action in this way, there is usually
no small parameter to allow us to estimate the error we are making.
One indirect
way to estimate the quality of a truncation relies on an analysis of
the cutoff scheme dependence. The effective action $\Gamma_k$
obviously depends on the choice of the cutoff function $R_k$. This
dependence is similar to the scheme dependence of the renormalized
effective action in perturbative QFT;
only physically observable quantities
derived from $\Gamma_k$ must be independent of $R_k$.
This provides an indirect check on the quality of the truncation.
For example, the critical exponents should be universal quantities
and therefore cutoff--independent. In concrete calculations, usually involving a
truncation of the action, critical exponents do depend on the
cutoff scheme, and the observed dependence can be taken as a
quantitative measure of the quality of the approximation.
Ultimately, there is no substitute for performing calculations with
truncations that contain more terms. Note that a good truncation is
not necessarily one for which the new terms are small, but one for which the
effect of the new terms on the old ones is small. In other words, in
search of a nontrivial FP, we want the addition of new terms not to
affect too much the FP value of the ``old'' couplings, nor the
``old'' critical exponents.

%%%%%%%%%%%%%%%%%%%%%%%%%%%%%%%%%%%%%%%%%%%%%%%%%%%%%%%%%%%%%%%%%%%%%%%%%%%%%%%%%%%

\section{Matter fields and cutoff schemes}
\label{Cutoffschemes}

In this section we illustrate the method that is used to compute the
trace in the r.h.s. of (\ref{ERGE}) in a gravitational setting
and to evaluate the beta functions of the gravitational couplings.
Quite generally, we will consider the contribution of fields whose inverse
propagator $\Gamma^{(2)}$ is a differential operator of the form
$\Delta=-\nabla^2+\mathbf{E}$, where $\nabla$ is a covariant
derivative, both with respect to the gravitational field and
possibly also with respect to other gauge connections coupled to the
internal degrees of freedom of the field, and $\mathbf{E}$ is a
linear map acting on the quantum field.
In general, $\mathbf{E}$ could contain mass terms or
terms linear in curvature. For example, in the case of a
nonminimally coupled scalar, $\mathbf{E}=\xi R$, where $\xi$ is a coupling.
A priori, nothing will
be assumed about the gravitational action and also the spacetime
dimension $d$ can be left arbitrary at this stage.

In order to write the ERGE we have to define the cutoff. For the
operator to be used in the definition of (\ref{cutoffterm}),
several possible choices suggest themselves.
Let us split $\mathbf{E}=\mathbf{E}_1+\mathbf{E}_2$, where
$\mathbf{E}_1$ does not contain any couplings and $\mathbf{E}_2$
consists only of terms containing the couplings.
We call a cutoff {\em of type I}, if $R_k$ is a function of
the ``bare Laplacian'' $-\nabla^2$, {\em of type II} if it
is a function of $-\nabla^2+\mathbf{E}_1$ and {\em of type III}
if it is a function of the full kinetic operator $\Delta=-\nabla^2+\mathbf{E}$.
The substantial difference between the first two types and the third is that in
the latter case, due to the running of the couplings, the spectrum
changes along the flow. For this reason these cutoffs are said to be
``spectrally adjusted'' \cite{Gies}.
\footnote{In (\ref{cutoffterm}) it was assumed for simplicity that the operator
$\calo$ appearing in the argument of the cutoff function is also the
operator whose eigenfunctions are used as a basis in the evaluation
of the functional trace. It is worth stressing that this need not be
the case, as discussed in Appendix A.}

Let us now restrict ourselves to the case when $\mathbf{E}_2=0$, i.e. the kinetic
operator does not depend on the couplings; then there is only a choice
between cutoffs of type I and II. The derivation of the beta functions is
technically simpler with a type II cutoff. In this case we choose a
real function $R_k$ with the properties listed in section II and
define a modified inverse propagator
\begin{equation}
\label{PkII}
P_k(\Delta)=\Delta+R_k(\Delta)\ .
\end{equation}
If the operator $\mathbf{E}$ does not contain couplings, using
(\ref{HKasymp}) the trace in the r.h.s. of the ERGE reduces simply to:
\begin{equation}
\label{expII}
{\rm Tr}\frac{\partial_t R_k(\Delta)}{P_k(\Delta)} =
\frac{1}{(4\pi)^{d/2}}\sum_{i=0}^\infty
Q_{\frac{d}{2}-i}\left(\frac{\partial_t R_k}{P_k}\right)B_{2i}(\Delta)
\end{equation}
where $B_{2i}(\Delta)$ are the heat kernel coefficients of the operator
$\Delta$  and the $Q$-functionals, defined in
(\ref{Qnpos},\ref{Qnneg}) are the analogs of momentum integrals in
this curved spacetime setting.
We have written $\partial_t R_k$ to denote the derivative with respect
to the explicit dependence of $R_k$ on $k$; when the argument of $R_k$
does not contain couplings this coincides with the total derivative $\frac{d}{dt}R_k$.

With a type I cutoff we use the same profile function $R_k$ but now
with $-\nabla^2$ as its argument. This implies the replacement of
the inverse propagator $\Delta$ by
\begin{equation}
\label{PkI}
\Delta+R_k(-\nabla^2)=P_k(-\nabla^2)+\mathbf{E}\ .
\end{equation}
Therefore the r.h.s. of the ERGE will now contain the trace ${\rm
Tr}\frac{\partial_t R_k(-\nabla^2)}{P_k(-\nabla^2)+\mathbf{E}}$.
Since $\mathbf{E}$ is linear in curvature, in the limit when
the components of the curvature tensor are uniformly much smaller
than $k^2$, we can expand
\begin{equation*}
\frac{\partial_t R_k}{P_k+\mathbf{E}}=
\sum_{\ell=0}^\infty (-1)^\ell \mathbf{E}^\ell
\frac{\partial_t R_k}{P_k^{\ell+1}}\ .
\end{equation*}
Each one of the
terms on the r.h.s. can then be evaluated in a way analogous to
(\ref{HKasymp}), so in this case we get a double series:
\begin{equation}
\label{expI}
{\rm Tr}\frac{\partial_t R_k(-\nabla^2)}{P_k(-\nabla^2)+\mathbf{E}}=
\frac{1}{(4\pi)^{d/2}}\sum_{i=0}^\infty \sum_{\ell=0}^\infty
Q_{\frac{d}{2}-i}\left(\frac{\partial_t R_k}{P_k^{\ell+1}}\right)
\int dx\,\sqrt{g} (-1)^\ell{\rm tr}\mathbf{E}^\ell b_{2i}(-\nabla^2)\ .
\end{equation}
In order to extract the beta functions of the gravitational
couplings one has to collect terms with the same monomials in
curvature. We will see an example of this shortly.

Before discussing specific examples, however, it is interesting to
consider the scheme--independent part of the trace. In general, on
dimensional grounds, the functionals
$Q_n\left(\frac{\partial_t R_k}{P_k^m}\right)$
appearing in (\ref{expII}) and (\ref{expI}) will be equal to
$k^{2(n-m+1)}$ times a number depending on the profile function. As
discussed in Appendix A, the integrals with $m=n+1$ are independent
of the shape of $R_k$. Thus, in even-dimensional spacetimes
with a cutoff of type II, and using (\ref{univzero}),
the coefficient of the term in the sum (\ref{expII})
with $i=\frac{d}{2}$ is $Q_0\left(\frac{\partial_t R_k}{P_k}\right)B_d(\Delta)=2 B_d(\Delta)$.
On the other hand with a type I cutoff,
using (\ref{univpos}), (\ref{univzero}) and (\ref{shifted})
the terms with $\ell=\frac{d}{2}-i$ add up to
\begin{eqnarray}
&& \sum_{\ell=0}^{d/2}
 Q_{\ell}\left(\frac{\partial_t R_k}{P_k^{\ell+1}}\right)
\int dx\,\sqrt{g}
(-1)^\ell{\rm tr}\mathbf{E}^\ell b_{2i}(-\nabla^2)\nonumber\\
& = & 2\int dx\,\sqrt{g}{\rm tr}\left[b_d(-\nabla^2)-\mathbf{E}
b_{d-2}(-\nabla^2)+\ldots
+\frac{(-1)^{d/2}}{(d/2)!}\mathbf{E}^{d/2} b_0(-\nabla^2)\right]\nonumber\\
&=& 2 B_d(-\nabla^2+\mathbf{E})\nonumber
\end{eqnarray}
Therefore, in addition to being independent of the shape of the cutoff function,
these coefficients are also the same using type I or type II cutoffs.

As an example we will now specialize to four-dimensional gravity
coupled to $n_S$ scalar fields, $n_D$ Dirac fields, $n_M$ gauge
(Maxwell) fields, all massless and minimally coupled:
\begin{equation}\label{matteraction}
 \Gamma_k(g_{\mu\nu},\phi,\psi,A_\mu)=
\int d^4x\,\sqrt{g}\left[
\frac{1}{2}\nabla_\mu\phi\nabla^\mu\phi
+\bar\psi\gamma^\mu\nabla_\mu\psi +\frac{1}{4}F^{\mu\nu}F_{\mu\nu}
\right]\ .
\end{equation}
In addition $\Gamma_k$ must contain a generic action for gravity
of the form (\ref{Gammak}), which we do not write.
(For the terms with four derivatives we use the parametrization
given in (\ref{actionansatz}) below.)
We shall compute here the contribution of these matter fields to the
gravitational beta functions. The contribution of the gravitational
field to its beta functions will be calculated in the next section
using the same methods; as we shall see, the details of the
calculation are technically more involved, but conceptually
there is no difference.

The field equation of each type of field defines a second order
differential operator $\Delta^{(A)}=-\nabla^2+\mathbf{E}^{(A)}$,
with $A=S,D,M, gh$ and
\begin{equation}
\mathbf{E}^{(S)}=0\ ;\qquad \mathbf{E}^{(D)}=\frac{R}{4}\ ;\qquad
\mathbf{E}^{(M)}={\rm Ricci}\ ;\qquad \mathbf{E}^{(gh)}=0\ .
\end{equation}
Here ``Ricci'' stands for the Ricci tensor regarded as a linear operator
acting on vectors: ${\rm Ricci}(v)_\mu=R_\mu{}^\nu v_\nu$. For the gauge fields we have
chosen the Lorentz gauge, and $\Delta^{(gh)}$ is the operator acting
on the scalar ghost. (It can be shown that the results do not depend
on the choice of gauge \cite{Barvinsky}.)

With a type II cutoff, for each type of field we define the modified
inverse propagator $P_k(\Delta^{(A)})=\Delta^{(A)}+R_k(\Delta^{(A)})$.
Then, the ERGE reduces simply to
\begin{eqnarray}
\label{matterergeII}
\frac{d\Gamma_k}{dt}&=&
\frac{n_S}{2}{\rm Tr}_{(S)}\left(\frac{\partial_t R_k(\Delta^{(S)})}{P_k(\Delta^{(S)})}\right)
-\frac{n_D}{2}{\rm Tr}_{(D)}\left(\frac{\partial_t R_k(\Delta^{(D)})}{P_k(\Delta^{(D)})}\right)\nonumber\\
&&+\frac{n_M}{2}{\rm Tr}_{(M)}\left(\frac{\partial_t R_k(\Delta^{(M)})}{P_k(\Delta^{(M)})}\right)
-{n_M}{\rm Tr}_{(gh)}\left(\frac{\partial_t R_k(\Delta^{(gh)})}{P_k(\Delta^{(gh)})}\right)\nonumber\\
&=& \frac{1}{2}\frac{1}{(4\pi)^2}\int\,d^4x\,\sqrt{g} \Biggl[
\left(n_S-4n_D+2n_M \right)Q_2\left(\frac{\partial_t R_k}{P_k}\right)\nonumber\\
&&+\frac{1}{ 6}R \left(n_S+2n_D-4n_M \right)Q_1\left(\frac{\partial_t R_k}{ P_k}\right)\nonumber\\
&&+\frac{1}{180}\Bigg( \left(3 n_S+18 n_D+36 n_M\right) C^2
-\left(n_S+11 n_D+62 n_M\right)E \nonumber\\
&&+5 n_S R^2 +12\left(n_S+n_D-3n_M\right)\nabla^2R\Bigg)
+\ldots \Biggr]\ ,
\end{eqnarray}
where $C^{2}$ is the square of Weyl's tensor and
$E=R_{\mu\nu\rho\sigma}R^{\mu\nu\rho\sigma}-4R_{\mu\nu}R^{\mu\nu}+R^2$
(in four dimensions $\chi=\frac{1}{32\pi^2}\int d^4x\,\sqrt{g}E$ is Euler's topological invariant).
The terms with zero and one power of $R$ depend on the profile function $R_k$
but using (\ref{univzero}) we see that the coefficients of the
four--derivative terms, {\it i.e.} the beta functions of $g^{(4)}_i$, are scheme--independent.

With type I cutoffs, the modified inverse propagators are
$\Delta^{(A)}+R_k(-\nabla^2)=P_k(-\nabla^2)+\mathbf{E}^{(A)}$
and the ERGE then becomes:
\begin{eqnarray}
\frac{d\Gamma_k}{dt}&=&
\frac{n_S}{2}{\rm Tr}_{(S)}\left(\frac{\partial_t
P_k(-\nabla^2)}{P_k(-\nabla^2)}\right)
-\frac{n_D}{2}{\rm Tr}_{(D)}\left(\frac{\partial_t R_k(-\nabla^2)}{P_k(-\nabla^2)
+\frac{R}{4}}\right)\nonumber\\
&&+\frac{n_M}{2}{\rm Tr}_{(M)}\left(\frac{\partial_t R_k(-\nabla^2)}{P_k(-\nabla^2)
+\mathrm{Ricci}}\right) -{n_M}{\rm Tr}_{(gh)}\left(\frac{\partial_t R_k(-\nabla^2)}{P_k(-\nabla^2)}
\right)\ .
\end{eqnarray}
Expanding each trace as in (\ref{HKasymp}), collecting terms with the same
number of derivatives of the metric, and keeping terms up to four
derivatives we get
\begin{eqnarray}
\label{matterergeI}
\frac{d\Gamma_k}{dt}&=&
\frac{1}{2}\frac{1}{(4\pi)^2}\int\,d^4x\,\sqrt{g} \Biggl[
\left(n_S-4n_D+2n_M \right)Q_2\left(\frac{\partial_t R_k}{P_k}\right)\nonumber\\
&&+\Biggl[\frac{1}{6}Q_1\left(\frac{\partial_t R_k}{ P_k}\right)n_S
-\left(\frac{2}{3}Q_1\left(\frac{\partial_t R_k}{ P_k}\right)
-Q_2\left(\frac{\partial_t R_k}{P_k^2}\right)\right)n_D\nonumber\\
&&+\left(\frac{1}{3}Q_1\left(\frac{\partial_t R_k}{ P_k}\right)
-Q_2\left(\frac{\partial_t R_k}{P_k^2}\right)\right)n_M\Biggr]R \nonumber\\
&&+\frac{1}{180}\Bigg( \left(3 n_S+18 n_D+36 n_M\right) C^2
-\left(n_S+11 n_D+62 n_M\right)E \nonumber\\
&&+5 n_S R^2 +12\left(n_S+n_D-3n_M\right)\nabla^2R\Bigg) +\ldots
\Biggr]\ .
\end{eqnarray}
We see that the terms linear in curvature, which contribute to the
beta function of Newton's constant, have changed.
However, the terms quadratic in curvature have
the same coefficients as before, confirming that the
beta functions of the dimensionless couplings are scheme--independent.

In order to have more explicit formulae, and in numerical work, one
needs to calculate also the scheme--dependent $Q$-functionals. This
requires fixing the profile $R_k$. In this paper we will mostly use
the so--called optimized cutoff (\ref{optimizedcutoff}) in which the
integrals are readily evaluated, see equations
(\ref{qoptpos},\ref{qoptzero},\ref{qoptneg}). This cutoff has the
very convenient property that $Q_{-n}\left(\frac{\partial_t R_k}{P_k}\right)=0$ for
$n\geq 1$. Thus, the sum over heat kernel coefficients on the r.h.s.
of (\ref{HKasymp}) terminates.
In particular, in four dimensions, there are no terms beyond those
that are explicitly written in (\ref{matterergeII}) or (\ref{matterergeI})
\footnote{this had also been observed in a different context in \cite{Connes}.}.
For more general cutoffs a calculation of beta functions for
curvature--polynomials of cubic and higher order would require the
knowledge of higher heat kernel coefficients.

Let us briefly comment on the spectrally adjusted (type III)
cutoffs. These only occur when the kinetic operator $\Delta$
contains couplings, for example a mass term or, in the case of a
scalar field, a nonminimal coupling of the form $\xi R$. In this
case the last factor in (\ref{ERGE}) is
\begin{equation*}
\frac{d R_k}{dt}=\partial_t R_k +\sum_i R'_k
\frac{\partial\mathbf{E}}{\partial g_i}\partial_t g_i\ ,
\end{equation*}
where $R'_k$ denotes the derivative of $R_k(z)$ with respect to $z$
and the sum extends over all couplings
(this assumes that the derivative of the operator appearing in $R_k$
commutes with the operator itself). This introduces
further nonlinearities in the system. Since the beta functions of
the couplings appear linearly in the r.h.s. of the equation, to
obtain the beta functions one has to solve a system of linear equations.

In a consistent truncation one would have to add to the terms in
(\ref{matterergeI}) or (\ref{matterergeII}) the contribution
due to the gravitational field. This will be done in the next section.
For the time being we observe that if the number of matter fields
is of order $N\to\infty$, this is the
dominant contribution and constitutes the leading order
of a $1/N$ expansion \cite{PercacciN}.
The matter contributions by themselves have a form that leads to a gravitational FP.
Comparing equation (\ref{dtGamma}) with equations (\ref{matterergeI}) or (\ref{matterergeII})
one can read off the beta functions, which all have the form
\begin{equation*}
\frac{d g^{(n)}_i}{dt}=a^{(n)}_i k^{4-n}\ ,
\end{equation*}
where $a^{(n)}_i$ are constants.
Then, the beta functions of the dimensionless variables
$\tilde g^{(n)}_i=k^{n-4}g^{(n)}_i$ are
\begin{equation}
\label{largenbetas}
\frac{d\tilde g^{(n)}_i}{dt}=(n-4)\tilde g^{(n)}_i+a^{(n)}_i\ .
\end{equation}
This simple flow has indeed a FP for all couplings.
For $n\not= 4$
\begin{equation}
\tilde g^{(n)}_{i*}=\frac{a^{(n)}_i}{4-n}\ ,
\end{equation}
in particular writing $g^{(0)}=2Z\Lambda$ and $g^{(2)}=-Z=-\frac{1}{16\pi G}$,
the flow of the cosmological constant and Newton's constant is given by
\begin{align}
\label{nlambdag}
\frac{d \tilde \Lambda}{d t}=&-2\tilde \Lambda+8\pi a^{(0)} \tilde G+16\pi a^{(2)}\tilde G\tilde\Lambda\ ,\cr
\frac{d \tilde G}{d t}=&2\tilde G+16\pi a^{(2)} \tilde G^2\ ,
\end{align}
which, for a type II cutoff, has a FP at
\begin{equation}
\tilde\Lambda_*=-\frac{3}{4}\frac{n_S-4n_D+2 n_M}{n_S+2n_D-4n_M}\ ,\ \
\tilde G_*=\frac{12\pi}{-n_S-2n_D+4n_M}\ .
\end{equation}
Note that the FP occurs for positive or negative
$\tilde\Lambda$ depending on whether there are more bosonic of
fermionic degrees of freedom. The FP value of $\tilde G$,
on the other hand, will be positive provided there are not too many
scalar fields.
For $n=4$, (\ref{largenbetas}) gives a logarithmic running
$$
g^{(4)}_i(k)=g^{(4)}_i(k_0)+a^{(4)}_i\ln(k/k_0)\ ,
$$
implying asymptotic freedom for the couplings $1/g^{(4)}_i$. This is the same
behavior that is observed in Yang--Mills theories and is in
accordance with earlier perturbative calculations
\cite{Smolin,Tomboulis}.
As noted, it follows from (\ref{qoptneg}) that with the optimized cutoff,
for $n>4$, $\tilde g^{(n)}_{i*}=0$.

\begin{figure}
[t]\center
{\resizebox{0.7 \columnwidth}{!}
{\includegraphics{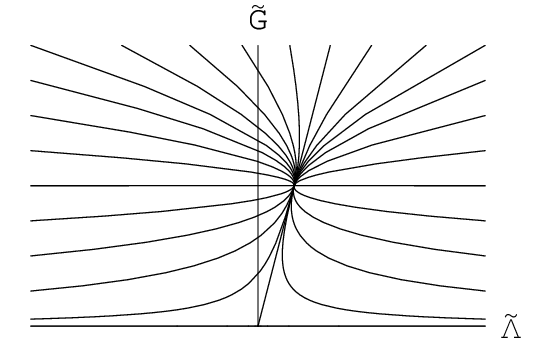}}
    \caption{\label{fig:flowmatter}The generic form of the flow
induced by matter fields.
}}
\end{figure}

The critical exponents at the nontrivial FP
are equal to the canonical dimensions of the $g^{(n)}$'s,
so $\Lambda$ and $G$ are UV--relevant (attractive),
$1/g^{(4)}_i$ are marginal and all the higher terms are UV--irrelevant.
Note that in perturbation theory $G$ is irrelevant.
At the nontrivial FP the quantum corrections conspire with the classical
dimensions of $\Lambda$ and $G$ to reconstruct the dimensions of $g^{(0)}$ and $g^{(2)}$.
This must happen because the critical exponents for $g^{(0)}$ and $g^{(2)}$
are equal to their canonical dimensions and the critical exponents are invariant under
regular coordinate transformations in the space of all couplings;
the trasformation between $\tilde G$ and $\tilde g^{(2)}$
is regular at the nontrivial FP, but it is singular at the Gau\ss ian FP.

This simple flow is exact in the limit $N\to\infty$, but is also a rough approximation
when graviton effects are taken into account, as we shall discuss in sections IV-G and VI.
It is shown in figure \ref{fig:flowmatter}.

%%%%%%%%%%%%%%%%%%%%%%%%%%%%%%%%%%%%%%%%%%%%%%%%%%%%%%%%%%%%%%%%%%%%%%%%%%%%%%%%%%%%%%%%%%%%
\medskip

\section{Einstein's theory}
\label{EH}

As a first step towards the inclusion of quantum gravitational effects, we discuss
in this section the RG flow for Einstein's gravity, with or without
cosmological constant.
This truncation has been extensively discussed before \cite{Reuter,Lauscher}.
Here we will extend those results in various directions.
Since the dependence of the results on the choice of gauge and
profile function $R_k$ has already been discussed in \cite{Lauscher,Fischer}
here we shall fix our attention on a particular gauge and profile function,
and analyze instead the dependence of the results on different ways of
implementing the cutoff procedure.
The simplicity of the truncation will allow us to compare the results
of different approximations and cutoff schemes, a luxury that is
progressively reduced going to more complicated truncations.

The theory is parametrized by the cosmological constant $\Lambda$
and Newton's constant $G=1/\bl(16\pi Z\br)$, so that we set
$g^{(0)}=2\Lambda Z$ and $g^{(2)}=-Z$ in equation (\ref{Gammak}). All higher
couplings are neglected. Then the truncation takes the form
\begin{equation}
\label{ehaction}
\Gamma_{k}=\int dx\sqrt{g}\,(2\Lambda Z-ZR(g))+S_{GF}+S_{\rm ghost}\ ,
\end{equation}
where $S_{GF}$ is a gauge--fixing term and $S_{\rm ghost}$ is the ghost action.
We decompose the metric into $g_{\mu\nu}=g_{\mu\nu}^{(B)}+h_{\mu\nu}$
where $g_{\mu\nu}^{(B)}$ is a background.
We will refer to the field $h_{\mu\nu}$ as the graviton,
even though it is not assumed to be a small perturbation.
We consider background gauges of the type:
\eq
\label{ehgaugefixing}
S_{GF}(g^{(B)},h)=\frac{Z}{2\alpha}\int dx\sqrt{g^{(B)}}\chi_{\mu}g^{(B)\mu\nu}\chi_{\nu}\ ,
\feq
where
$$
\chi_{\nu}=\nabla^{\mu}h_{\mu\nu}-\frac{1+\rho}{d}\nabla_{\nu}h\ .
$$
All covariant derivatives are with respect to the background metric.
In the following all metrics will be background metrics, and we will
omit the superscript $(B)$ for notational simplicity.
In this section we will restrict ourselves to the de Donder gauge
with parameters $\alpha=1$, $\rho=\frac{d}{2}-1$,
which leads to considerable simplification.
The inverse propagator of $h_{\mu\nu}$, including the gauge fixing term,
can be written in the form
\begin{equation*}
\frac{1}{2}\int dx\,\sqrt{g}h_{\mu\nu}\Gamma_k^{(2)\mu\nu\rho\sigma}h_{\rho\sigma}
\end{equation*}
containing the minimal operator:
\begin{equation}
\label{e1}
\Gamma^{(2)\mu\nu}_{k\ \rho\sigma}
=Z
\left[K_{\rho\sigma}^{\mu\nu}(-\nabla^2-2\Lambda)+U_{\rho\sigma}^{\mu\nu}\right]\ ,
\end{equation}
where
\footnote{These definitions coincide with those of \cite{Reuter} except that the $\Lambda$ term has been
removed from $\mathbf{U}$.}
\begin{eqnarray}
K_{\rho\sigma}^{\mu\nu}& = &\frac{1}{2}\left(\delta_{\rho\sigma}^{\mu\nu}-\frac{d}{2}P_{\rho\sigma}^{\mu\nu}\right)\ ;\qquad
\delta_{\rho\sigma}^{\mu\nu}=\frac{1}{2}\left(\delta_{\rho}^{\mu}\delta_{\sigma}^{\nu}+\delta_{\sigma}^{\mu}\delta_{\rho}^{\nu}\right)\ ;\qquad
P_{\rho\sigma}^{\mu\nu}  =  \frac{1}{d}g^{\mu\nu}g_{\rho\sigma}\ ;\nonumber\\
U_{\rho\sigma}^{\mu\nu} & = &
R\,K_{\rho\sigma}^{\mu\nu}
+\frac{1}{2}\left(g^{\mu\nu}R_{\rho\sigma}+R^{\mu\nu}g_{\rho\sigma}\right)
-\delta_{(\rho}^{(\mu}R_{\sigma)}^{\nu)}
-R^{(\mu}{}_{(\rho}{}^{\nu)}{}_{\sigma)}\ .\nonumber
\end{eqnarray}
In the following we will sometimes suppress indices for notational clarity;
we will use boldface symbols to indicate linear operators on the space of symmetric tensors.
For example, the objects defined above will be denoted $\mathbf{K}$, $\mathbf{1}$, $\mathbf{P}$, $\mathbf{U}$.
Note that $\mathbf{P}$ and $\mathbf{1}-\mathbf{P}$
are projectors onto the trace and tracefree parts in the space of symmetric tensors:
$h_{\mu\nu}=h^{(TF)}_{\mu\nu}+h^{(T)}_{\mu\nu}$ where
$h^{(T)}_{\mu\nu}=P_{\mu\nu}^{\rho\sigma}h_{\rho\sigma}=\frac{1}{d}g_{\mu\nu}h$.
Using that $\mathbf{K}=\frac{1}{2}\left((\mathbf{1}-\mathbf{P})+\frac{2-d}{2}\mathbf{P}\right)$,
if $d\not=2$ we can rewrite equation (\ref{e1}) in either of the following forms:
\begin{eqnarray}
\label{usefulforms}
\mathbf{\Gamma}_{k}^{(2)}
&=&Z\,\mathbf{K}(-\nabla^2-2\Lambda\mathbf{1}+\mathbf{W})\nonumber\\
&=&\frac{Z}{2}\left[(\mathbf{1-P})
\left(-\nabla^2-2\Lambda\mathbf{1}+2\mathbf{U}\right)
-\frac{d-2}{2}\mathbf{P}
\left(-\nabla^2-2\Lambda\mathbf{1}-\frac{4}{d-2}\mathbf{U}\right) \right]
\end{eqnarray}
where we have defined
\begin{equation*}
W_{\rho\sigma}^{\mu\nu}=2U_{\rho\sigma}^{\mu\nu}
-\frac{(d-4)}{2(d-2)}(R_{\rho\sigma}g^{\mu\nu}+g_{\rho\sigma}R^{\mu\nu}-Rg_{\rho\sigma}g^{\mu\nu})\ .
\end{equation*}
Note that the overall sign of the second term in the second line
of (\ref{usefulforms}) is negative when $d>2$. This is the famous problem of
the unboundedness of the Euclidean Einstein--Hilbert action.
We shall see shortly how this is dealt with in the ERGE.
Later on, we will need the traces:
\begin{eqnarray}
\label{traces}
\textrm{tr}\,\mathbf{1}&=&\frac{d(d+1)}{2}\ ;\ \
\textrm{tr}\,\mathbf{P}=1\ ;\ \
\textrm{tr}\,\left(\mathbf{1}-\mathbf{P}\right)=\frac{d^2+d-2}{2}\ ;\ \
\textrm{tr}\,\mathbf{W}=\frac{d(d-1)}{2}R\ ;\ \
\nonumber\\
\textrm{tr}\,\mathbf{W}^2&=&3 R_{\mu\nu\rho\sigma}R^{\mu\nu\rho\sigma}
+\frac{d^2-8d+4}{d-2}R_{\mu\nu}R^{\mu\nu}
+\frac{d^3-5d^2+8d+4}{2(d-2)}R^2
\ .
\end{eqnarray}
The ghost action is:
$$
S_{\rm ghost}=-\int\sqrt{g}\,
\bar C_\mu\left(-\nabla^2\delta_{\,\,\nu}^{\mu}-R_{\,\,\nu}^{\mu}\right)C^\nu\ .
$$
On the $d$-dimensional sphere we can write
$$
\mathbf{U}=\frac{1}{2}\left[\bl(\mathbf{1-P}\br)\frac{d^{2}-3d+4}{d(d-1)}R
-\mathbf{P}\frac{d-2}{2}\frac{d-4}{d}R\right]\ .
$$
Then, using the second line of (\ref{usefulforms}), we have
\begin{equation}
\label{invpropsphere}
\mathbf{\Gamma}_{k}^{(2)}
=\frac{Z}{2}\left[(\mathbf{1-P})\left(-\nabla^2-2\Lambda+\frac{d^{2}-3d+4}{d\bl(d-1\br)}R\right)
-\frac{d-2}{2}\mathbf{P}\left(-\nabla^2-2\Lambda+\frac{d-4}{d}R\right)\right]\ .
\end{equation}
We will now discuss separately various types of cutoff schemes.

\subsection{Cutoff of type Ia}

This is the scheme that was used originally in \cite{Reuter}. It is defined by the
cutoff term
\begin{equation}
\label{cutoffIa}
 \Delta S_{k}[h_{\mu\nu}]=
\frac{1}{2}\int dx\,\sqrt{g}\, h_{\mu\nu}R_{k}(-\nabla^2)^{\mu\nu\rho\sigma}\,h_{\rho\sigma}
-\int dx\,\sqrt{g}\, \bar C_\mu R_k^{(gh)}(-\nabla^2){}^\mu{}_\nu C^\nu\ ,
\end{equation}
where
\begin{eqnarray}
\mathbf{R}_{k}(-\nabla^2)&=&Z\mathbf{K}R_{k}(-\nabla^2)\nonumber\\
R_k^{(gh)}(-\nabla^2){}^\mu{}_\nu&=&\delta_{\,\,\nu}^{\mu}R_{k}(-\nabla^2)\,.
\end{eqnarray}
for gravitons and ghosts respectively.
Defining the anomalous dimension by
\eq
\eta=\frac{1}{Z}\frac{dZ}{dt}\,,
\feq
we then have
\eq
\frac{d \mathbf{R}_k}{dt}=Z\mathbf{K}\left[\partial_t{R}_{k}(-\nabla^2)+\eta
R_{k}(-\nabla^2)\right]\ . \feq
The calculation in \cite{Reuter} proceeded as follows.
The background metric is chosen to be that of Euclidean de Sitter space.
The modified inverse propagator is obtained from (\ref{invpropsphere})
just replacing $-\nabla^2$ by $P_k(-\nabla^2)$.
Using the properties of the projectors, its inversion is trivial:
\begin{equation}
\left(\mathbf{\Gamma}_{k}^{(2)}+{\bf R}_k\right)^{-1}
=\frac{2}{Z}
\left[({\bf 1-P})\frac{1}{P_k-2\Lambda+\frac{d^2-3d+4}{d(d-1)}R}
-\frac{2}{d-2}{\bf P}\frac{1}{P_k-2\Lambda+\frac{d-4}{d}R}\right]
\end{equation}
Decomposing in the same way the term $\frac{d}{dt}\mathbf{R}_k$,
multiplying and tracing over spacetime indices one obtains
\begin{equation}
\frac{d\Gamma_k}{dt}=
\frac{1}{2}\textrm{Tr}({\bf 1-P})\frac{\partial_t R_k+\eta
R_k}{P_k-2\Lambda+\frac{d^2-3d+4}{d(d-1)}R}
+\frac{1}{2}\textrm{Tr}{\bf P}\frac{\partial_t R_k+\eta
R_k}{P_k-2\Lambda+\frac{d-4}{d}R}
-\textrm{Tr}\delta^{\mu}_{\nu}\frac{\partial_t
R_k}{P_k-\frac{R}{d}}\ .\nonumber\\
\end{equation}
One can now expand to first order in $R$,
use the traces (\ref{traces}) and formula (\ref{HKasymp}) to obtain:
\begin{eqnarray}
\label{eqeh5}
\frac{d\Gamma_k}{dt} &=& \frac{1}{(4\pi)^{d/2}} \int
dx\,\sqrt{g} \Biggl\lbrace \frac{d\bl(d+1\br)}{4}Q_{\frac{d}{2}}
\left( \frac{\partial_t R_k+\eta R_k}{P_k-2\Lambda}\right) -d
Q_{\frac{d}{2}}\left(\frac{\partial_t R_k}{P_k}\right)
\nonumber\\
&& +\left[
\frac{d\bl(d+1\br)}{24}Q_{\frac{d}{2}-1}\left(\frac{\partial_t
R_k+\eta R_k}{P_k-2\Lambda}\right)
-\frac{d}{6} Q_{\frac{d}{2}-1}\left(\frac{\partial_t R_k}{P_k}\right)\right.\nonumber\\
&&-\left.\frac{d\bl(d-1\br)}{4}Q_{\frac{d}{2}}\left(\frac{\partial_t
R_k+\eta R_k}{\bl(P_k-2\Lambda\br)^2}\right)
-Q_{\frac{d}{2}}\left(\frac{\partial_t R_k}{P_k^2}\right)\right]R+O(R^2)
\Biggr\rbrace\ .
\end{eqnarray}
This derivation highlights two noteworthy facts.
The first is that the negative sign of the kinetic term for the trace
part of $h$ is immaterial.
With the chosen form for the cutoff,
any prefactor multiplying the kinetic operator in the inverse propagator
cancels out between the two factors in the r.h.s. of the ERGE.
The second fact, which we will exploit in the following,
is that the singularity occurring in the kinetic operator for
the trace part in $d=2$ (see equation (\ref{usefulforms})) is actually made harmless by a hidden
factor $d-2$ occurring in $\mathbf{U}$.
So, the final result (\ref{eqeh5})
is perfectly well defined also in two dimensions.

On the other hand, an issue that is sometimes raised in connection
with this calculation is background dependence. The calculations in
section III were done without choosing a specific background, so the
question arises whether the same can be done here. The answer is
positive, provided we do not decompose the field $h_{\mu\nu}$ into
tracefree and trace parts, and we use for the inverse propagator the
form given in the first line of (\ref{usefulforms}). Then, the
modified inverse propagator for gravitons is
\begin{equation}
\label{invpropgrav}
\mathbf{\Gamma}_k^{(2)}+\mathbf{R}_k
= Z\mathbf{K}\left(P_k(-\nabla^2)-2\Lambda\mathbf{1}+\mathbf{W}\right)\ .
\end{equation}
On a general background it is impossible to invert
$\mathbf{\Gamma}_k^{(2)}+\mathbf{R}_k$ exactly, but remembering that
$\bf W$ is linear in curvature we can expand to first order:
\begin{eqnarray}
\left(\mathbf{\Gamma}_k^{(2)}+\mathbf{R}_k\right)^{-1}
&=& \frac{1}{P_k-2\Lambda}\left[ {\bf
1}-\frac{1}{P_k-2\Lambda}{\bf W}+O(R^2) \right]
\nonumber\\
\left(\Gamma^{(2)}_{C\bar{C}}{}^\mu{}_\nu +R_k^{(gh)}{}^\mu{}_\nu\right)^{-1}&=&
\frac{1}{P_k}\left[ \delta^{\mu}_{\nu}
+\frac{1}{P_k}R^{\mu}{}_{\nu}+O(R^2)\right]\ .
\end{eqnarray}
Then the ERGE becomes, up to terms of higher order in curvature,
\begin{eqnarray}
\frac{d \Gamma_k}{dt} &=&
\frac{1}{2}\textrm{Tr}\frac{\partial_t R_k+\eta R_k}
{P_k-2\Lambda}\left[ {\bf 1}-\frac{1}{P_k-2\Lambda} {\bf W}\right]
-\textrm{Tr}\frac{\partial_t R_k}{P_k}\left[ \delta^{\mu}_{\nu}
+\frac{1}{P_k}R^{\mu}{}_{\nu}\right]\ .\nonumber
\end{eqnarray}
From here, using (\ref{HKasymp}) one arrives again at (\ref{eqeh5}).
This alternative derivation explicitly highlights the
background independence of the results.

We are now ready to extract the beta functions.
The first line of (\ref{eqeh5}) gives the beta function of $2Z\Lambda$, while the
other two lines give the beta function of $-Z$. Note the appearance
of the beta function of $Z$ in the $\eta$ terms on the r.h.s. In a
perturbative one--loop calculation such terms would be absent;
they are a result of the ``renormalization group improvement''
implicit in the ERGE.
The beta functions can be written in the form
\begin{eqnarray}
\label{generalvz}
\frac{d}{dt}\left(\frac{2\Lambda}{16\pi G}\right)&=&\frac{k^d}{16\pi}(A_1+A_2\eta)\nonumber\\
\label{generalz} -\frac{d}{dt}\left(\frac{1}{16\pi G}\right)&=&
\frac{k^{d-2}}{16\pi}(B_1+B_2\eta)
\end{eqnarray}
where $A_1$, $A_2$, $B_1$ and $B_2$ are dimensionless functions of $\Lambda$, $k$
and of $d$ which, by dimensional analysis, can also be written as functions
of $\tilde\Lambda=\Lambda k^{-2}$ and $d$.
One can solve these equations for $\frac{d\tilde\Lambda}{dt}$ and $\frac{d\tilde G}{dt}$,
obtaining
\begin{eqnarray}
\label{generallambdag}
\frac{d\tilde\Lambda}{dt}&=&
-2\tilde\Lambda+\tilde G\,\frac{A_1+ 2B_1\tilde\Lambda+\tilde G(A_1
B_2-A_2B_1)}{2(1+B_2\tilde
G)}\ ,\nonumber\\
\frac{d\tilde G}{dt}&=&(d-2)\tilde G+\frac{B_1\tilde
G^2}{1+B_2\tilde G}\ .
\end{eqnarray}
The corresponding perturbative one loop beta functions are obtained
by neglecting the $\eta$ terms in (\ref{generalvz}),
\ie\ setting $A_2=B_2=0$, and expanding $A_1$ and $B_1$ in $\tilde\Lambda$.
The leading term is
\begin{eqnarray}
\label{pertehflow}
\frac{d\tilde\Lambda}{dt}&=&
-2\tilde\Lambda+\frac{1}{2}A_1(0)\tilde G+ B_1(0)\tilde G\tilde\Lambda\ ,\nonumber\\
\frac{d\tilde G}{dt}&=&(d-2)\tilde G+B_1(0)\tilde G^2\ ,
\end{eqnarray}
where $A_1$ and $B_1$ are evaluated at $\tilde\Lambda=0$.
This flow has the same structure of the one written in (\ref{nlambdag});
we will refer to it as the ``perturbative Einstein--Hilbert flow''.
We will discuss its solution in section IV G.

The explicit form of the coefficients appearing in (\ref{generalvz}),
with the optimized cutoff, is
\begin{eqnarray}
A_1&=& \frac{16\,{\pi }\,(  d -3+ 8\,{{\tilde\Lambda }} ) }
    {{(4\pi) }^{\frac{d}{2}}\Gamma(\frac{d}{2})\,( 1 - 2\,{\tilde\Lambda } )}
\nonumber\\
A_2&=& \frac{16\pi\,( d+1)}
    {{(4\pi) }^{\frac{d}{2}}(d+2) \,{\Gamma}(\frac{d}{2})\,( 1 - 2\,{{\tilde\Lambda }} ) }
\nonumber\\
B_1&=&\frac{-4\pi(
-d^3+15 d^2-12 d+48
+(2 d^3-14 d^2-192)\tilde\Lambda
+(16 d^2+192)\tilde\Lambda^2)}
      {3{(4\pi) }^{\frac{d}{2}}\,d\,{\Gamma}(\frac{d}{2})\,( 1 - 2\,\tilde\Lambda)^2}
\nonumber\\
B_2&=&\frac{4\pi\,( d^2 - 9\,d +14 -
      2\,(d+ 1) \,(  d +2) \,\tilde\Lambda  ) }
      {3\,{(4\pi) }^{\frac{d}{2}}( d+2) \,
    {\Gamma}(\frac{d}{2})\,{( 1 - 2\,\tilde\Lambda) }^2}\ .\nonumber
\end{eqnarray}
A similar form of the beta functions had been given in \cite{LitimFPs}
in another gauge.
For the sake of clarity we write here the beta functions in four
dimensions:
\begin{eqnarray}
\label{betasIa}
\beta_{\tilde\Lambda}
&=&-2\tilde\Lambda+\frac{\tilde G}{6\pi}
\frac{3-4 \tilde\Lambda-12 \tilde\Lambda ^2-56 \tilde\Lambda ^3
+\frac{107-20 \tilde\Lambda}{12 \pi}\tilde G}
{   (1-2 \tilde\Lambda )^2 -  \frac{1+10\tilde \Lambda}{12\pi}\tilde G}
\nonumber\\
\beta_{\tilde G}
&=& 2\tilde G-\frac{\tilde G^2}{3\pi}
\frac{11-18 \tilde\Lambda +28 \tilde\Lambda^2}{(1-2 \tilde\Lambda )^2-\frac{1+10 \tilde\Lambda}{12 \pi} \tilde G }\ .
\end{eqnarray}
Note the nontrivial denominators, which in a series expansion
could be seen as resummations of infinitely many terms of perturbation theory.
They are the result of the ``RG improvement'' in the ERGE.

\subsection{Cutoff of type Ib}

This type of cutoff was introduced in \cite{Dou}. The fluctuation
$h_{\mu\nu}$ and the ghosts are decomposed into their different spin
components according to
\eq
\label{decomposition}
h_{\mu\nu}=h^T_{\mu\nu}+\nabla_{\mu}\xi_{\nu}
+\nabla_{\nu}\xi_{\mu}+\nabla_{\mu}\nabla_{\nu}\sigma-\frac{1}{
d}g_{\mu\nu}\nabla^2\sigma+\frac{1}{d}g_{\mu\nu}h. \feq and
\begin{equation}
\label{ghostdecomposition}
C^{\mu}=c^T{}^{\mu}+\nabla^{\mu} c\ \ , {\bar C}_{\mu}={\bar
c}^T_{\mu}+\nabla_{\mu} {\bar c}\ ,
\end{equation}
where $h^T_{\mu\nu}$ is tranverse and traceless, $\xi$ is a transverse vector,
$\sigma$ and $h$ are scalars, $c^T$ and $\bar c^T$ are transverse vectors, and $c$ and $\bar c$ are scalars.
These fields are subject to the following differential constraints:
\begin{equation*}
h_{\mu}^{T\mu}=0\ ; \qquad\nabla^{\nu}h_{\mu\nu}^{T}=0\ ;
\qquad\nabla^{\nu}\xi_{\nu}=0\ ; \nabla^\mu \bar c^{T}_{\mu}=0\ ;
\nabla_\mu c^{T\mu}=0\ .
\end{equation*}
Using this decomposition can be advantageous in some cases because
it can lead to a partial diagonalization of the kinetic operator and
it allows an exact inversion. This is the case for example when the
background is a maximally symmetric metric. In this section we will
therefore assume that the background is a sphere;
this is enough to extract exactly and unambiguously
the beta functions of the cosmological constant and Newton's
constant. Then the ERGE (\ref{ERGE}) can be written down for
arbitary gauge $\alpha$ and $\rho$. We refer to \cite{Dou} for more
details of the calculation. In the gauge $\alpha=1$ and without making
any approximation, the inverse propagators of the individual components are
\begin{eqnarray}
\label{kinIb}
\Gamma^{(2)}_{h^T_{\mu\nu}h^T_{\alpha\beta}}&=&
\frac{Z}{2}\left[-\nabla^2+\frac{d^2-3d+4}{d(d-1)}R-2\Lambda\right]\delta^{\mu\nu,\alpha\beta}\nonumber\\
\Gamma^{(2)}_{\xi_{\mu}\xi_{\nu}}&=&
Z\left(-\nabla^2-\frac{R}{d}\right)\left[-\nabla^2+\frac{d-3}{d}R-2\Lambda\right]g^{\mu\nu}\nonumber\\
\Gamma^{(2)}_{hh}&=&
-Z\frac{d-2}{4d}\left[-\nabla^2+\frac{d-4}{d}R-2\Lambda\right]\nonumber\\
\Gamma^{(2)}_{\sigma\sigma}&=&
Z\frac{d-1}{2d}\bl(-\nabla^2\br)\left(-\nabla^2-\frac{R}{d-1}\right)\left[-\nabla^2+\frac{d-4}{d}R-2\Lambda\right]\nonumber\\
\Gamma^{(2)}_{{\bar c}^T_{\mu}{c}^T_{\nu}}&=&
\left[\nabla^2+\frac{R}{d}\right]g^{\mu\nu}\nonumber\\
\Gamma^{(2)}_{{\bar c}{c}}&=&
-\nabla^2\left[\nabla^2+\frac{2}{d}R\right]
\end{eqnarray}
The change of variables (\ref{decomposition}) and (\ref{ghostdecomposition}) leads to
Jacobian determinants involving the operators
\begin{equation*}
J_V = -\nabla^2-\frac{R}{d},\,\, J_S =
-\nabla^2\bl(-\nabla^2-\frac{R}{d-1}\br),\,\, J_c = -\nabla^2
\end{equation*}
for the vector, scalar and ghost parts.
%Due to the appearance of derivatives in (\ref{decomposition}),
The inverse propagators (\ref{kinIb}) contain four derivative terms.
In \cite{Dou,Lauscher} this was avoided by making the field redefinitions
\begin{eqnarray}
\label{redefinitions}
\xi_{\mu}&\rightarrow&\sqrt{-\nabla^2-\frac{R}{d}}\,\xi_{\mu},\,\,
\sigma\rightarrow\sqrt{-\nabla^2}\sqrt{-\nabla^2-\frac{R}{d-1}}\sigma.
\end{eqnarray}
At the same time, such redefinitions also eliminate the Jacobians.
These field redefinitions work well for truncations containing up to
two powers of curvature, but cause poles for higher truncations as
the heat kernel expansion will involve derivatives of the trace
arguments. Therefore, in later sections we will not perform the
field redefinitions, but treat the Jacobians instead as further
contribution to the ERGE by exponentiating them, introducing
appropriate auxiliary fields and a cutoff on these variables.
Here we describe the result of performing the field redefinitions.
The ERGE is
\begin{eqnarray}
\label{withred}
 \frac{d \Gamma_k}{dt}
&=& \frac{1}{2} \textrm{Tr}_{(2)} \frac{\partial_t R_k+\eta
R_k}{P_k-2\Lambda+\frac{d^2-3 d+4}{d(d-1)}R}
+\frac{1}{2} \textrm{Tr}'_{(1)}
\frac{\partial_t R_k+\eta R_k}{P_k-2  \Lambda+\frac{d-3}{d}R}\nonumber\\
&&+\frac{1}{2} \textrm{Tr}_{(0)} \frac{   \partial_t R_k+\eta R_k
}{P_k - 2 \Lambda + \frac{d-4}{d} R}
+\frac{1}{2} \textrm{Tr}''_{(0)} \frac{   \partial_t R_k+\eta
R_k}{P_k - 2 \Lambda + \frac{d-4}{d} R}\nonumber\\
&&- \textrm{Tr}_{(1)} \frac{\partial_t R_k}{P_k-\frac{R}{d}}
-\textrm{Tr}'_{(0)} \frac{\partial_t R_k}{ P_k- \frac{2R}{d}}\ .
\end{eqnarray}
The first term comes from the spin--2, transverse traceless
components, the second from the spin--1 transverse vector, the third
and fourth from the scalars $h$ and  $\sigma$.
The last two contributions come from the transverse and
longitudinal components of the ghosts. A prime or a double prime
indicate that the first or the first and second eigenvalues have to
be omitted from the trace. The reason for this is explained in Appendix B.

Expanding the denominators to first order in $R$, but keeping
the exact dependence on $\Lambda$ as in the case of a type Ia
cutoff, and using the formula (\ref{HKasymp}), one obtains
\begin{eqnarray}
\label{eqEHIb} \frac{d \Gamma_k}{dt} &=&\frac{1}{(4\pi)^{d/2}}\int
dx\,\sqrt{g}\Biggl\lbrace
\frac{d(d+1)}{4}Q_{\frac{d}{2}}\left(\frac{\partial_t R_k+\eta
R_k}{P_k-2 \Lambda }\right)
-d Q_{\frac{d}{2}}\left(\frac{\partial_t R_k}{P_k} \right)
\nonumber\\
&&+ R\left[
-\frac{d^4-2d^3-d^2-4d+2}{4d(d-1)}Q_{\frac{d}{2}}\left(\frac{\partial_t
R_k+\eta R_k}{(P_k-2 \Lambda)^2 }\right)
-\frac{d+1}{d} Q_{\frac{d}{2}}\left(\frac{\partial_t R_k}{P_k^2}\right)
\right.\\
&&+\left.\frac{d^4-13d^2-24d+12}{24d(d-1)}Q_{\frac{d}{2}-1}\left(\frac{\partial_t
R_k+\eta R_k}{P_k-2 \Lambda }\right) -\frac{d^2-6}{6d}
Q_{\frac{d}{2}-1}\left(\frac{\partial_t R_k}{P_k} \right) \right]
+O(R^2)
\Biggl\rbrace .\nonumber
\end{eqnarray}
In principle in two dimensions one has to subtract the contributions
of some excluded modes. However, using the results in Appendix B,
the contributions of these isolated modes turn out to cancel. Thus,
the ERGE is continuous in the dimension also at $d=2$.

The beta functions have again the form (\ref{generallambdag});
the coefficients $A_1$ and $A_2$ are the same as for the type Ia cutoff
but now the coefficients $B_1$ and $B_2$ are
\begin{eqnarray}
B_1&=&4\pi
\Bigl(d(d-1)(d^3-15 d^2-36)+24
-2(d^5-8d^4-5d^3-72d^2-36d+96)\tilde\Lambda
\nonumber\\
&&
-16(d-1)(d^3+6d+12)\tilde\Lambda^2\Bigr)
\Big/
3(4\pi)^{\frac{d}{2}}d^2(d-1)\Gamma\left(\frac{d}{2}\right)(1-2\tilde\Lambda)^2
\nonumber\\
B_2&=&4\pi\frac{ d(d^4-10d^3 +11d^2-38d+12)  -
      2(d +2)(d^4- 13\,d^2 -24d+12) \,\tilde\Lambda }
      {3(4\pi)^{\frac{d}{2}}\,
    ( 2+d) ( d -1) d^2\,{\Gamma}(\frac{d}{2})
    ( 1 - 2\tilde\Lambda)^2}
\nonumber
\end{eqnarray}
In four dimensions, the beta functions are
\begin{eqnarray}
\label{betasIbwfr}
\beta_{\tilde\Lambda}
&=&-2\tilde\Lambda+\frac{1}{24\pi}
\frac{(12-33 \tilde{\Lambda }+20 \tilde{\Lambda}^2-200 \tilde{\Lambda }^3)\tilde{G}
+\frac{467-572 \tilde{\Lambda }}{12\pi} \tilde{G}^2}
{    (1-2 \tilde{\Lambda })^2-\frac{ 29-9 \tilde{\Lambda}}{72\pi}\tilde{G}}
\nonumber\\
\beta_{\tilde G}
&=&2\tilde G-\frac{1}{24\pi}
\frac{
  (105-212 \tilde{\Lambda }+200 \tilde{\Lambda }^2)\tilde{G}^2}
{  (1-2\tilde{\Lambda })^2
-\frac{29-9 \tilde{\Lambda }}{72\pi}\tilde{G}}\ .
\end{eqnarray}

In order to appreciate the numerical differences between this
procedure and the one where the fields $\xi_\mu$ and $\sigma$
are not redefined as in (\ref{redefinitions}),
we report in Appendix D the results of the alternative formulation.

\subsection{Cutoff of type II}

Let us define the following operators acting on gravitons and on ghosts:
\begin{eqnarray}
\label{e2}
\Delta_{2} &=& -\nabla^2+\mathbf{W}\\
\Delta_{(gh)}&=& -\nabla^2-\mathrm{Ricci}\ .
\end{eqnarray}
The traces of the $\mathbf{b}_2$--coefficients of the heat--kernel expansion for these operators are
\begin{eqnarray}
\textrm{tr}\mathbf{b}_2(\Delta_{2}) &=&
\textrm{tr}\bl(\frac{R}{6}\mathbf{1}-\mathbf{W}\br)=\frac{d(7-5d)}{12}R\nonumber\\
\textrm{tr}\mathbf{b}_2(\Delta_{gh}) &=&
\textrm{tr}\bl(\frac{R}{6}\mathbf{1}+\mathrm{Ricci}\br)=\frac{d+6}{6}R\ .\nonumber
\end{eqnarray}
The type II cutoff is defined by the choice
\begin{eqnarray*}
\mathbf{R}_{k}&=&Z\mathbf{K}R_{k}(\Delta_2)\nonumber\\
R_k^{(gh)}{}^\mu{}_\nu&=&\delta_{\,\,\nu}^{\mu}R_{k}(\Delta_{(gh)})\ ,
\end{eqnarray*}
which results in
\begin{eqnarray}
\mathbf{\Gamma}^{(2)}_k+\mathbf{R}_k
&=&
Z\mathbf{K}\left(P_k(\Delta_{2})-2\Lambda \right)\nonumber\\
\label{invpropghost} \Gamma^{(2)}_{C\bar C}+R^{(gh)}_k
&=&P_{k}(\Delta_{(gh)})
\end{eqnarray}
and
\begin{equation}
\frac{d{\mathbf{R}}_{k}}{dt}=Z\mathbf{K}
\left(\partial_t{R}_{k}(\Delta_{2})+\eta R_{k}(\Delta_{2})\right) \ .\nonumber
\end{equation}
Collecting all terms and evaluating the traces leads to
\begin{eqnarray}
\label{eqEHII} \frac{d{\Gamma}_k}{dt}\!  &=&
\frac{1}{2}\textrm{Tr} \frac{\partial_t R_k(\Delta_{2})+\eta
R_k(\Delta_{2})}{P_k(\Delta_{2})-2\Lambda} -
\textrm{Tr}\frac{\partial_t
R_k(\Delta_{(gh)})}{P_k(\Delta_{(gh)})}
\nonumber\\
&=& \frac{1}{(4\pi)^{d/2}}\int dx\,\sqrt{g}\left\lbrace
\frac{d(d+1)}{4}Q_{\frac{d}{2}}\left(\frac{\partial_t R_k+\eta
R_k}{P_k-2\Lambda} \right)-d\, Q_{\frac{d}{2}}\left(\frac{\partial_t
R_k}{P_k}\right)
\right.\nonumber\\
&&\left.
+\left[\frac{d(7-5d)}{24}\,Q_{\frac{d}{2}-1}\left(\frac{\partial_t
R_k+\eta R_k}{P_k-2\Lambda}\right)
-\frac{d+6}{6}Q_{\frac{d}{2}-1}\left(\frac{\partial_t
R_k}{P_k}\right)\right]R+O(R^2)\right\rbrace\ .
\end{eqnarray}
The beta functions are again of the form (\ref{generallambdag}),
and the coefficients $A_1$ and $A_2$ are the same as in the case
of the cutoffs of type I. The coefficients $B_1$ and $B_2$ are now
\begin{eqnarray}
B_1&=&-\frac{4\pi( 5d^2 -3d+24- 8( d+6)\tilde\Lambda)}
  {3(4\pi)^{\frac{d}{2}}{\Gamma}(\frac{d}{2})( 1 - 2\,{{\tilde\Lambda }} ) }
\nonumber\\
B_2&=&-\frac{4\pi( 5d -7) }
  {3(4\pi)^{\frac{d}{2}}{\Gamma}(\frac{d}{2})( 1 - 2\tilde\Lambda ) }\nonumber
\end{eqnarray}
In four dimensions, the beta functions are
\begin{eqnarray}
\label{betasII}
\beta_{\tilde\Lambda}
&=&-2\tilde\Lambda+\frac{1}{6 \pi}
\frac{ (3-28 \tilde\Lambda+84 \tilde\Lambda ^2-80 \tilde\Lambda ^3 )\tilde  G+ \frac{191-512 \tilde\Lambda}{12\pi}\tilde G^2}
{  (1-2
  \tilde \Lambda ) (  1-2\tilde \Lambda -\frac{13}{12 \pi}\tilde G)}
\nonumber\\
\beta_{\tilde G}
&=& 2\tilde G-\frac{1}{3 \pi }
\frac{ (23-20 \tilde\Lambda )\tilde G^2}{  (1-2 \tilde\Lambda )-\frac{13}{12 \pi}\tilde G}\ .
\end{eqnarray}

\subsection{Cutoff of type III}

Finally we discuss the spectrally adjusted, or type III cutoff. This
consists of defining the cutoff function as a function of the whole
inverse propagator $\Gamma_k^{(2)}$, only stripped of the overall
wave function renormalization constants. In the case of the graviton,
$\mathbf{\Gamma}_k^{(2)}=Z\mathbf{K}(\Delta_2-2\Lambda\mathbf{1})$
while for the ghosts $\Gamma^{(2)}_{C\bar C}=\Delta_{gh}$, where
$\Delta_2$ and $\Delta_{gh}$ were defined in (\ref{e2}). Type III
cutoff is defined by the choice
\eq
\mathbf{R}_{k}=Z\mathbf{K}R_k(\Delta_{2}-2\Lambda)
\feq
for gravitons, while for ghosts it is the same as in the case of type II
cutoff. Since the operator in the graviton cutoff now contains the
coupling $\Lambda$, the derivative of the graviton cutoff now
involves an additional term:
\begin{eqnarray}
\frac{d\mathbf{R}_k}{dt} &=&Z\mathbf{K}\left(
\partial_t{R}_{k}(\Delta_{2}-2\Lambda)+\eta R_{k}(\Delta_{2}-2\Lambda)-2 R_k'(\Delta_{2}-2\Lambda)\partial_t\Lambda
\right)
\end{eqnarray}
where $R_k'$ denotes the partial derivative of $R_k(z)$ with respect
to $z$. Note that the use of the chain rule in the last term is only
legitimate if the $t$-derivative of the operator appearing as the
argument of $R_k$ commutes with the operator itself. This is the
case for the operator $\Delta_2-2\Lambda$, since its $t$-derivative
is proportional to the identity. The modified inverse propagator is then simply
\begin{equation*}
\mathbf{\Gamma}^{(2)}_k+\mathbf{R}_k
=Z\mathbf{K}P_k(\Delta_{2}-2\Lambda)
\end{equation*}
for gravitons, while for ghosts it is again given by equation (\ref{invpropghost}).
Collecting,
\begin{equation}
\label{typethree}
\frac{d\Gamma_k}{dt}\!  =
\frac{1}{2}\textrm{Tr}
\frac{\partial_t R_k(\Delta_{2}-2\Lambda)+\eta R_k(\Delta_{2}-2\Lambda)
-2 R_k'(\Delta_{2}-2\Lambda)\partial_t\Lambda}{P_k(\Delta_{2}-2\Lambda)} -
\textrm{Tr}\frac{\partial_t
R_k(\Delta_{(gh)})}{P_k(\Delta_{(gh)})}\ .
\end{equation}
The traces over the ghosts are exactly as in the case of a cutoff of type II.
As in previous cases, one should now proceed to evaluate the trace over the tensors using
equation (\ref{HKasymp}) and the heat kernel coefficients
of the operator $\Delta_2-2\Lambda$.
However, the situation is now more complicated
because the heat kernel coefficients $B_{2k}(\Delta_2-2\Lambda)$ contain terms
proportional to $\Lambda^k$ and $\Lambda^{k-1}R$, all of which contribute
to the beta functions of $2\Lambda Z$ and $-Z$.
This is in contrast to the calculations with cutoffs of type I and II,
where only the first two heat kernel coefficients contributed
to the beta functions of $2\Lambda Z$ and $-Z$.
In order to resum all these contributions, one can proceed as follows.
We define the function $W(z)=\frac{\partial_t R_k(z)+\eta R_k(z)-2 R_k'(z)\partial_t\Lambda}{P_k(z)}$
and the function $\bar W(z)=W(z-2\Lambda)$.
It is shown explicitly in the end of Appendix A (equation (\ref{qfunctions}) and following)
that $\mathrm{Tr}W=\mathrm{Tr}\bar W$.
Then, the terms without $R$ and the terms linear in $R$ (which give the
beta functions of $2\Lambda Z$ and $-Z$ respectively)
correspond to the first two lines in (\ref{tracedelta}).
In this way we obtain
\begin{eqnarray}
\label{gammatypethree}
\frac{d{\Gamma}_k}{dt}\!\!
&=& \!\!\frac{1}{(4\pi)^{d/2}}\int dx\,\sqrt{g}\left\lbrace
\frac{d(d+1)}{4}\sum_{i=0}^\infty \frac{(2\Lambda)^i}{i!}
Q_{\frac{d}{2}-i}\left(\frac{\partial_t R_k+\eta R_k-2\partial_t\Lambda R'_k}{P_k} \right)
-dQ_{\frac{d}{2}}\left(\frac{\partial_t R_k}{P_k}\right)
\right.\nonumber\\
&&\left.
\!\!\!\!\!\!\!\!\!\!\!\!\!\!\!\!\!\!\!\!+\frac{d(7-5d)}{24}R\,
\sum_{i=0}^\infty \frac{(2\Lambda)^i}{i!}
Q_{\frac{d}{2}-1-i}\left(\frac{\partial_t R_k+\eta R_k-2\partial_t\Lambda R'_k}{P_k}\right)
-\frac{d+6}{6}Q_{\frac{d}{2}-1}\left(\frac{\partial_t R_k}{P_k}\right)R\right\rbrace\ .
\end{eqnarray}
The remarkable property of the optimized cutoff is that in even dimensions the sums in those
expressions contain only a finite number of terms;
in odd dimensions the sum is infinite but can still be evaluated analytically.
Using the results (\ref{qoptpos},\ref{qoptzero},\ref{qoptneg},\ref{qoptposR},\ref{qoptzeroR},\ref{qoptnegR},\ref{qoptone})
the first sum in (\ref{gammatypethree}) gives
\begin{equation}
\label{firstsum} \frac{1}{(4\pi)^{d/2}}\frac{d+1}{2}
\frac{(k^2+2\Lambda)^{d/2}}{\Gamma(d/2)}
\left(2+\frac{\eta}{\frac{d}{2}+1}\frac{k^2+2\Lambda}{k^2}+2\frac{\partial_t\Lambda}{k^2}\right)
\int dx\,\sqrt{g}
\end{equation}
whereas the second sum gives
\begin{equation}
\label{secondsum} \frac{1}{(4\pi)^{d/2}}\frac{d(7-5d)}{24}\,
\frac{(k^2+2\Lambda)^{\frac{d-2}{2}}}{\Gamma(d/2)}
\left(2+\frac{\eta}{d/2}\frac{k^2+2\Lambda}{k^2}+2\frac{\partial_t\Lambda}{k^2}\right)
\int dx\,\sqrt{g}R
\end{equation}
This resummation can actually be done also with other cutoffs.
An alternative derivation of these formulae, based on the proper time form of the ERGE
is given in Appendix C.

The beta functions cannot be written in the form (\ref{generallambdag})
anymore, because of the presence of the derivatives of $\Lambda$ on the
right hand side of the ERGE. Instead of (\ref{generalvz}) we have
\begin{eqnarray}
\label{newgeneralvz}
\frac{d}{dt}\left(\frac{2\Lambda}{16\pi G}\right)&=&
\frac{k^d}{16\pi}(A_1+A_2\eta+A_3\partial_t\tilde
\Lambda)\ ,\nonumber\\
-\frac{d}{dt}\left(\frac{1}{16\pi G}\right)&=&
\frac{k^{d-2}}{16\pi}(B_1+B_2\eta+B_3\partial_t\tilde \Lambda)\ ,
\end{eqnarray}
where
\begin{eqnarray}
A_1&=&\frac{16\pi ( -4 +
      (d +1) ( 1 + 2\tilde\Lambda )^{\frac{d}{2}+1}) }
      {(4\pi)^{\frac{d}{2}}\Gamma(   \frac{d}{2})}
\nonumber\\
A_2&=&\frac{16\pi(d+1)
    ( 1 + 2\tilde\Lambda )^{\frac{d}{2}+1}}
    {(4\pi)^{\frac{d}{2}}(d +2) \Gamma(\frac{d}{2})}
\nonumber\\
A_3&=& \frac{16\pi(d+1) (1+2 \tilde{\Lambda })^{\frac{d}{2}}}
{(4\pi)^{\frac{d}{2}}\Gamma \left(\frac{d}{2}\right)}
\nonumber\\
B_1&=& \frac{4\pi(-4(d+6)+d(7-5d)(1+2\tilde\Lambda)^{\frac{d}{2}}) }
      {3(4\pi)^{\frac{d}{2}}\Gamma(\frac{d}{2})}
\nonumber\\
B_2&=&\frac{4\pi(7 - 5d)
    {( 1 + 2\tilde\Lambda) }^{\frac{d}{2}}}
    {3(4\pi)^{\frac{d}{2}}\Gamma(\frac{d}{2})}
\nonumber\\
B_3&=& \frac{4\pi d (7-5 d)  (1+2 \tilde{\Lambda })^{\frac{d}{2}-1}}
{3(4\pi)^{\frac{d}{2}}\Gamma \left(\frac{d}{2}\right)}
\nonumber
\end{eqnarray}
Solving (\ref{newgeneralvz}) for $d\tilde\Lambda/dt$ and $d\tilde G/dt$
gives
\begin{eqnarray}
\frac{d\tilde\Lambda}{dt}&=& -2\tilde\Lambda
+\frac{(A_1+2(B_1-A_3)\tilde\Lambda-4B_3\tilde\Lambda^2)\tilde G
+(A_1B_2-A_2B_1+2(A_2B_3-A_3B_2)\tilde\Lambda)\tilde G^2}
{2+(2B_2-A_3-2B_3\tilde\Lambda)\tilde G+(A_2B_3-A_3B_2)\tilde
G^2}\nonumber
\\
\label{newtonIII}
\frac{d\tilde G}{dt}&=& (d-2)\tilde G+
\frac{2(B_1-2B_3\tilde\Lambda)\tilde G^2+(A_1B_3-A_3B_1)\tilde G^3}
{2+(2B_2-A_3-2B_3\tilde\Lambda)\tilde G+(A_2B_3-A_3B_2)\tilde G^2}\ .
\end{eqnarray}
In four dimensions, the beta functions are
\begin{eqnarray}
\label{betasIII}
\beta_{\tilde\Lambda}
&=&-2\tilde\Lambda+\frac{1}{6\pi}
\frac{   (3+14 \tilde\Lambda+8 \tilde\Lambda ^2 )\tilde G+  \frac{(1+2 \tilde\Lambda )^2}{12\pi}
 (191-60 \tilde\Lambda-260 \tilde\Lambda ^2 )\tilde G^2}{1 -\frac{1}{12 \pi}
(43+120 \tilde\Lambda+68 \tilde\Lambda ^2 )\tilde G+\frac{65}{72\pi^2} (1+2 \tilde\Lambda )^4\tilde G^2}
\nonumber\\
\beta_{\tilde G}
&=& 2\tilde G-\frac{1}{3\pi}
\frac{(23+26 \tilde\Lambda )\tilde G^2- \frac{ 51+152 \tilde\Lambda +100 \tilde\Lambda ^2}{\pi}\tilde G^3}
{1-\frac{1}{12 \pi }  (43+120 \tilde\Lambda+68 \tilde\Lambda ^2 )\tilde G
+\frac{65}{72 \pi ^2} (1+2 \tilde\Lambda )^4\tilde G^2}\ .
\end{eqnarray}

\subsection{Gravity without cosmological constant}

\begin{figure}
{\resizebox{1\columnwidth}{!}
{\includegraphics{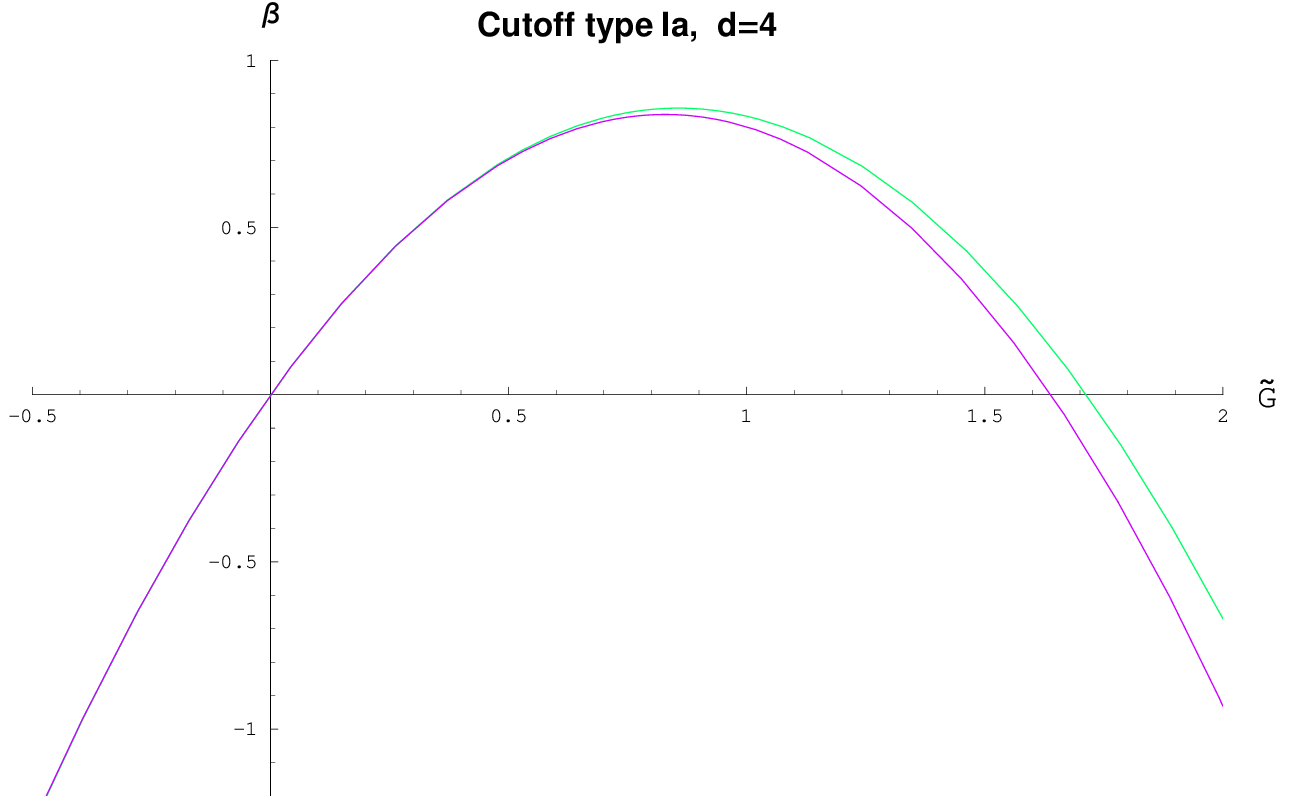} \includegraphics{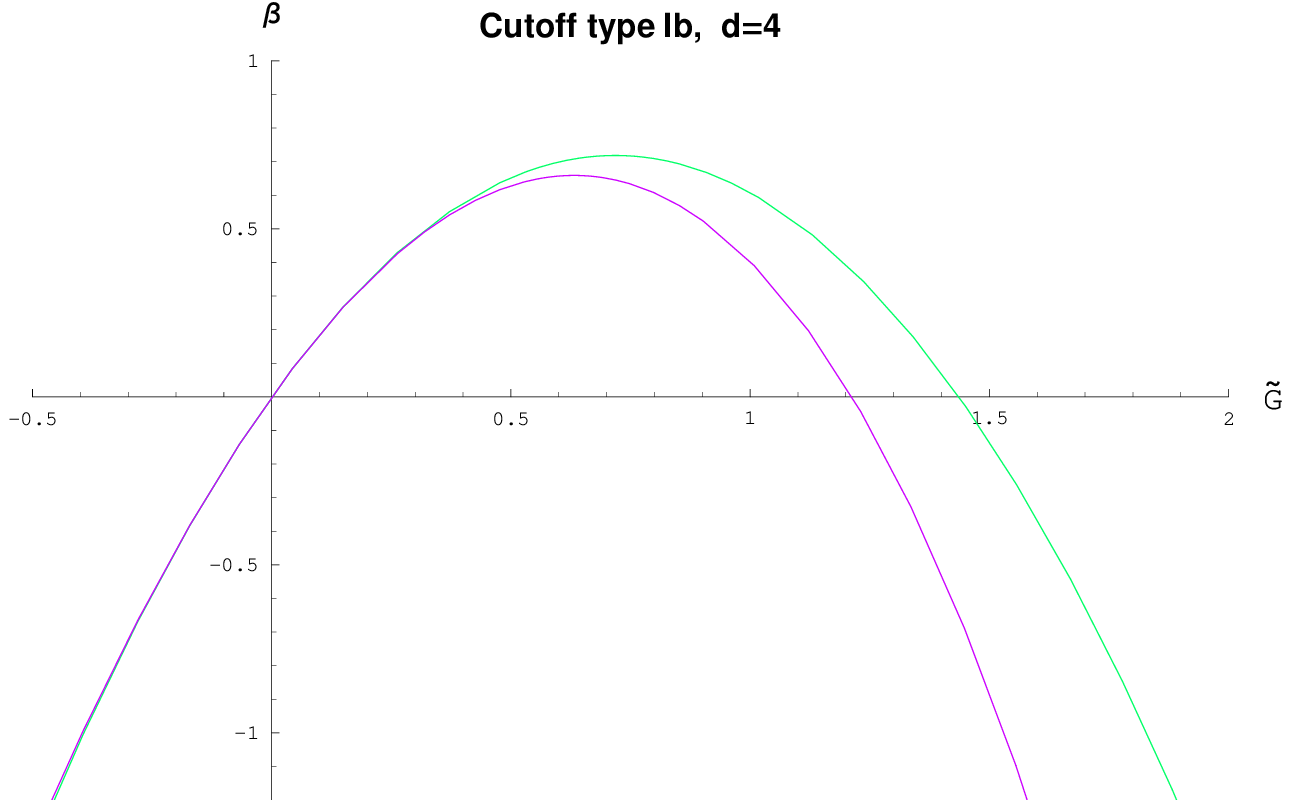}}
    \caption{\label{fig:betaI}The beta function of $\tilde G$ with $\tilde\Lambda=0$ and
cutoffs of type Ia and Ib. The perturbative one loop result in light gray, the RG--improved
one in darker color.}}
\end{figure}

\begin{figure}
%[t]\center
{\resizebox{1\columnwidth}{!}
{\includegraphics{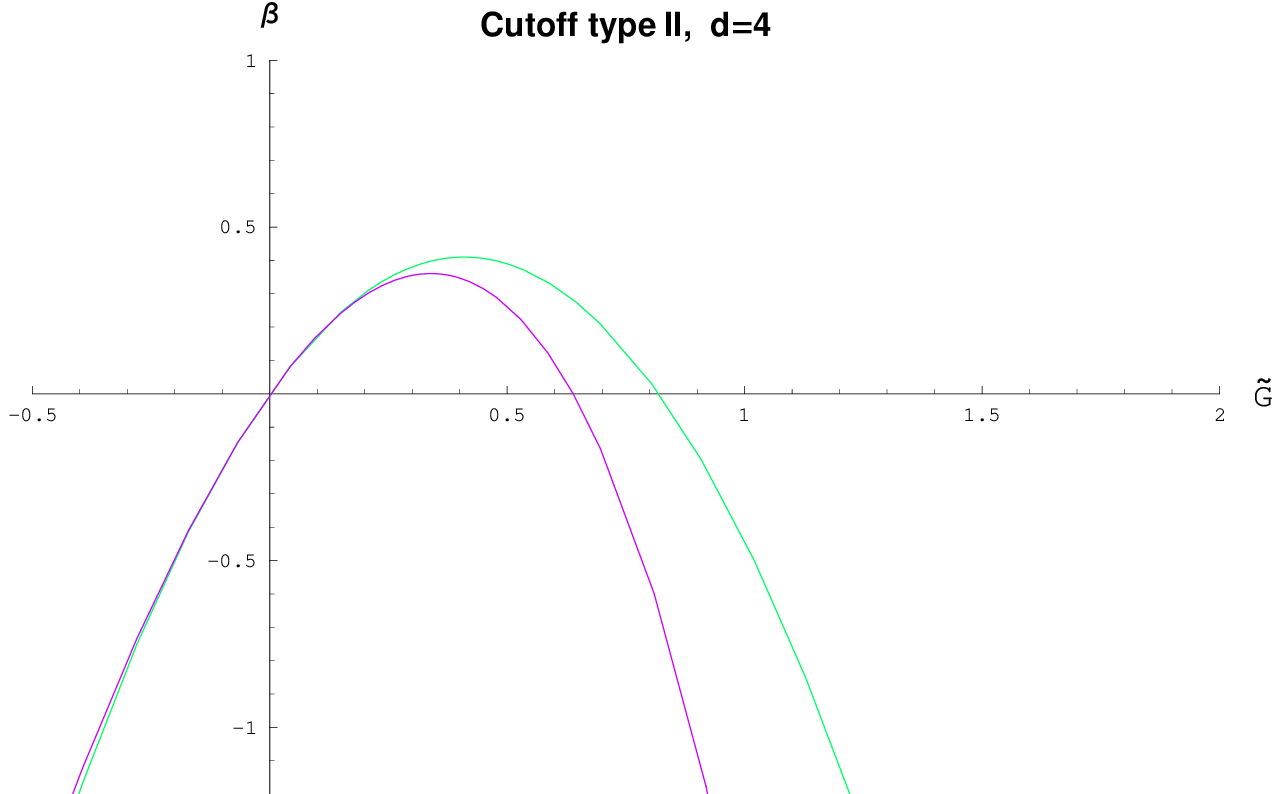} \includegraphics{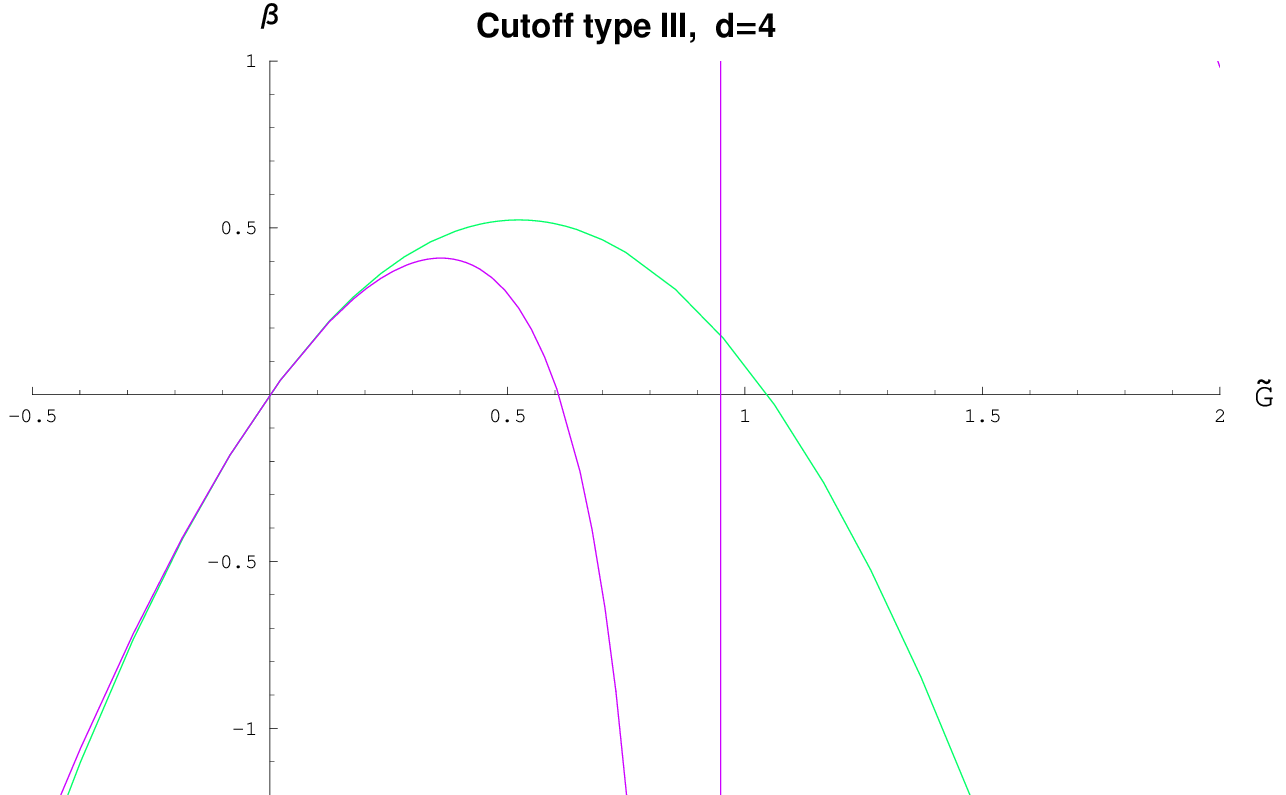}}
    \caption{\label{fig:betaII}The beta function of $\tilde G$ with $\tilde\Lambda=0$ and
cutoffs of type II and III. The perturbative one loop result in light gray,
the RG--improved one in darker color.
In the case of cutoff type III the vertical line is the asymptote of the RG--improved beta function.
}}
\end{figure}

Before discussing the general case, it is instructive to consider the
case $\Lambda=0$.
In terms of the dimensionless coupling $\tilde G$ the beta functions
with cutoffs of type I and II have the form (\ref{generallambdag}),
where the constants $B_1$ and $B_2$ are evaluated at $\tilde\Lambda=0$.
The cutoff of type III leads instead to the more complicated beta function
(\ref{newtonIII}), with $\tilde\Lambda=0$.
The beta function of $\tilde G$ in four dimensions is shown in figures
\ref{fig:betaI} and \ref{fig:betaII} (blue, dark lines) for different cutoff types.
It always has a Gau\ss ian FP in the origin and a nontrivial FP at $\tilde
G_*=-\frac{d-2}{B_1+(d-2)B_2}$. The Gau\ss ian FP is always
UV--repulsive (positive slope) whereas the non--Gau\ss ian FP is
UV--attractive. In a theory with a single coupling constant such as
this, the slope of the beta function at the nontrivial FP is related
to $\nu$, the mass critical exponent.
For type I and II cutoffs it is given by
\begin{equation}
\vartheta=\frac{1}{\nu}=
-\frac{\partial\beta_{\tilde G}}{\partial\tilde G}\Bigg|_*=
(d-2)\left(1+(d-2)\frac{B_2}{B_1}\right)\ .
\end{equation}
The leading, classical term is universally equal to $d-2$, the correction is scheme--dependent.


It is instructive to compare these ``RG improved'' results to the
respective ``one loop approximations''.
As discussed in section 2, the ``unimproved'' perturbative one--loop beta functions
are obtained by neglecting the derivatives of the couplings occurring in the r.h.s.
of the ERGE, in practice setting $B_2=0$:
\begin{equation}
\label{oneloopforg}
\beta_{\tilde G}|_{\textrm 1 loop}=(d-2)\tilde G+B_1 \tilde G^2\ .
\end{equation}
These beta functions are shown as the gray, light lines in figures
\ref{fig:betaI} and \ref{fig:betaII}. Since $B_1<0$, they are
inverted parabolas, with a Gau\ss ian FP in the origin and a
nontrivial FP at $\tilde G_*=-(d-2)/B_1$. The slope at the
nontrivial FP is always the opposite of the one at the Gau\ss ian
FP, and therefore equal to $-2$. Note that both (\ref{betaepsilon})
and (\ref{betabjerrum}) are of this form, for specific values of the constant $B_1$.
\begin{table}
\begin{center}
\begin{tabular}{l|l|l|l|l|l|l}  %\hline
{\rm Approximation}     & \ $B_1(0)$\ &\ $B_2(0)$\ &\ $\tilde G_*$\ &\ $\vartheta$\ \\
\hline
$\epsilon$ exp., leading order  & & &0.158&2.000\\
Ia  &$-\frac{11}{3\pi}$&$-\frac{1}{12\pi}$&1.639&2.091\\
Ia - 1 loop & & &    1.714      & 2.000 \\
Ib with f.r.  &$-\frac{35}{8\pi}$&$-\frac{29}{72\pi}$&1.213&2.368\\
Ib with f.r. - 1 loop & & &   1.436      & 2.000 \\
Ib without f.r.  &$-\frac{131}{45\pi}$&$-\frac{61}{720\pi}$&2.040&2.116\\
Ib without f.r. - 1 loop & & &   2.158    & 2.000\\
II  &$-\frac{23}{3\pi}$&$-\frac{13}{12\pi}$&0.954&2.565\\
II - 1 loop & & & 0.820     & 2.000\\
III &$-\frac{23}{3\pi}$&$-\frac{13}{12\pi}$& 0.424 & 3.870\\
III - 1 loop & & & 0.820   & 2.000
\end{tabular}
\end{center}
\caption{The fixed point for Einstein's theory in $d=4$ without cosmological constant.
The leading beta function for the $\epsilon$ expansion is derived in section IVF;
the one for the cutoff of type Ib without field redefinitions is given in Appendix D.}
\label{table1}
\end{table}
Because $B_1<B_2<0$, the ``RG improved'' FP occurs always at smaller values of $G_*$
than the corresponding perturbative one.
The Gau\ss ian FP is always UV--repulsive
(positive slope) whereas the non--Gau\ss ian FP is UV--attractive.
In table I we report the numerical values of the coefficients $B_1$ and $B_2$
and the position of the FP and the critical exponent in four dimensions.

One can see from figure \ref{fig:betaI} that for cutoffs of type I
the one--loop approximation is quite good up to the nontrivial FP
and a little beyond. For larger values of $\tilde G$ the effect of
the denominator in (\ref{generallambdag}) becomes important; the
beta function deviates strongly from the one--loop approximation and
has a negative pole at $\tilde G=-1/B_2>\tilde G_*$.
When one considers type II
and III cutoffs, the RG improved beta functions deviate from the
perturbative one loop beta functions sooner and the effect is
stronger; in the case of the type III cutoff the pole occurs before
the one loop beta function has the zero. We see that the RG
improvement leads to stronger effects if more terms of the operator
are taken into the definition of the cutoff. It is interesting to
observe that the RG improvement always brings the nontrivial fixed
point closer to the perturbative regime.
Since at low energies $\tilde G$ is close to zero, the region
of physical interest is $0<\tilde G<\tilde G_*$.
Thus, the pole in the beta functions at finite
values of $\tilde G$ should not worry us.

\subsection{The $\epsilon$ expansion.}

\begin{figure}
%[t]\center
{\resizebox{0.7\columnwidth}{!}
{\includegraphics{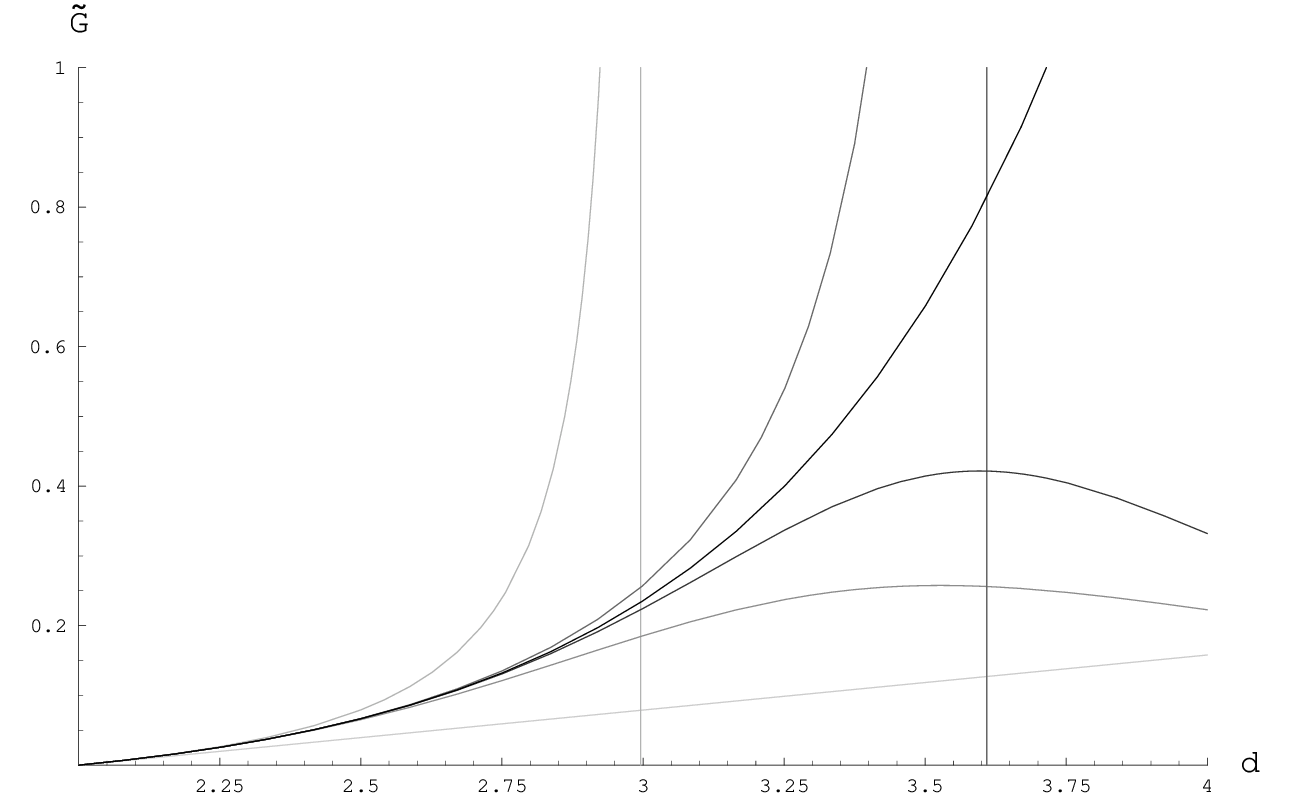}}
    \caption{\label{fig:epsilon}The position of $\tilde G_*$ as a function of $d$
in the one loop approximation with cutoff of type Ia (central line in black).
The other lines are the first five orders of the $\epsilon$ expansion
(order $n$ means the beta function has been expanded to order $n$ in $\epsilon$.)
The zeroth order is the light gray straight line. Higher orders are represented by
darker shades of gray. Order 1 and order 3 have singularities at $d\approx 2.998$
and $d\approx 3.609$.
}}
\end{figure}

Before discussing the four dimensional case, it is useful and instructive to
consider the situation in arbitrary dimensions.
In particular, this will allow us to compare the results of the ERGE with those obtained
in the $\epsilon$--expansion.
We have seen within the approximations of section 2 that in $d$ dimensions the
beta function of the dimensionless coefficient of the $R^{d/2}$ term is scheme--independent.
Therefore in two dimensions one expects the beta function of Newton's constant,
or at least its leading term, to be scheme--independent.
This is confirmed by formula (\ref{oneloopforg}) for the one loop beta function,
and the results listed in sections 4.1 to 4.4:
\begin{equation}
\label{famous}
-B_1\big|_{d=2}=\frac{38}{3}
\end{equation}
for all types of cutoff.
On the other hand the coefficient $B_2$ is scheme--dependent.
This mirrors the well known fact that in perturbation theory the leading
term of the beta functions of dimensionless couplings is scheme--independent
and higher loop corrections are not.

The beta function (\ref{generallambdag}) with $\tilde\Lambda=0$
can be solved exactly for any $d$.
The nontrivial FP occurs at:
\begin{equation}
\label{puregfp}
\tilde G_*=-\frac{d-2}{B_1+(d-2)B_2}\ .
\end{equation}
Knowing the solution in any dimension we can now check a
posteriori how good the $\epsilon$--expansion is. To this end we
have to expand the beta function in powers of $\epsilon=d-2$ and
look for FPs of the approximated beta functions.

The leading term of the $\epsilon$ expansion consists in retaining
only the constant, scheme--independent term (\ref{famous}). Then the beta
function is given by equation (\ref{betaepsilon}) and the fixed
point occurs (for any cutoff type) at
$$
\tilde G_*=\frac{3}{38}\epsilon\ .
$$
We see from table I that this value is quite small compared to the direct calculations
in four dimensions.
For the higher orders of this expansion
it is necessary to specialize the discussion to a specific
type of cutoff. For definiteness we will consider the case of a type Ia cutoff,
for which (setting $\tilde\Lambda$ to zero)
\begin{equation}
\label{ddependence} B_1=
\frac{4\pi\left(d^3-15d^2+12d-48\right)}{3(4\pi)^{d/2}d\Gamma (\frac{d}{2})}\ .
\end{equation}
To order $\epsilon$ the beta function is
$$
\beta_{\tilde G}=-\frac{38}{3}\tilde G^2
+\left[\tilde G+
\left(\frac{1}{3}-\frac{19\gamma}{3}+\frac{38\log 2}{3}+\frac{19\log\pi}{3}\right)
\tilde G^2\right]\epsilon+O(\epsilon^2)\ .
$$
Whereas to leading order the solution exists for all $d$, to this
order the solution has a singularity for $\epsilon\approx2.99679$.
The occurrence of such singularities at finite values of $\epsilon$
is expected. The one loop solution together with some of its
approximants is plotted in figure \ref{fig:epsilon} for dimensions
$2<d<4$. We see that when the beta function is expanded to even
order in $\epsilon$ ($n=0,2,4$ in the figure) the $\epsilon$
expansion significantly underestimates the value of $\tilde G_*$ in
$d=4$ whereas for odd order ($n=1,3$ in the figure) it has a
positive pole at some value of the dimension.
On the other hand from equations (\ref{puregfp}) and (\ref{ddependence}) one sees
that for large $d$, $\tilde G_*$ grows faster than exponentially.

\begin{figure}
%[t]\center
{\resizebox{1\columnwidth}{!}
{\includegraphics{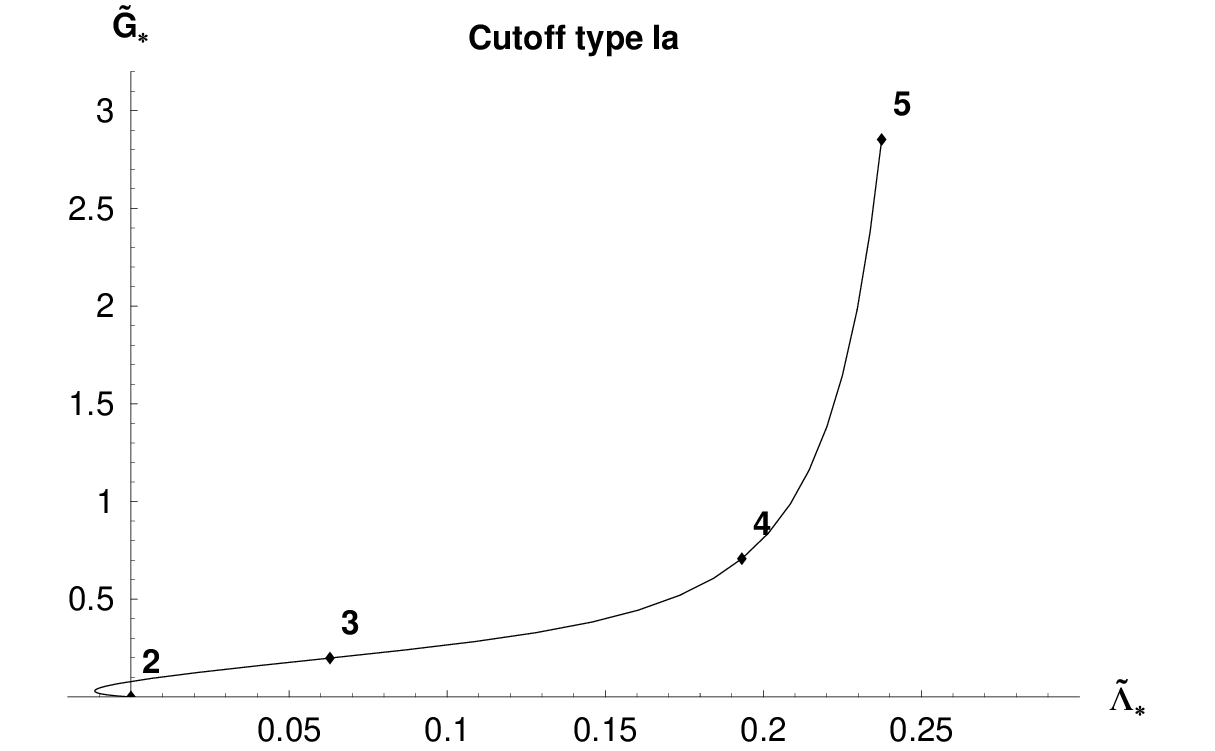}\includegraphics{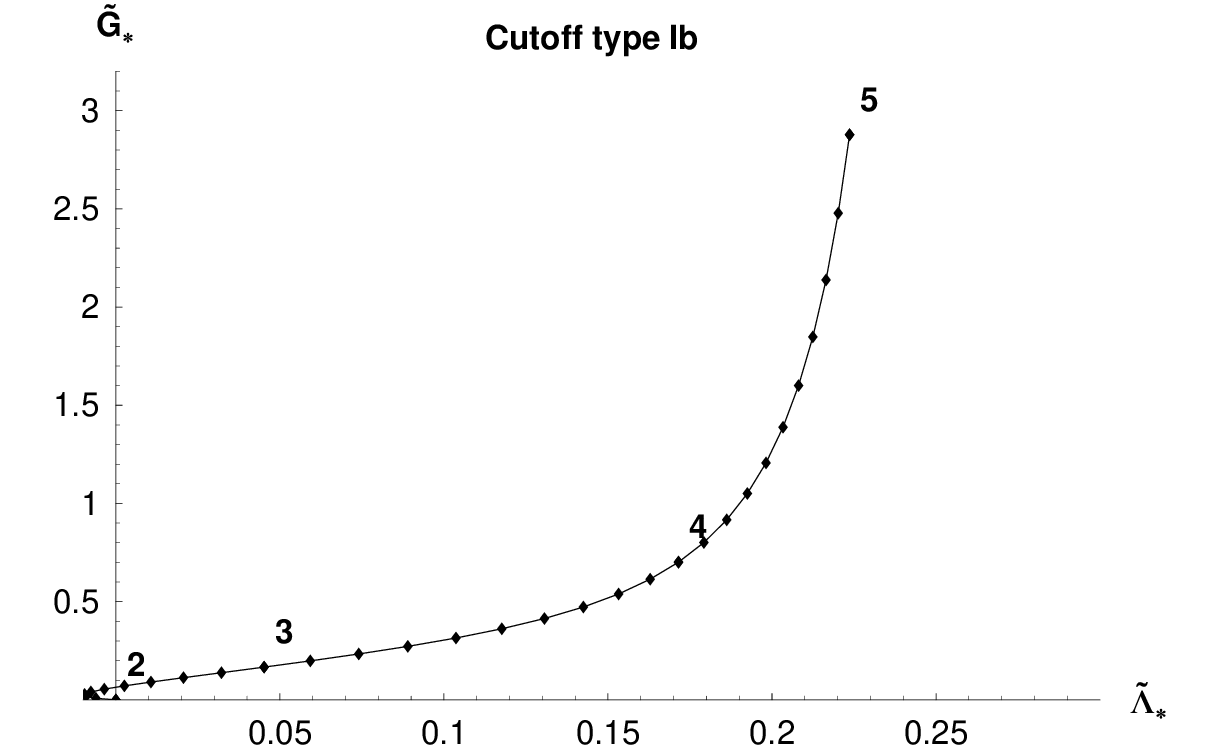}}
    \caption{\label{fig:1.1.5}The position of the nontrivial fixed point as a function of $d$
for cutoffs of type Ia (left panel) and Ib (right panel).
}}
\end{figure}

\begin{figure}
%[t]\center
{\resizebox{1\columnwidth}{!}
{\includegraphics{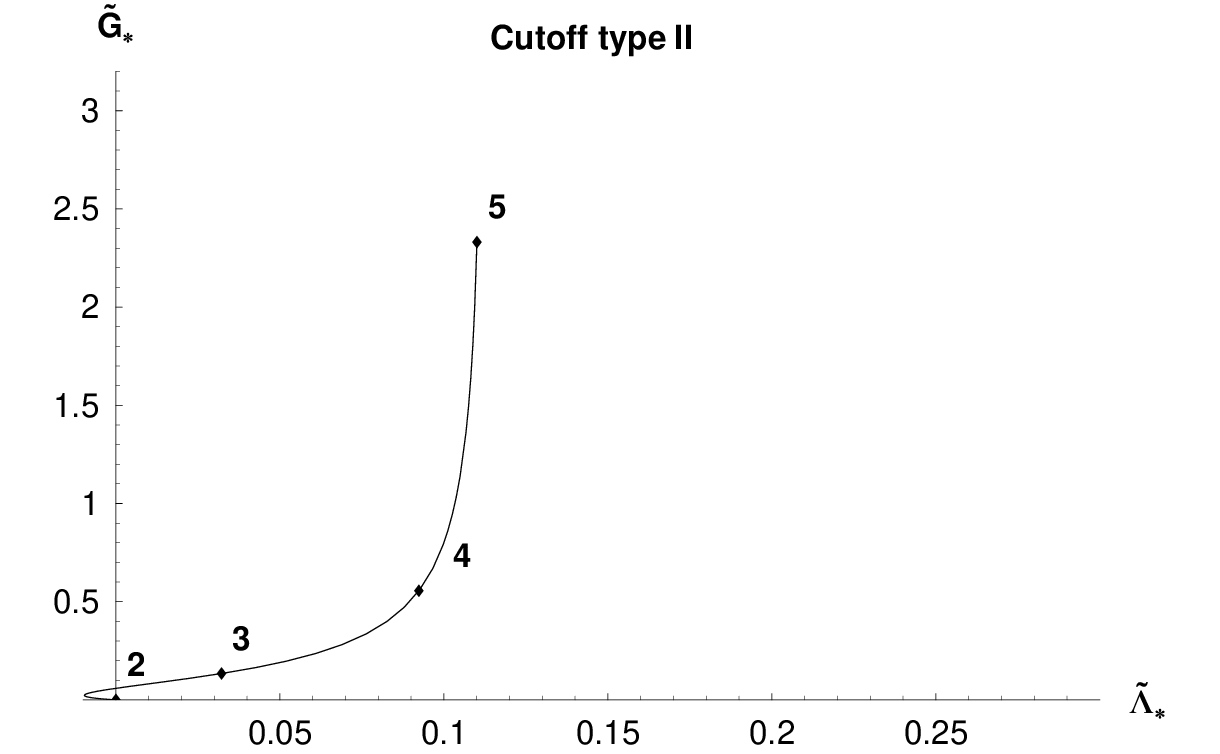}\includegraphics{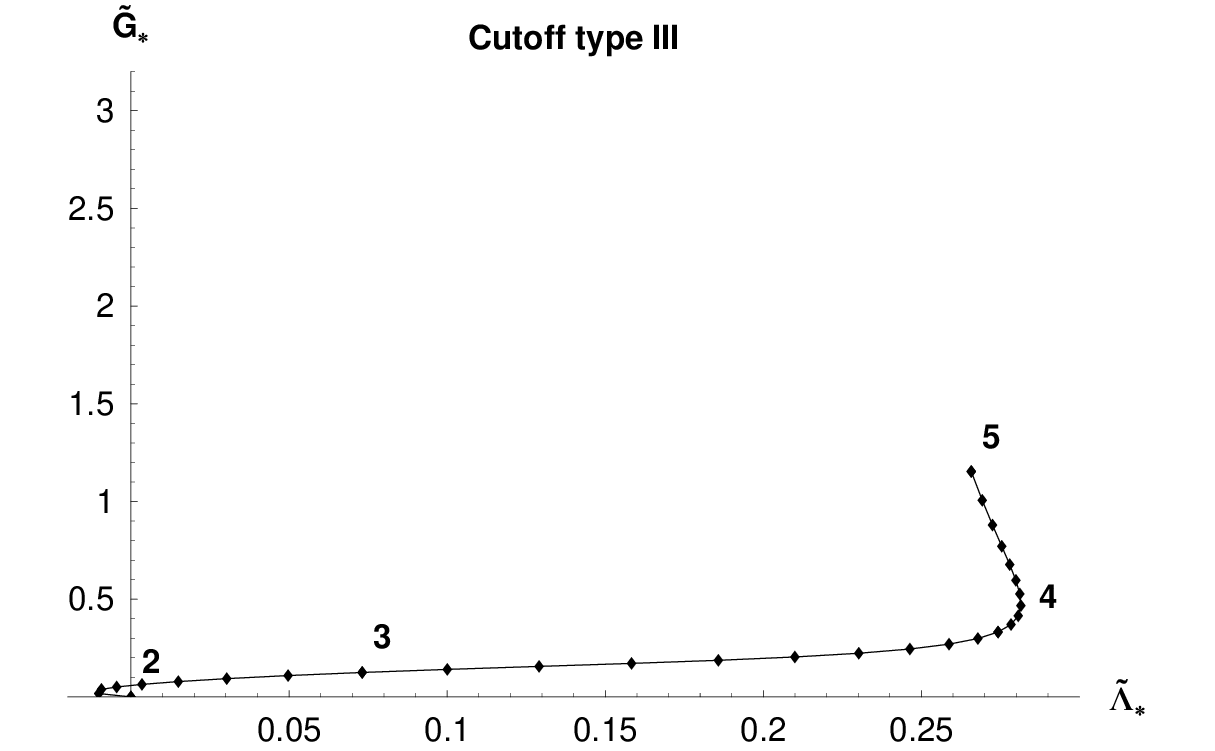}}
    \caption{\label{fig:1.1.6}The position of the nontrivial fixed point as a function of $d$
for cutoffs of type II (left panel) and III (right panel).
}}
\end{figure}

We conclude this discussion by mentioning that when the $\epsilon$
expansion is used in presence of a cosmological constant, there are
several FPs and even for $\epsilon$ very small they have
negative $\tilde\Lambda_*$. Thus the $\epsilon$ expansion is not
very helpful in the presence of $\tilde\Lambda$. One can solve exactly the
equations $\frac{d}{dt}\tilde\Lambda=0$ and $\frac{d}{dt}\tilde G=0$ for
arbitrary $d$ and plot the position of the FP in the
$\tilde\Lambda$--$\tilde G$ plane as a function of $d$. This is
shown in figures \ref{fig:1.1.5} and \ref{fig:1.1.6}. The fixed
point is in the origin at $d=2$; as $d$ grows, $\tilde G_*$ grows
monotonically while $\tilde\Lambda_*$ is initially negative, then
becomes positive. For moderately large dimensions (of order $10$)
$\tilde G_*$ becomes very large (of the order $10^6$) while $\tilde\Lambda_*<1/2$ always.

\subsection{Four dimensions}

Let us now consider Einstein's theory with cosmological constant in $d=4$.
The beta functions for $\tilde\Lambda$ and $\tilde G$ for various cutoff types have
been given in equations (\ref{betasIa},\ref{betasIbwfr},\ref{betasII},\ref{betasIII}).
All of these beta functions admit a trivial (Gau\ss ian) FP at
$\tilde\Lambda=0$ and $\tilde G=0$ and a nontrivial FP at positive
values of $\tilde\Lambda$ and $\tilde G$.
Let us discuss the Gau\ss ian FP first.
As usual, the perturbative critical exponents are equal to
$2$ and $-2$, the canonical mass dimensions of $\Lambda$ and $G$.
However, the corresponding eigenvectors
are not aligned with the $\tilde\Lambda$ and $\tilde G$ axes.
It is instructive to trace the origin of this fact.
Since it can be already clearly seen in perturbation theory,
we consider the perturbative Einstein--Hilbert flow (\ref{pertehflow}).
The linearized flow is given by the matrix
\begin{equation}
\label{linmatrix}
M=\left(
\begin{array}{cc}
\frac{\partial\beta_{\tilde\Lambda}}{\partial\tilde\Lambda}
&\frac{\partial\beta_{\tilde\Lambda}}{\partial\tilde G}\\
\frac{\partial\beta_{\tilde G}}{\partial\tilde\Lambda}
&\frac{\partial\beta_{\tilde G}}{\partial\tilde G}\\
\end{array}
\right)
=
\left(
\begin{array}{cc}
-2+B_1\tilde G+\frac{1}{2}\tilde G\frac{\partial A_1}{\partial\tilde\Lambda}
+\tilde\Lambda\tilde G\frac{\partial B_1}{\partial\tilde\Lambda}
&\frac{1}{2}A_1+B_1\tilde\Lambda\\
\tilde G^2\frac{\partial B_1}{\partial\tilde\Lambda}
&2+2B_1\tilde G\\
\end{array}
\right)\ .
\end{equation}
At the Gaussian FP this matrix becomes
\begin{equation}
M=\left(\begin{array}{cc}
-2 & \frac{1}{2}A_1(0) \\
0 & 2\\
\end{array}\right)\ ,
\end{equation}
which has the canonical dimensions of $\Lambda$ and $G$ on the diagonal,
as expected. However, the eigenvectors do not point along the $\Lambda$ and $G$ axes.
At the Gaussian FP the ``attractive'' eigenvector
is in the direction $(1,0)$ but
the `repulsive'' one is in the direction $(A_1(0)/4,1)$.
The slant is proportional to $A_1(0)$ and can therefore
be seen as a direct consequence of the running of the vacuum energy.
This fact has a direct physical consequence:
it is not consistent to study the ultraviolet limit of gravity neglecting the cosmological constant.
One can set $\tilde\Lambda=0$ at some energy scale, but if $\tilde G\not=0$,
as soon as one moves away from that scale the
renormalization group will generate a nontrivial cosmological constant.
This fact persists when one considers the renormalization group improved flow.

\begin{figure}
%[t]\center
{\resizebox{1\columnwidth}{!}
{\includegraphics{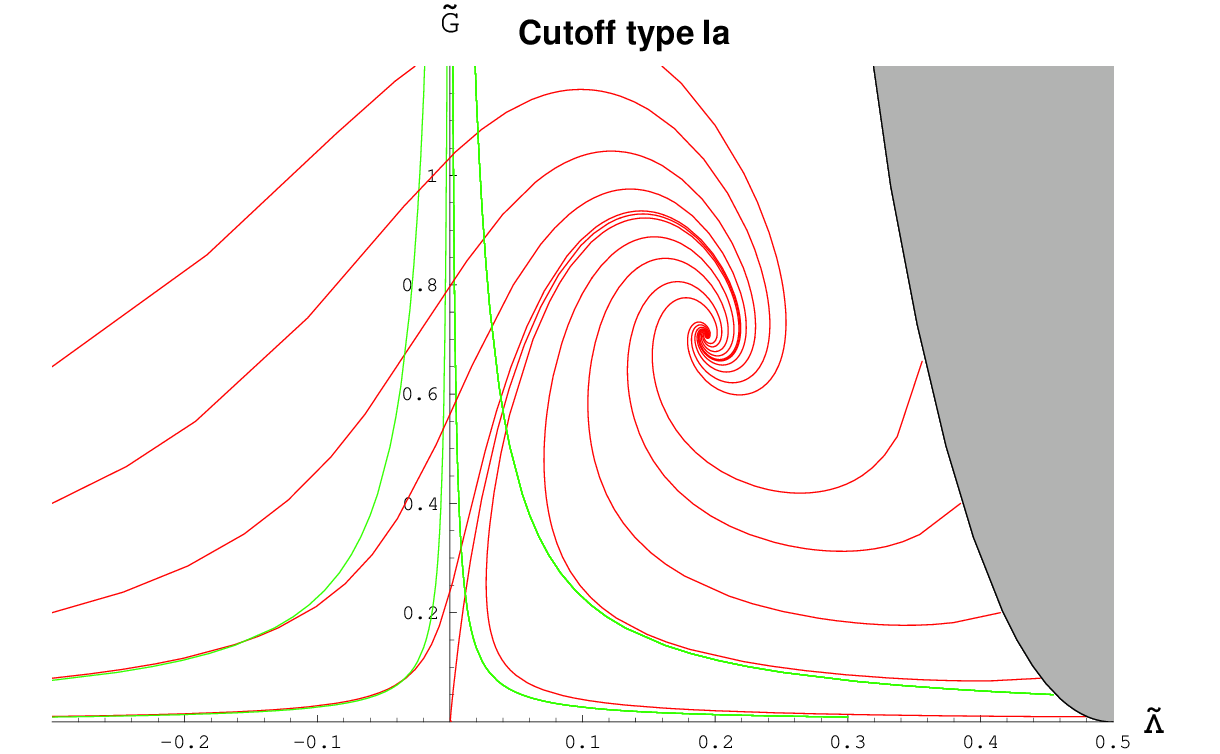} \includegraphics{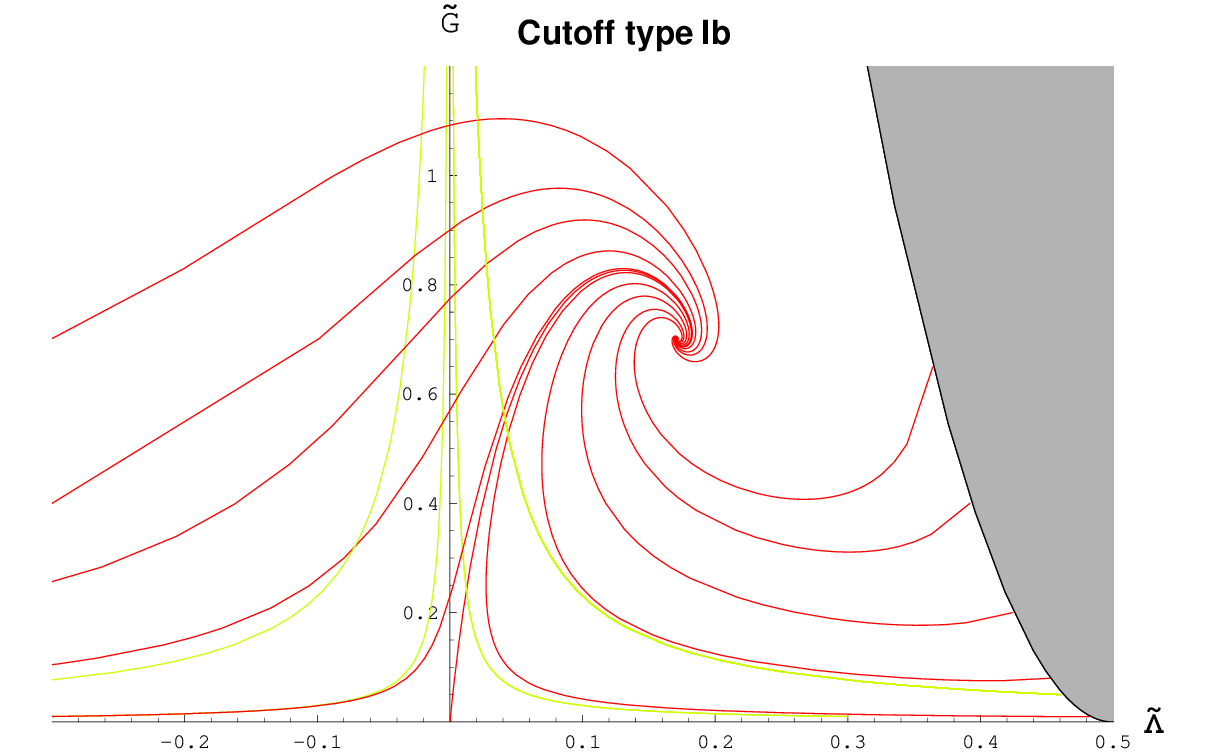}}
    \caption{\label{fig:flowI}The flow near the perturbative region
with cutoffs of type Ia and Ib. The boundary of the shaded region
is a singularity of the beta functions.
The curves in light color are ``classical'' trajectories
with constant $\tilde\Lambda\tilde G$.
}}
\end{figure}

\begin{figure}
%[t]\center
{\resizebox{1\columnwidth}{!}
{\includegraphics{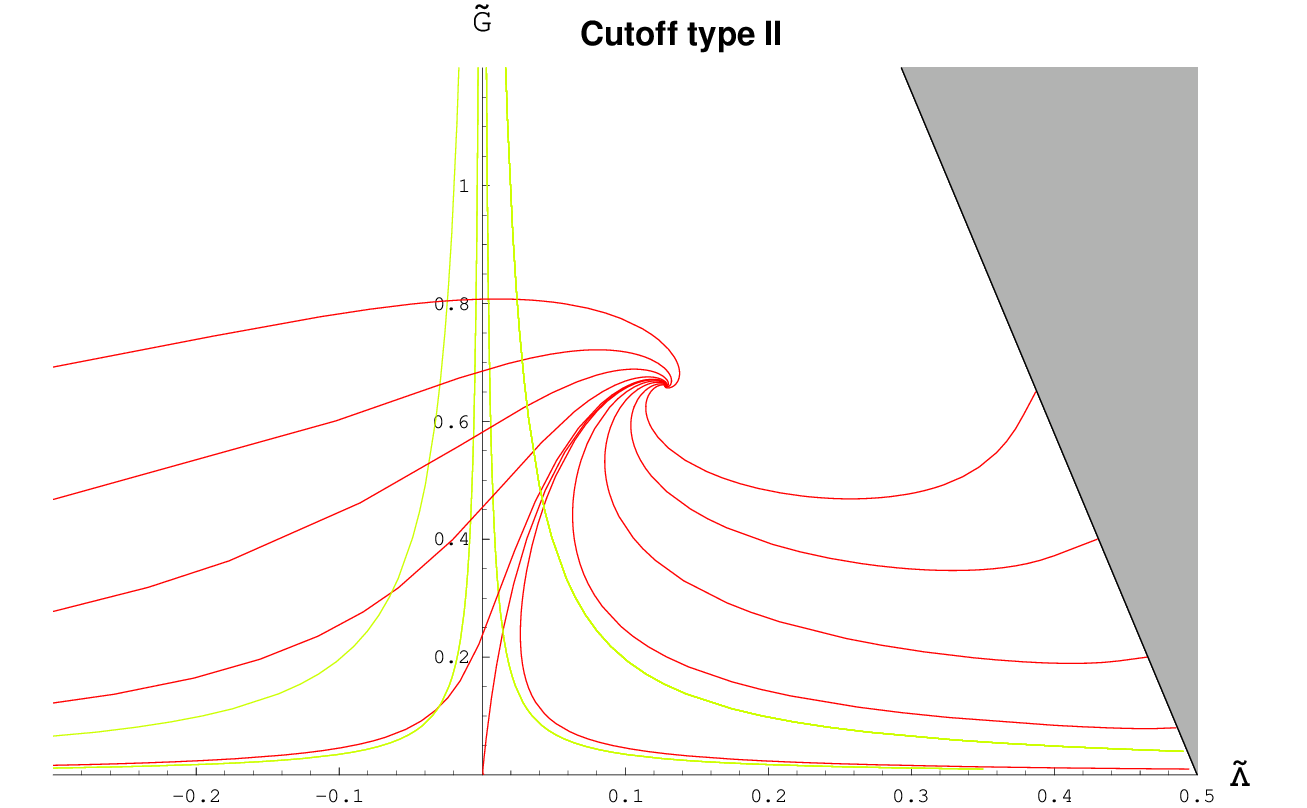} \includegraphics{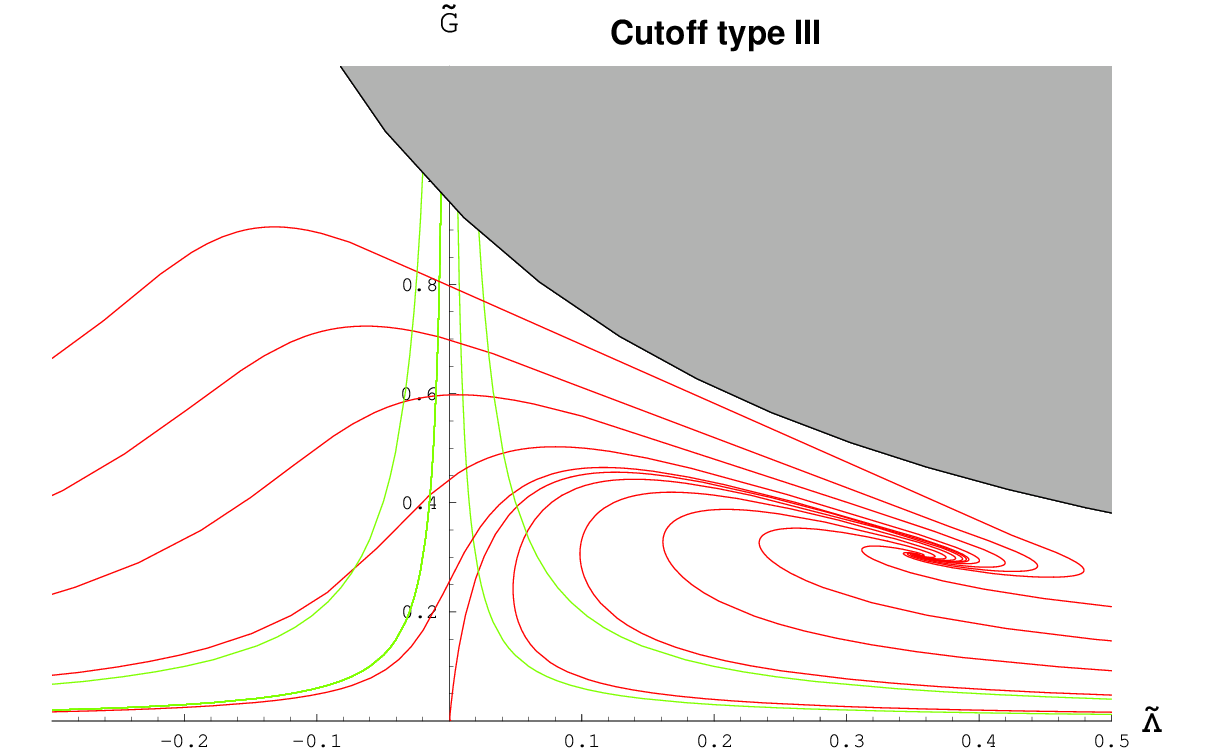}}
    \caption{\label{fig:flowII}The flow near the perturbative region
with cutoffs of type II and III. The boundary of the shaded region
is a singularity of the beta functions.}}
\end{figure}

Let us now come to the nontrivial FP.
We begin by making for a moment the drastic approximation
of treating $A_1$ and $B_1$ as constants, independent of $\tilde\Lambda$
(this is the leading term in a series expansion in $\tilde\Lambda$).
Thus we consider again the perturbative Einstein--Hilbert flow (\ref{pertehflow}).
In this approximation the flow can be solved exactly:
\begin{align}
\label{exactsol}
\tilde \Lambda(t)&=
\frac{(2\tilde\Lambda_0-\frac{1}{4}A_1\tilde G_0(1-e^{4t}))e^{-2t}}
{2+B_1\tilde G_0(1-e^{2t})}\ ,\nonumber\\
\tilde G(t)&=
\frac{2 \tilde G_0 e^{2t}}
{2+B_1\tilde G_0(1-e^{2t})}\ .
\end{align}
The FP would occur at $\tilde\Lambda_*=-A_1/4B_1$, $\tilde
G_*=-2/B_1$, at which point the matrix (\ref{linmatrix}) becomes
\begin{equation}
M=\left(\begin{array}{cc}
-4 & -\frac{1}{4}A_1 \\
0 & -2\\
\end{array}\right)\ .
\end{equation}
It has real critical exponents $2$ and $4$, equal to the canonical dimensions of the
constants $g^{(0)}=2Z\Lambda$ and $g^{(2)}=-Z$.
This should not come as a surprise, since the linearized flow matrix for
the couplings $g^{(0)}$ and $g^{(2)}$ is diagonal, with eigenvalues equal to their
canonical dimensions, and the eigenvalues are invariant under regular
coordinate transformations in the space of the couplings.
So we see that a nontrivial UV--attractive FP in the $\tilde\Lambda$--$\tilde G$ plane
appears already at the lowest level of perturbation theory.
It has the form shown in figure \ref{fig:flowmatter}.

All the differences between the perturbative Einstein--Hilbert flow and the exact flow
are due to the dependence of the constants $A_1$ and $B_1$ on $\tilde\Lambda$,
and in more accurate treatments to the RG improvements incorporated in the flow through the
functions $A_2$, $B_2$, $A_3$, $B_3$.
Such improvements are responsible for the nonpolynomial form of the beta functions.
In all these calculations the critical exponents at the nontrivial FP
always turn out to be a complex conjugate pair, giving rise to a spiralling flow.
The real part of these critical exponents is positive,
corresponding to eigenvalues of the linearized flow matrix with
negative real part. Therefore, the nontrivial FP is always UV-attractive
in the $\tilde\Lambda$--$\tilde G$ plane.
Conversely, an infinitesimal perturbation away from the FP will
give rise to a renormalization group trajectory that flows towards lower
energy scales away from the nontrivial FP.
Among these trajectories there is a unique one that connects the nontrivial
FP in the ultraviolet to the Gau\ss ian FP in the infrared.
This is called the ``separatrix''.

\begin{table}
\begin{center}
\begin{tabular}{l|l|l|l|l}
%\hline
{\rm Scheme}    &   $\tilde\Lambda_*$   &$\tilde G_*$ &$\tilde\Lambda_*\tilde G_*$& $\vartheta$  \\
\hline
Ia      &   0.1932   & 0.7073& 0.1367& 1.475$\pm$3.043i  \\
Ia - 1 loop & 0.1213  & 1.1718& 0.1421& 1.868$\pm$1.398i \\
Ib with fr     &   0.1715  & 0.7012& 0.1203& 1.689$\pm$ 2.486i \\
Ib with fr - 1 loop &0.1012     & 1.1209& 0.1134& 1.903$\pm$ 1.099 i  \\
Ib without fr     & 0.1677    &0.7204 &0.120804 &1.794 $\pm$2.393  i \\
Ib without fr - 1 loop &0.1012     & 1.1209& 0.1134& 1.903$\pm$ 1.099 i  \\
II      &   0.0924  & 0.5557& 0.0513& 2.425$\pm$1.270i \\
II - 1 loop &   0.0467  & 0.7745& 0.0362& 2.310$\pm$ 0.382 i \\
III      &  0.2742   & 0.3321& 0.0910& 1.752$\pm$2.069 i \\
III - 1 loop &0.0840     &0.7484 & 0.0628& 1.695$\pm$ 0.504 i
\end{tabular}
\end{center}
\caption{The nontrivial fixed point for Einstein's theory in $d=4$ with cosmological constant. Note that for cutoff Ib the one-loop result agrees with and without field redefinitions. This is because the Jacobians which have to be taken into account when field redefinitions are performed do not contain any couplings. In the case without field redefinitions instead, the unabsorbed factors multiply couplings and hence lead to additional terms with derivatives of couplings in the trace. 
\label{table2}}
\end{table}

One noteworthy aspect of the flow equations in the Einstein--Hilbert
truncation is the existence of a singularity of the beta functions.
In section IV E, when we neglected the cosmological constant, they
appeared at some value $\tilde G_c>\tilde G_*$. Now, looking at
equations
(\ref{betasIa},\ref{betasIbwfr},\ref{betasII},\ref{betasIII}), we
see that there are always choices of $\tilde\Lambda$ and $\tilde G$
for which the denominators vanish. The singularities are the
boundaries of the shaded regions in figures \ref{fig:flowI} and
\ref{fig:flowII}. Of course the flow exists also beyond these
singularities but those points cannot be joined continuously to the
flow in the perturbative region near the Gau\ss ian FP, which we
know to be a good description of low energy gravity. When the
trajectories emanating from the nontrivial FP approach
these singularities, they reach it at finite values of $t$ and the
flow cannot be extended to $t\to -\infty$. The presence of these
singularities can be interpreted as a failure of the
Einstein--Hilbert truncation to capture all features of infrared
physics and it is believed that they will be avoided by considering
a more complete truncation. Let us note that for cutoffs of type I
and II the singularities pass through the point $\tilde\Lambda=1/2$,
$\tilde G=0$. Thus, there are no regular trajectories emanating from
the nontrivial FP and reaching the region $\tilde\Lambda>1/2$.
However, for type III cutoffs the shaded region is not attached to
the $\tilde\Lambda$ axis and there are trajectories that avoid it,
reaching smoothly the region $\tilde\Lambda>1/2$.

In table \ref{table2} we collect the main features of the UV--attractive FP
for the Einstein--Hilbert truncation with cosmological constant
for the different cutoff schemes.

%%%%%%%%%%%%%%%%%%%%%%%%%%%%%%%%%%%%%%%%%%%%%%%%%%%%%%%%%%%%%%%%%%%%%%%%%%%%%%%%%%%%%%

\section{Ultraviolet divergences}

The Einstein--Hilbert truncation does not give a closed set of flow equations,
in the sense that the beta functions of the higher couplings, which have been
neglected in the previous section, are not zero.
So, if we assume that the higher couplings vanish at some initial scale,
they will immediately appear as one integrates the flow equations.
Before discussing truncations that involve higher derivative terms,
it will be instructive to see, using the ERGE, how such terms are generated
in Einstein's theory and how this is related to the issue of
ultraviolet divergences in perturbation theory.

A cautionary remark is in order here.
In perturbation theory, the divergences appear in the formulae
relating bare and renormalized couplings.
We recall that in our approach we never talk of the bare action;
instead, we follow the flow of the renormalized action $\Gamma_k$ as $k\to\infty$.
In this limit divergences can appear.
However, the limit of $\Gamma_k$ for $k\to\infty$ cannot be simply
identified with the bare action.
Exploring the relation between these two functionals would require
introducing an ultraviolet cutoff. We are not going to do this here.
In the following we will simply compare the divergences of $\Gamma_k$
to the perturbative ones.

In the perturbative approach to quantum gravity, the analysis of
ultraviolet divergences plays a central role.
This issue is not so prominent in the modern literature on asymptotic safety,
but this does not mean that divergences do not occur.
In an asymptotically safe theory, the asymptotic behavior of every
quantity is dictated simply by dimensional analysis.
The dimensionless ``couplings in cutoff units'' $\tilde g_i$ defined in (\ref{couplings})
tend to constant values, so the dimensionful couplings $g_i$ must
run like $k^{d_i}$. The couplings with positive mass dimension
diverge, and those with negative mass dimension go to zero. So, for
example, near the nontrivial FP in the Einstein--Hilbert truncation
discussed in the previous section, the graviton wave function
renormalization $Z=(16\pi G)^{-1}$ diverges quadratically and the
vacuum energy $2\Lambda Z$ diverges quartically at the FP.

This matches the powerlike divergences that one
encounters in perturbation theory when one uses an ultraviolet cutoff.
However, in the Wilsonian context these divergences have a different
physical meaning. The parameter $k$ has not been introduced in the
functional integral as an UV regulator, rather as an IR cutoff, and
in any physical application $k$ corresponds to some externally
prescribed scale.
So, in the Wilsonian approach the divergences would seem to acquire
almost a physical character: they are a manifestation of the dependence
of the couplings on an external parameter, and it should not be too
surprising that if an input parameter is allowed to tend to infinity
also some output could tend to infinity.

At a deeper level, however, one should take into account the fact that a
dimensionful quantity does not have an intrinsic value and therefore
cannot be observable. In order to give a value to a dimensionful
quantity $q$, one has to specify a unit $u$, and the result of any
measurement gives only a value for the dimensionless ratio $q/u$.
Near a FP, the divergence of a coupling $g_i$
with positive mass dimension for $k\to\infty$ is just
a restatement of the fact that $g_i$, measured in units of $k$, tends to a constant.
If we choose another unit $u$, since $u$ is also ultimately expressible
in terms of other couplings, it will also be subject to RG flow.
Then, the limit $q(k)/u(k)$ may tend
asymptotically to zero, to a finite limit or to infinity depending
on the behavior of $u$.
This highlights that the divergence of a dimensionful coupling cannot
have a direct physical meaning.
The only intrinsic (unit-independent) statement that one can make
about a dimensionful quantity is whether it is zero, positive or
negative.

Furthermore, only dimensionless functions of the couplings have a
chance of being observable. It is only for such combinations that
the theory is required to give unambiguous answers, \ie\ it is only
such combinations that one could expect to be scheme--independent.
Now, very often the choice of $k$ which is appropriate to a specific
experimental setup is not entirely unambiguous. Rather,
$k$ sets a characteristic scale of the problem and is usually known
only up to a factor of order one (see for example the discussion of
equation (\ref{betabjerrum}) or \cite{Hewett} for some concrete
examples in a gravitational context). The reason why $k$ is
nevertheless a useful quantity in practice is that dimensionless
functions of $k$ tend to depend weakly on $k$ and so an uncertainty
of order one in the value of $k$ produces only a very small
uncertainty in the value of the observable (think for example of the
logarithmic running of gauge coupling constants). However, if a
coupling $g_i$ is dimensionful, the corresponding dimensionless
variable $\tilde g_i$ depends strongly on $k$, so one should not always
expect the value of $\tilde g_i$ at a given scale to be precisely
defined. In particular, one should not expect the value of $\tilde g_i$
at the FP to be scheme--independent.

These expectations are confirmed in the previous treatment of gravity
in the Einstein--Hilbert truncation. In two dimensions Newton's
constant is dimensionless; its value at the nontrivial FP
(namely zero) and the slope of the beta function are scheme--independent.
In four dimensions the position of the nontrivial FP in
the $\tilde\Lambda$--$\tilde G$ plane is scheme--dependent. However, for
all cutoff schemes that have been tried so far $\tilde\Lambda$ and
$\tilde G$ are always positive: the existence of the FP and the sign
of the couplings are robust features of the theory.
In fact, in order for $\tilde\Lambda_*$ and $\tilde G_*$ to be zero,
one would have to find a cutoff function such that the $Q$-functionals
in front of the heat kernel coefficients $B_2$ and $B_0$ are zero.
Since $P_k$ and $\partial_t R_k$ are positive functions,
one sees from (\ref{Qnpos}) that no such choice exists.

The dimensionless combination
$\Lambda G$ is related to the on-shell effective action and is
known to be gauge--independent \cite{Kawai:1989yh}. Numerical studies
have also shown that the value of $\Lambda G$ at the FP is only very
weakly dependent on the cutoff function, much less so than
the values of $\tilde\Lambda$ and $\tilde G$. It is expected that this
residual weak dependence is only an effect of the truncation. In fact,
it has been argued in \cite{Lauscher}
that the weakness of this dependence is a sign that the
Einstein--Hilbert truncation must be stable against the inclusion of
further terms in the truncation.
We will see in section VII that this is indeed the case.

The UV behavior of dimensionless couplings
(\ie\ those that are marginal in power counting)
requires some additional clarification.
According to the preceding discussion, they have a chance of being
physically measurable and the existence of a FP requires that they have
a finite limit. On the other hand in perturbation theory they
generically present logarithmic divergences. How can these two
behaviors be reconciled? It is necessary here to distinguish two
possibilities: the limit could be finite and nonzero, or it could be
zero. If in a certain theory all couplings have the former behavior,
then there cannot be any logarithmic divergences.
On the other hand if the coefficient $g$ of some operator diverges
logarithmically, its inverse will go to zero.
So if the coupling is the inverse of $g$, it is asymptotically free.
This is what happens in Yang--Mills theories, where the
(square of the) asymptotically free Yang-Mills
coupling is the inverse of the coefficient of $F^2$.

In the derivative expansion of four dimensional gravity it is the
terms with four derivatives of the metric that have dimensionless
coefficients. We can parametrize this part of the action as follows:
\begin{equation}
\label{actionansatz}
\sum_i g^{(4)}_i\calo^{(4)}_i[g_{\mu\nu}]
=\int d^4x\,\sqrt{g}\left[
\frac{1}{2\lambda}C^{2}+\frac{1}{\xi}R^{2}-\frac{1}{\rho}E
+\frac{1}{\tau}\nabla^{2}R\right]\,
\end{equation}
where we use the notation introduced in (\ref{matterergeII}).
The question then arises, what is the asymptotic
behavior of these couplings, in particular what is the behavior of
$\lambda$ and $\xi$? This issue can be addressed at various levels,
the most basic one being: if we start from Einstein's theory,
do we encounter divergences proportional to these terms?

It was shown early on by 't Hooft and Veltman \cite{Veltman} using dimensional
regularization that (neglecting total derivatives)
the one loop effective action contains the
following simple pole divergence
\begin{equation}
\label{thooftveltman} \frac{1}{\epsilon}\int d^4x\,\sqrt{g}
\left[\frac{7}{20}R_{\mu\nu}R^{\mu\nu}+\frac{1}{120}R^2\right]\ .
\end{equation}
Can this result be seen within the ERGE? Let us return to the
Einstein--Hilbert truncation. In the previous section we have
expanded the r.h.s. of the ERGE using the heat kernel formula (\ref{HKasymp})
and retained only the first two terms, which are sufficient to give
the beta functions of the cosmological constant and Newton's contant.
Keeping the same inverse propagators,
we can now consider the next terms in the heat kernel expansion, which
will give the beta functions of $\lambda$, $\xi$, $\rho$, $\tau$ or
more precisely the dependence of these beta functions on Newton's
constant and on the cosmological constant.
We begin by considering a type II cutoff;
the terms $O(R^2)$ which were not computed in (\ref{eqEHII}) are
\begin{equation}
\label{quadraticterms} \int d^4x\,\sqrt{g} \left[
\frac{1}{2}Q_0\left(\frac{\partial_t R_k+\eta R_k}{P_k-2\Lambda}\right)
\mathrm{tr}b_4(\Delta_2)
-Q_0\left(\frac{\partial_t R_k}{P_k}\right)
\mathrm{tr}b_4(\Delta_{(gh)}) \right]\ .
\end{equation}
The $b_4$ coefficients for the relevant operators can be computed
using equation (\ref{b4}) and the traces in (\ref{traces}):
\begin{eqnarray}
\label{betasrsquared}
\mathrm{tr}b_4(\Delta_{2}) &=&
\frac{7}{12}C^2+\frac{35}{36}R^2+\frac{17}{36}E-\frac{2}{3}\nabla^2 R\\
\mathrm{tr}b_4(\Delta_{gh}) &=&
\frac{7}{60}C^2+\frac{13}{36}R^2-\frac{8}{45}E+\frac{3}{10}\nabla^2 R\ .
\end{eqnarray}
In order to compare to the 't Hooft--Veltman calculation we use the
one loop approximation to the ERGE, which as explained in section II
consists in neglecting $\eta$, and we also set $\Lambda=0$ in
(\ref{quadraticterms}). This gives a contribution to the ERGE equal to
\begin{eqnarray}
\label{thooftflow} \frac{d \Gamma_k}{dt}\Biggr|_{\sim R^2}&=&
\frac{1}{16\pi^2}\int d^4x\,\sqrt{g}
\left[\frac{7}{20}C^2+\frac{1}{4}R^2+\frac{149}{180}E-\frac{19}{15}\nabla^2 R\right]\nonumber\\
&=&\frac{1}{16\pi^2}\int d^4x\,\sqrt{g}
\left[\frac{7}{10}R_{\mu\nu}R^{\mu\nu}+\frac{1}{60}R^2+\frac{53}{45}E-\frac{19}{15}\nabla^2 R\right]\ ,
\end{eqnarray}
where in the last step we have used the identity
$C^2=E+ 2 \left(R_{\mu\nu}R^{\mu\nu}-\frac{1}{3}R^2\right)$.
From here one can directly read off the beta functions.
When one then solves for the flow, in the limit of large $k$
each of these couplings diverges logarithmically,
with a coefficient that can be read off (\ref{thooftflow}).
Recalling that for 't Hooft and Veltman $\frac{1}{\epsilon}$ corresponds to
$\frac{1}{8\pi^2}\log\Lambda_{UV}$, where $\Lambda_{UV}$ is an UV cutoff,
we find agreement with their result in the topologically trivial case.
On the other hand, if we assume $R_{\mu\nu}=0$,
which is the on shell condition in perturbation theory without cosmological constant,
(\ref{thooftflow}) agrees
with the one loop divergence computed in \cite{Gibbons:1978ac}.
This provides an independent check on the coefficient of the Euler term.
Furthermore, our calculation shows that all these terms,
being proportional to $Q_0\left(\frac{\partial_t R_k}{P_k}\right)$,
are independent of the choice of the profile function $R_k$.
We have also verified that calculating these terms with
cutoffs of type Ia and III leads to the same results.
Thus, these divergences are indeed independent of the cutoff scheme
\footnote{In these calculations we stick to the de Donder gauge
with $\alpha=1$.}.

In the case of cutoffs of type Ib there is a subtlety that needs some
clarification. With these cutoffs (whether one performs a field redefinition,
as in section IV B, or not, as in Appendix D) it is only possible to
perform the calculation on Euclidean de Sitter space (a 4--sphere).
This provides a check on a single
combination of the terms appearing in (\ref{actionansatz}).
Specializing to the sphere and using that the volume of the sphere is $384\pi^2/R^2$,
(\ref{thooftflow}) becomes
\begin{equation}
\label{christoph} \frac{d \Gamma_k}{dt}\Biggr|_{\sim R^2}=\frac{419}{45}\
.
\end{equation}
Using type Ib cutoffs one has to pay special attention to the contribution
of some of the lowest modes in the traces. This is due to the fact that
the traces over vector and scalar modes may have a prime or a double prime,
meaning that some modes have to be left out.
When evaluating the ERGE without redefining the fields $\xi_\mu$ and $\sigma$,
as in equation (D1),
the isolated modes give an overall contribution -14, which adds up
to the contribution of the rest of the spectrum, which is equal to $\frac{1049}{45}$,
to give the correct result.
This is an important consistency check on the expression (D1).
A similar result holds for the calculation when the fields $\xi_\mu$ and $\sigma$
are redefined, as in equation (\ref{eqEHIb}).

It is interesting to consider also the case when $\Lambda\not=0$.
In the case of a type II cutoff, it appears from (\ref{quadraticterms}), with $\eta=0$,
and using (\ref{qoptpos},\ref{qoptzero},\ref{qoptneg}),
that the logarithmic divergence will be the same as in the case $\Lambda=0$.
This is due to the fact that expanding the fraction in $\Lambda$, terms containing $\Lambda$
give rise to power--like divergences, so only the leading, $\Lambda$--independent term contributes to
the logarithmic divergence.
The same will be the case for type I cutoffs, since again $\Lambda$ only appears in denominators.
On the other hand for a type III cutoff, (\ref{quadraticterms})
should be replaced by
\begin{equation}
\int d^4x\,\sqrt{g} \left[
\frac{1}{2}Q_0\left(\frac{\partial_t R_k}{P_k}\right)
\mathrm{tr}b_4(\Delta_2-2\Lambda\mathbf{1})
-Q_0\left(\frac{\partial_t R_k}{P_k}\right)
\mathrm{tr}b_4(\Delta_{(gh)}) \right]\ .
\end{equation}
Then, assuming $R_{\mu\nu}=\Lambda g_{\mu\nu}$,
which is the on shell condition in perturbation theory with cosmological constant,
\begin{equation}
\frac{d \Gamma_k}{dt}\Biggr|_{\sim R^2}= \frac{1}{16\pi^2}\int d^4x\,\sqrt{g}
\left[\frac{53}{45}R_{\mu\nu\rho\sigma}R^{\mu\nu\rho\sigma}
-\frac{58}{5}\Lambda^2\right]\ .
\end{equation}
On the 4--sphere this gives, instead of (\ref{christoph})
\begin{equation}
\label{duffs} \frac{d \Gamma_k}{dt}\Biggr|_{\sim R^2}=-\frac{571}{45}\ .
\end{equation}
These results agree with those obtained in \cite{Christensen:1979iy}.
Note therefore that the $\Lambda$--dependent contributions to the
logarithmic divergences {\it are scheme--dependent}.
This should not come as a surprise, in view of the discussion above.

We observe here for future reference that certain authors {\it subtract} from
the ghost term the contribution of the ten lowest eigenvalues of $-\nabla^2$
on vectors, which correspond to the ten Killing vectors of $S^4$
\cite{allen,polchinski,taylor}, see also \cite{vassilevich,wipf} for a discussion.
This amounts to putting a prime on the ghost determinant
(e.g. putting an extra prime on the last two terms in (\ref{withred})).
This can be motivated by the observation that Killing vectors
generate global symmetries, and global symmetries are not gauge transformations.
Since the ghost contribution has a minus sign, with a cutoff of type II
this corresponds to {\it adding} to (\ref{duffs}) the term $10\frac{\partial_t R_k}{P_k}$
evaluated on the lowest eigenvalue of the operator $-\nabla^4-\frac{R}{4}$,
which is equal to zero. This is equal to $10\frac{\partial_t R_k(0)}{P_k(0)}=20$.
Thus, (\ref{duffs}) would be replaced by
\begin{equation}
\label{venezianos} \frac{d \Gamma_k}{dt}\Biggr|_{\sim R^2}=\frac{329}{45}\
.
\end{equation}
We will return to this issue in section VII, when we define the
ERGE for $f(R)$--gravity.

Having found the expected agreement with earlier one loop calculations
based on the Einstein-Hilbert action, the next level of sophistication
would be to include the terms in (\ref{actionansatz}) in the truncation,
as is required in a more accurate approximation to the exact flow.
We will discuss this in the next section.
In the rest of this section we shall discuss the possible appearance
of divergences that are cubic or of higher order in curvature,
keeping the kinetic operator that comes from the Einstein--Hilbert action.

Among higher powers of curvature,
of particular interest is the term cubic in the Riemann tensor.
In perturbation theory, the divergence
(\ref{thooftveltman}) can be absorbed into a redefinition of the
metric and therefore does not affect the $S$ matrix: pure Einstein
theory is one--loop renormalizable. The first divergence in the
effective action that cannot be eliminated by a field redefinition
in perturbation theory is proportional to
$R_{\mu\nu}{}^{\rho\sigma}R_{\rho\sigma}{}^{\alpha\beta}R_{\alpha\beta}{}^{\mu\nu}$.
The coefficient of this term was calculated in \cite{Sagnotti} at two loops.
Can this divergence be seen in the Einstein--Hilbert truncation of the ERGE,
in the same way as we have seen the 't Hooft--Veltman divergence? Let $g$ be the
coefficient of this operator in the Lagrangian.
In the one loop approximation, neglecting $\Lambda$ and using a type II cutoff,
the beta function of $g$ will be proportional to
$Q_{-1}\left(\frac{\partial_t R_k}{P_k}\right)=k^{-2}\tilde Q$ where
$\tilde Q$ is a scheme--dependent dimensionless number.
As explained in Appendix A, one can choose
the cutoff scheme in such a way that $\tilde Q=0$ (this is the case,
for example, with the optimized cutoff used in this paper).
Then, the coupling $\tilde g$  will have a FP at $\tilde g_*=0$
\footnote{if we chose another cutoff such that $\tilde Q\not=0$,
$g$ will go asymptotically to zero; however, as discussed before,
this is a unit--dependent statement and does not have much physical meaning.}.
These facts are not surprising: they are a reflection of the fact that the
ERGE has the structure of a one loop RG equation,
and of the absence of a Riemann-cube divergence at one loop in perturbation theory.
The Goroff--Sagnotti counterterm can only be seen in perturbation theory
at two loops.
On the other hand, if we truncate the ERGE at higher order, for example
including all terms cubic in curvature, it is expected that the beta
function of $g$, though scheme--dependent, cannot be set to zero just by
a choice of cutoff. There are then two possibilities.
The arguments given in the end of section III suggest that a
FP will exist for all terms in the derivative expansion,
including the Riemann--cube term; in this case the Goroff--Sagnotti
divergence would be an artifact of perturbation theory.
Alternatively, it is possible that the FP will cease to exist when the
Riemann--cube term (or some other high order term) is added to the truncation.
It is not yet known whether this is the case or not;
in the conclusions we give some reasons why we believe in the former alternative.

%%%%%%%%%%%%%%%%%%%%%%%%%%%%%%%%%%%%%%%%%%%%%%%%%%%%%%%%%%%%%%%%%%%%%%%%%%%%%%

\section{Curvature squared truncations.}

In the preceding section we have seen that starting from the Einstein--Hilbert action,
the RG flow will generate terms quadratic in curvature.
In particular, we have discussed the way in which these terms diverge
logarithmically when $k\to\infty$.
It is therefore not consistent to neglect these terms,
and a more accurate treatment
will take into account the contribution of these terms to the
r.h.s. of the ERGE. In other words, we should include the terms
(\ref{actionansatz}) in the truncation.
The resulting theory is perturbatively renormalizable \cite{Stelle}
but has problems with unitarity.
Furthermore, in the perturbative treatment the cosmological
constant is set to zero, something we know can only be imposed at a given scale.
The corresponding Wilsonian calculation has been done within the one loop approximation,
and has been briefly reported in \cite{Codello}.
Here we review and extend those results.

We take the action $\Gamma_k$ as the sum of the Einstein--Hilbert
action (\ref{ehaction}) and the terms (\ref{actionansatz}). The
linearized wave operator is now a complicated fourth order operator.
In order to simplify its form, following
\cite{Julve,AvramidiBarvinsky,Shapiro,Buchbinder}
it is convenient to choose a gauge fixing of the form
\begin{equation*}
S_{GF}=\int d^4x\sqrt{g}\,\chi_{\mu}Y^{\mu\nu}\chi_{\nu}
\end{equation*}
where
$\chi_{\nu}=\nabla^{\mu}h_{\mu\nu}+\beta\nabla_{\nu}h$
(all covariant derivatives are with respect to the background metric) and
\begin{equation*}
Y^{\mu\nu}=\frac{1}{\alpha}\left[g^{\mu\nu}\nabla^{2}
+\gamma\nabla^{\mu}\nabla^{\nu}-\delta\nabla^{\nu}\nabla^{\mu}\right]\ .
\end{equation*}
The ghost action contains the term
\begin{equation*}
S_c=\int d^4x\sqrt{g}\,\bar{C}_{\nu}(\Delta_{gh})^\nu{}_\mu C^{\mu}
\end{equation*}
where
\begin{equation*}
(\Delta_{gh})^\nu{}_\mu=-\delta_{\mu}^{\nu}\nabla^2-(1+2\beta)\nabla_{\mu}\nabla^{\nu}
+R^\nu{}_\mu\ ,
\end{equation*}
as well as a ``third ghost'' term
\begin{equation*}
S_b=\frac{1}{2}\int d^4x\sqrt{g}\,b_{\mu}Y^{\mu\nu}b_{\nu}\ ,
\end{equation*}
due to the fact that the gauge averaging operator $\mathbf{Y}$
depends nontrivially on the metric. We follow earlier authors
\cite{Shapiro} in choosing the gauge fixing parameters $\alpha$,
$\beta$, $\gamma$ and $\delta$ in such a way that the quadratic part
of the action is:
\begin{equation}
\frac{1}{2}\int d^4x\sqrt{g}\ \delta g\mathbf{K}\mathbf{\Delta}^{(4)}\delta g\ ,
\end{equation}
where
\begin{equation}
\mathbf{\Delta}^{(4)}=(-\nabla^2)^2+\mathbf{V}^{\rho\lambda}\nabla_{\rho}\nabla_{\lambda}+\mathbf{U}\ .
\end{equation}
For details of the operators $\mathbf{K}$, $\mathbf{V}$ and
$\mathbf{U}$ we refer the reader to \cite{Shapiro} whose notation we
mostly follow. We choose type III cutoffs:
\begin{equation*}
\mathbf{R}_k^g(\mathbf{\Delta}^{(4)})= \mathbf{K} R_k^{(4)}(\mathbf{\Delta}^{(4)})\ ;\quad
R_k^c(\mathbf{\Delta}_{(gh)})^\mu{}_\nu=\delta^\mu_\nu R_k^{(2)}(\mathbf{\Delta}_{(gh)})\ ;\quad
R_k^b(\mathbf{Y})^{\mu\nu}=g^{\mu\nu}R_k^{(2)}(\mathbf{Y})\ .
\end{equation*}
For higher derivative operators we will use a generalized optimized cutoff
of the form $R_k^{(n)}(z)=(a k^n-z)\theta(a k^n-z)$,
with $a=1$ unless otherwise stated.

We restrict ourselves to the one--loop approximation, as explained in previous sections.
Furthermore, only the contributions of the heat kernel coefficients up to $B_4$
will be taken into account.
In this way, we will essentially neglect all RG improvements.
The beta functions of the dimensionless couplings appearing in (\ref{actionansatz}) turn out to be:
\begin{align}
\label{rsqkj}
\beta_{\lambda} & =  -\frac{1}{(4\pi)^{2}}\frac{133}{10}\lambda^{2}\ ,\nonumber\\
\beta_{\xi} & =  -\frac{1}{(4\pi)^{2}}\left(10\lambda^2-5\lambda\xi+\frac{5}{36}\xi^2\right)\ ,\nonumber\\
\beta_{\rho} & =  \frac{1}{(4\pi)^{2}}\frac{196}{45}\rho^2\lambda\ .
\end{align}
They form a closed system and agree with those calculated in
dimensional regularization
\cite{AvramidiBarvinsky,Shapiro,Buchbinder}. It is convenient
to define new couplings $\omega$ and $\theta$ by
$\xi=-3\lambda/\omega$ and $\rho=\lambda/\theta$. In this way
$1/\lambda$ is the overall strength of the terms quadratic in
curvatures, and $\theta$ and $\omega$ give the relative strength of
the different invariants. Then, the beta functions become
\begin{align}
\beta_{\omega} & =  -\frac{1}{(4\pi)^{2}}\frac{25 + 1098\,\omega+ 200\,\omega^2}{60}\lambda\ ,\nonumber\\
\beta_{\theta} & =  \frac{1}{(4\pi)^{2}}\frac{7(56-171\,\theta)}{90}\lambda\ .
\end{align}
The coupling $\lambda$ has the usual logarithmic approach to
asymptotic freedom, while the other two couplings have the FP values
$\omega_*\approx(-5.467,-0.0228)$ and $\theta_*\approx0.327$. Of the
two roots for $\omega$, the first turns out to be UV--repulsive, so
the second has to be chosen
\cite{AvramidiBarvinsky,Shapiro,Buchbinder}.

The beta functions of $\tilde\Lambda$ and $\tilde G$ are:
\begin{align}
\label{piero}
\beta_{\tilde \Lambda} & =
-2\tilde\Lambda
+\frac{1}{(4\pi)^{2}}\left[
\frac{1+20\omega^2}{256\pi\tilde G\omega^2}\lambda^2
+\frac{1+86\omega+40\omega^2}{12\omega}\lambda\tilde\Lambda\right]
-\frac{1+10\omega}{64\pi^2\omega}\lambda
+\frac{2\tilde G}{\pi}
-q(\omega)\tilde G \tilde\Lambda\ ,\nonumber\\
\beta_{\tilde G} & =  2\tilde G
-\frac{1}{(4\pi)^{2}}\frac{3+26\omega-40\omega^2}{12\omega}\lambda\tilde G
-q(\omega) \tilde G^2
\end{align}
where $q(\omega)=(83+70\omega+8\omega^2)/18\pi$. The first two terms
in each beta function exactly reproduce the results of
\cite{AvramidiBarvinsky,Shapiro,Buchbinder}, the remaining
ones are new. The origin of the new terms will be discussed below.

To study the flow of $\tilde\Lambda$ and $\tilde G$,
we set the remaining variables to their FP values $\omega=\omega_*$, $\theta=\theta_*$,
and $\lambda=\lambda_*=0$.
Then, the flow takes the form of the perturbative Einstein--Hilbert flow
(\ref{generallambdag}),
with $A_1=4/\pi$, $B_1=-q_*=-q(\omega_*)\approx -1.440$, $A_2=B_2=0$ (see figure \ref{fig:flowmatter}).
It has two FPs: the Gaussian FP at $\tilde{\Lambda}=\tilde{G}=0$ and another one at
$$
\tilde{\Lambda}_{*}=\frac{1}{\pi q_*}\approx 0.221\ ,\ \ \ \ \
\tilde{G}_{*}=\frac{2}{q_*}\approx 1.389\ .
$$
Like all one loop flows with constant $A_1$ and $B_1$,
the critical exponents are 2 and 4. This, however,
is due to the approximation. If we could take into account the
contribution of the heat kernel coefficients $B_6$, $B_8$ etc., they
would contribute terms of order $\tilde\Lambda^2$ and higher to the
beta functions. We expect that these terms would produce a complex
conjugate pair of critical exponents, and the corresponding
spiralling flow. In this connection see also \cite{Niedermaierrev}.

In the preceding calculation we have used, besides the truncation to
four--derivative terms, also the one loop approximation, and
contributions coming from the heat kernel coefficients
$B_6$ and higher have been neglected. These are also the
approximations made in earlier perturbative calculations
\cite{Julve,AvramidiBarvinsky,Shapiro,Buchbinder}, so it is
instructive to understand the origin of the
additional terms in (\ref{piero}), which are essential
in generating the nontrivial FP. The beta functions were
originally derived as coefficients of $1/\epsilon$ poles in
dimensional regularization, which correspond to logarithmic
divergences in the effective action. In a heat kernel derivation
these terms are given by the $B_4$ coefficient. In the old
calculations only these terms were retained. The new terms that we
find come from the $B_2$ and $B_0$ coefficients, which in a
conventional calculation of the effective action would correspond to
quadratic and quartic divergences. Such terms are discarded in
dimensional regularization, but we see that proceeding in this way
one would throw away essential physical information. In order to
keep track of this information in dimensional regularization one
would have to take into account the contribution of the pole
occurring in dimension 2. This can be done with the $2+\epsilon$
expansion, and we have already discussed the way in which the FP
then appears. The situation is
entirely analogous to the Wilson--Fisher FP in three dimensional
scalar theory, which can be seen either using a cutoff
regularization or, if dimensional regularization is used, in the
$4-\epsilon$ expansion  \cite{Wilson:1971dc}. It is appropriate to
stress once more that our ``Wilsonian'' calculation of the beta
functions does not require any UV regularization. Accordingly, there
are no regularization/renormalization ambiguities; the only
ambiguity is in the choice of the cutoff procedure, but we have
already seen that no choice of $R_k$ could remove the $B_2$ and $B_0$ terms.

Having discussed the beta functions in the higher derivative truncation of pure gravity,
it is a simple exercise to add to them the contributions of minimally coupled matter fields,
which were discussed in section III.
From (\ref{rsqkj}) and (\ref{matterergeII}) we find
\begin{align}
\beta_{\lambda} & =  -\frac{1}{(4\pi)^{2}}\frac{133}{10}\lambda^{2}-2\lambda^2 a^{(4)}_\lambda\ ,\nonumber\\
\beta_{\xi} & =  -\frac{1}{(4\pi)^{2}}\left(10\lambda^2-5\lambda\xi+\frac{5}{36}\xi^2\right)-\xi^2 a^{(4)}_\xi\
,\nonumber\\
\beta_{\rho} & =  \frac{1}{(4\pi)^{2}}\frac{196}{45}\rho^2\lambda-\rho^2 a^{(4)}_\rho\ ,
\end{align}
where
\begin{align}
a^{(4)}_\lambda&= \frac{1}{2880\pi^2}\left(\frac{3}{2}n_S+9n_D+18n_M\right)\ ,\nonumber\\
a^{(4)}_\xi&= \frac{1}{2880\pi^2}\frac{5}{2}n_S\ ,\nonumber\\
a^{(4)}_\rho&= \frac{1}{2880\pi^2}\left(-\frac{1}{2}n_S-\frac{11}{2}n_D-31n_M\right)\ .
\end{align}
For these couplings, the new terms simply change the direction and speed of the logarithmic approach
to asymptotic freedom.
In particular, the ratios of the couplings approach asymptotically the following FP values
\begin{align*}
\omega_*&=
\frac{1}{200}\left(\!-549-960\pi^2 a^{(4)}_\lambda\pm\!
\sqrt{921600\pi^4(a^{(4)}_\lambda)^2\!+1054080 \pi^2 a^{(4)}_\lambda\!-576000
a^{(4)}_\xi\pi^2\!+296401}\right)\\
\theta_*&=\frac{8}{9}\frac{49+180\pi^2 a^{(4)}_\rho}{133+320\pi^2 a^{(4)}_\lambda}
\end{align*}
On the other hand, the beta functions of $\tilde\Lambda$ and $\tilde G$
are modified by the addition of $8\pi a^{(0)}\tilde G+32\pi a^{(2)}\tilde\Lambda\tilde G$
and  $32\pi a^{(2)}\tilde G^2$ respectively, where, using a type II optimized cutoff,
\begin{align}
a^{(0)}&= \frac{1}{32\pi^2}\left(n_S-4n_D+2n_M\right)\ ,\nonumber\\
a^{(2)}&= \frac{1}{96\pi^2}\left(n_S+2n_D-4n_M\right)\ .
\end{align}
Let us consider again the flow in the $\tilde\Lambda$--$\tilde G$ plane.
Putting $\omega$ and $\theta$ to their FP values and taking the
limit $\lambda\to 0$, the flow takes again the form (\ref{generallambdag}),
with the coefficients
$A_1=4/\pi+16\pi a^{(0)}$,
$B_1=-q(\omega_*)+16\pi  a^{(2)}$,
$A_2=B_2=0$.

This calculation is of some interest for the following reason.
Recall that whereas applying perturbation theory to pure Einstein theory,
terms proportional to
$R_{\mu\nu}R^{\mu\nu}$ and $R^2$ can be absorbed by a field redefinition,
this is no longer so when matter fields are present \cite{Veltman}.
Thus, gravity coupled to matter is nonrenormalizable already at one loop.
One may fear that the FP ceases to exist when one includes
in the truncation terms that correspond to nonrenormalizable divergences
in the perturbative treatment of Einstein's theory.
The preceding calculation shows, at least at one loop, that this is not the case.
We will further comment on the significance of this point in the conclusions.

%%%%%%%%%%%%%%%%%%%%%%%%%%%%%%%%%%%%%%%%%%%%%%%%%%%%%%%%%%%%%%%%%%%%%%%%%%%%%%%

\section{Truncation to polynomials in R}

Following the scheme outlined in the previous two sections, we now
extend the formalism to handle truncations
in the form of the so--called  ``$f(R)$--gravity'' theories,
where the Lagrangian density in (\ref{Gammak}) is a function
of the Ricci scalar only. Such theories rose
strong interest recently in cosmological applications \cite{fofr}.
At one loop, the quantization of these theories has been discussed
in \cite{Cognola}. Here we analyze
the RG flow of this type of theories where $f$ is a polynomial of
order $n\leq 8$.

The (Euclidean) action is approximated by
\begin{equation}
\label{fofr}
\Gamma_k[\Phi]= \sum_{i=0}^n g_i(k) \int d^d x\,\sqrt{g} R^i
+S_{GF}+S_{gh}\ ,
\end{equation}
where $\Phi=\{h_{\mu\nu},C_{\mu},\bar C_{\nu}\}$ and the last two terms
correspond to the gauge fixing and the ghost sector \cite{Reuter,Dou}.
The gauge fixing will have the general form
\begin{equation}
\label{fofrgaugefixing}
S_{GF}=\frac{1}{2}\int d^dx\sqrt{\bar
g}\,\chi_{\mu}{G}^{\mu\nu}\chi_{\nu}
\end{equation}
where
$\chi_{\nu}=\nabla^{\mu}h_{\mu\nu}-\frac{1+\rho}{d}\nabla_{\nu}h^{\mu}_{\,\,\mu}$
and $ G_{\mu\nu}=\left( \alpha +\beta\nabla^2 \right) g_{\mu\nu}$.
In the following four subsections we give the second variation of the action and the ghost action,
then we insert these expressions into the ERGE and finally we discuss the choice of gauge.

\subsection{Truncation Ansatz}

The second variation of $\int{\rm d}^d x\sqrt{g}f\bl(R\br)$ gives
\begin{eqnarray}
 \delta^2\int d^d x\,\bl(\sqrt{g}f\bl(R\br)\br) &=&
        \int d^d x\,\sqrt{g}\Biggl\lbrace\frac{1}{2} f''\bl(R\br)
        \bl(h_{\alpha\beta}\nabla^{\alpha}\nabla^{\beta}\nabla^{\mu}\nabla^{\nu}h_{\mu\nu}-
          2
          h\nabla^2\nabla^{\alpha}\nabla^{\beta}h_{\alpha\beta}
\right.\nonumber\\
&&\left.
          +h(\nabla^2)^2 h
          -2\nabla^{\alpha}\nabla^{\beta}h_{\alpha\beta}R_{\mu\nu}h^{\mu\nu} +
          2\nabla^2 h
          R_{\mu\nu}h^{\mu\nu}+R_{\mu\nu}R_{\alpha\beta}h^{\mu\nu}h^{\alpha\beta}\br)\nonumber\\
          &&+
    \frac{1}{2}f'\bl(R\br)\bl(h_{\alpha\beta}\nabla^2 h^{\alpha\beta}
+h\nabla_{\mu}\nabla_{\nu}h^{\mu\nu}
    -4h_{\alpha\beta}\nabla^{\alpha}\nabla^{\mu} h_{\mu}^{\beta}+2
    h_{\alpha\beta}\nabla^{\mu}\nabla^{\alpha} h_{\mu}^{\beta}
\right.\nonumber\\
&&\left.
    -h\nabla^2 h + 4 R_{\mu\nu}h^{\mu\beta}h^{\nu}_{\beta}-
       2 R_{\mu\nu}h^{\mu\nu}h \br)+
    f\bl(R\br)\bl(-\frac{1}{4}h_{\mu\nu} h^{\mu\nu}+ \frac{1}{8} h^2\br)\Biggl\rbrace\ .\nonumber
\end{eqnarray}
All covariant derivatives are with respect
to the background metric, the trace $h^{\mu}_{\,\,\mu}$ is abbreviated as $h$,
prime denotes derivative with respect to $R$.
It is already assumed here that the curvature tensor of the background metric
is covariantly constant.
To this one has to add the gauge fixing terms (\ref{fofrgaugefixing}).
In order to diagonalize the complete expression, we choose a (Euclidean) de Sitter background and
decompose into tensor, vector and scalar parts as we did in section IVB for the
cutoff of type Ib.
Then one obtains for the tensor part
\begin{equation}
\label{grop}
\Gamma^{(2)}_{h^T_{\mu\nu}h^T_{\alpha\beta}}=
-\frac{1}{2}\left[f'\bl(-\nabla^2-\frac{2\bl(d-2\br)}{d\bl(d-1\br)}R\br)+f\right]\delta^{\mu\nu,\alpha\beta}\ ,
\end{equation}
for the vector part
\begin{eqnarray}
\label{xiquadratic}
\Gamma^{(2)}_{\xi_{\mu}\xi_{\nu}}&=&
\bl(-\nabla^2-\frac{R}{d}\br)\left[(\alpha+\beta\nabla^2)\bl(-\nabla^2-\frac{R}{d}\br)+\frac{2R}{d}f'-f\right]
g^{\mu\nu} \, ,
\end{eqnarray}
and for the scalar part (which contains a nontrivial mixing between $h$ and $\sigma$)
\begin{eqnarray}
\label{scalarpart}
\Gamma^{(2)}_{hh}
&=& \frac{d-2}{4d}\left[
\frac{4(d-1)^2}{d(d-2)}f''\bl(-\nabla^2-\frac{R}{d-1}\br)^2+\frac{2(d-1)}{d}f'\bl(-\nabla^2-\frac{R}{d-1}\br)
-\frac{2R}{d}f'+f \right] \nonumber\\
&&-\frac{\rho^2}{d^2}\left[\alpha+\beta\bl(\nabla^2+\frac{R}{d}\br)\right]\nabla^2
\nonumber\\
\Gamma^{(2)}_{h\sigma}
&=&
\frac{d-1}{d^2}\left[(d-1)f''\bl(-\nabla^2-\frac{R}{d-1}\br)
\right.
\nonumber\\
&&
\left.
+\frac{d-2}{2}f'+\rho\bl(\alpha+\beta\bl(\nabla^2+\frac{R}{d}\br)\br)\right]
\nabla^2\bl(\nabla^2+\frac{R}{d-1}\br)
\nonumber\\
\Gamma^{(2)}_{\sigma\sigma}
&=&
\frac{d-1}{2d}\left[
\frac{2(d-1)}{d}f''\nabla^2\bl(\nabla^2+\frac{R}{d-1}\br)-\frac{d-2}{d}f'\nabla^2
+\frac{2R}{d}f'-f
\right.
\nonumber\\
&&
\left.+\frac{2(d-1)}{d}\bl(-\nabla^2-\frac{R}{d-1}\br)\bl(
\alpha+\beta\bl(\nabla^2+\frac{R}{d}\br)\br)\right]\nabla^2\bl(\nabla^2+\frac{R}{d-1}\br)\ .
\end{eqnarray}
We have dropped the subscript $k$ for typographical clarity.

\subsection{Ghost terms}

The Fadeev--Popov ghost consists of two parts.
It is calculated in the usual way from the variation of the gauge condition
and the generators of gauge transformations by
\eq
S_c=\int d^dx\sqrt{g}\,{\bar C}_{\mu}G^\mu{}_\rho\frac{\delta \chi^{\rho}}{\delta
\epsilon^{\nu}}C^{\nu}.
\feq
From the infinitesimal gauge transformation of the gravitational field,
$\delta h_{\mu\nu}=\nabla_{\mu}\epsilon_{\nu}+\nabla_{\nu}\epsilon_{\mu}$,
the variation of the gauge condition under gauge transformations is given by
\begin{eqnarray}
\delta
\chi_{\nu}&=&\nabla^{\mu}\delta h_{\mu\nu}-\frac{1+\rho}{d}\nabla_{\nu}\delta h\nonumber\\
&=&
\nabla^2\epsilon_{\nu}+R_{\mu\nu}\epsilon^{\mu}+\frac{d-2-2\rho}{d}\nabla_{\nu}\nabla_{\mu}\epsilon^{\mu}.
\end{eqnarray}
This gives
\eq
S_c=\int d^dx\sqrt{g}\, {\bar C}_\mu\bl(\alpha+\beta\nabla^2 \br)
\left[\delta^\mu_\nu\nabla^2+R^\mu{}_\nu+\frac{\left(d-2-2\rho\right)}{d}\nabla^{\mu}\nabla_{\nu}\right]
C^{\nu}
\feq
where ${\bar C}_\mu$ and $C^{\mu}$ are the ghost and anti-ghost fields.

As we want to treat also higher-derivative gravity, it is natural to assume
the operator $G^\mu{}_\nu$ can contain derivatives (for $\beta\neq0$).
In that case, one obtains a nonconstant square root of a determinant
in the Fadeev-Popov procedure which on exponentiation gives rise
to the so--called third ghost term \cite{Buchbinder}
\eq
S_b=\frac{1}{2}\int d^dx\sqrt{\bar g}\, b^{\mu} G_{\mu\nu}
b^{\nu}.
\feq
For $\beta=0$ this gives just a constant factor
which can be absorbed in the normalization of the functional integral.
The full ghost action is then $S_{gh}=S_c+S_b$.
The ghost, anti-ghost, and third ghost fields are also decomposed
into transverse and longitudinal parts as in (\ref{ghostdecomposition}).
The operators acting on these fields are:
\begin{eqnarray}
\label{ghop}
\Gamma^{(2)}_{\bar c^T_\mu c^T_\nu} &=&
\bl(\alpha+\beta\nabla^2\br)\bl(\nabla^2+\frac{R}{d}\br)g^{\mu\nu} \nonumber\\
\Gamma^{(2)}_{\bar c c}&=&
-\frac{2\bl(d-1-\rho\br)}{d} \left(\alpha +\beta
\bl(\nabla^2+\frac{R}{d}\br) \right)
\left(\nabla^2+\frac{1}{d-1-\rho}R\right)\nabla^2 \nonumber\\
\Gamma^{(2)}_{b^T_{\mu}b^T_{\nu}}&=&
\bl(\alpha+\beta\,\nabla^2\right)g^{\mu\nu} \nonumber\\
\Gamma^{(2)}_{bb}&=&-\left(\alpha+\beta\bl(\nabla^2+\frac{R}{d}\br)\right)\nabla^2\ .
\end{eqnarray}
Finally, the decomposition of the ghosts gives rise to
Jacobians involving the operators
\eq
J_c=-\nabla^2\ ;\qquad J_b=-\nabla^2\ .
\feq

\subsection{Inserting into the ERGE}

We choose a cutoff of type Ib.
The inverse propagators (\ref{grop},\ref{xiquadratic},\ref{scalarpart},\ref{ghop})
are all functions of $-\nabla^2$.
Then, for each type of tensor components,
the (generally matrix-valued) cutoff function $\mathbf{R_k}$ is chosen
to be a function of $-\nabla^2$ such that
\begin{equation}
\mathbf{\Gamma}^{(2)}(-\nabla^2)+\mathbf{R_k}(-\nabla^2)
=\mathbf{\Gamma}^{(2)}(P_k)\ ,\label{eq:9}
\end{equation}
where $P_k$ is defined as in (\ref{PkI}) for some profile function $R_k$.
Inserting everything into the ERGE (\ref{ERGE}) gives:
\begin{eqnarray}
\label{faerge} \frac{d \Gamma_k}{dt} & = &
\frac{1}{2}\textrm{Tr}_{(2)}\,
\frac{\frac{d}{dt}\mathbf{R}_{h^Th^T}}{\mathbf{\Gamma}^{(2)}_{h^T
h^T}+\mathbf{R}_{h^T h^T}} +\frac{1}{2}\textrm{Tr}'_{(1)}\,
\frac{\frac{d}{dt}R_{\xi\xi}}{\Gamma^{(2)}_{\xi\xi}+R_{\xi\xi}}
+\nonumber \\
 &  &
+\frac{1}{2}\textrm{Tr}''_{(0)}\, \left(
\begin{array}{cc}
\Gamma^{(2)}_{hh}+R_{hh}& \Gamma^{(2)}_{h\sigma}+R_{h\sigma}\nonumber \\
\Gamma^{(2)}_{\sigma h}+R_{\sigma h} & \Gamma^{(2)}_{\sigma\sigma}+R_{\sigma\sigma}
\end{array}
\right)^{-1}
\left(
\begin{array}{cc}
\frac{d}{dt}R_{hh} & \frac{d}{dt}R_{h\sigma}\nonumber \\
\frac{d}{dt}R_{\sigma h} & \frac{d}{dt}R_{\sigma\sigma}
\end{array}
\right)
\nonumber \\
&&+\frac{1}{2}\sum_{j=0,1}
\frac{\frac{d}{dt}R_{hh}(\lambda_j)}{\Gamma^{(2)}_{hh}(\lambda_j)+R_{hh}(\lambda_j)}
\nonumber \\
%ghostpart
&&-\textrm{Tr}_{(1)}\frac{\frac{d}{dt}R_{\bar c^T
c^T}}{\Gamma^{(2)}_{\bar c^T c^T}+R_{\bar c^T c^T}}
  -\textrm{Tr}'_{(0)}\frac{\frac{d}{dt}R_{\bar cc}}{\Gamma^{(2)}_{\bar c c}+R_{\bar cc}}\nonumber \\
&&
+\frac{1}{2}\textrm{Tr}_{(1)}\frac{\frac{d}{dt}R_{b^Tb^T}}{\Gamma^{(2)}_{b^Tb^T}+R_{b^Tb^T}}
+\frac{1}{2}\textrm{Tr}'_{(0)}\frac{\frac{d}{dt}R_{bb}}{\Gamma^{(2)}_{bb}+R_{bb}}\nonumber\\
%Jacobians
&&-\frac{1}{2}\textrm{Tr}'_{(1)}\frac{\frac{d}{dt}R_{J_V}}{J_V+{R}_{J_V}}
-\frac{1}{2}\textrm{Tr}''_{(0)}\frac{\frac{d}{dt}R_{J_S}}{J_S+{R}_{J_S}}\nonumber\\
&&+\textrm{Tr}'_{(0)}\frac{\frac{d}{dt}R_{J_c}}{J_c+{R}_{J_c}}
-\frac{1}{2}\textrm{Tr}'_{(0)}\frac{\frac{d}{dt}R_{J_b}}{J_b+{R}_{J_b}}
\,.\label{eq:8}
\end{eqnarray}
The first three lines contain the contribution from the metric
fluctuation $h_{\mu\nu}$, which has been decomposed into its
irreducible parts according to (\ref{decomposition}). Note that the
trace over the scalar components is doubly primed, since the first
two modes of the $\sigma$ field do not contribute to $h_{\mu\nu}$.
However, the first two modes of $h$ do contribute, and their
contribution to the trace is added separately in the third line. The
fourth and fifth lines contain the contributions of the ghosts and
the third ghost, each decomposed into transverse and longitudinal
parts. Note that the first mode of the scalar (longitudinal) parts
is omitted, as it does not contribute to $C^\mu$ and $b^\mu$
respectively. The sixth line is the contribution of the Jacobians of
the transformation (\ref{decomposition}). These have to carry the
same number of primes as the fields in (\ref{decomposition}). The
same is valid for the traces over the Jacobians resulting from the
split into vector and scalar parts of the ghost and third ghost
field, given in the last line. Eliminating the Jacobians by a
further field redefinition, as in section IVB, would produce some
technically undesirable poles in the heat kernel expansion. For this
reason we shall proceed as in Appendix D and explicitly retain the
Jacobians.

In equation (\ref{faerge}) we have used the convention for the primes
that was used in section IV B and in Appendix D.
This differs from the calculation we did in \cite{CPR} in having less primes
in the traces over the ghosts.
We will discuss this point in some detail in the next section.

\subsection{Discussion of gauge choices}

In the gauge fixing term (\ref{fofrgaugefixing}) we have allowed in principle
three gauge fixing parameters $\rho$, $\beta$, and $\alpha$.
The choice $\rho=1$, $\beta=0$ corresponds to the familiar de Donder gauge;
in the discussion of the Einstein--Hilbert truncation we have further chosen
the gauge fixing parameter $\alpha=Z$, which produces a minimal kinetic operator.
Other values of $\alpha$ have been treated in \cite{Lauscher}.
To avoid the issue of the RG running of $\alpha$, the limit $\alpha/Z\to\infty$
is sometimes invoked.
A gauge fixing with $\beta\not=0$ contains terms with four derivatives and is
natural in higher derivative gravity.
We will only discuss gauge choices where either $\alpha$ or $\beta$
are nonzero, and not both simultaneously. We will call these ``$\alpha$ gauges''
and ``$\beta$ gauges'' respectively.
Note that $\rho$ and $\beta$ are dimensionless but $\alpha$ has dimension of mass squared.
There are two ways of turning it into a dimensionless parameter:
the first is to proceed as with all other couplings and define $\tilde\alpha=\alpha k^{-2}$;
then we can set $a=1/\tilde\alpha$. The second is to proceed as in (\ref{ehgaugefixing})
for the Einstein--Hilbert truncation and set $\alpha=Z/a$.
In the following, when we use $\alpha$ gauges we will always adopt the first method;
the second method yields similar results (up to a rescaling of $a$) and will not be reported.
We will also use the definition $b=1/\beta$. We will always neglect the RG running
of the dimensionless gauge parameters $\rho$, $a$ and $b$.

To reach the highest degree of simplification, a convenient gauge choice
is to set $\rho=0$ and then  take either $\beta=0$ and $\alpha\rightarrow \infty$
or $\alpha=0$ and $\beta\rightarrow \infty$.
We will now show explicitly how this simplification works in the $\beta$ gauges,
treating tensor, vector and scalar components separately.
To write the formulae in a more compact form, in this section we will denote
$\Delta=-\nabla^2$ and we define the following shorthands
\begin{equation*}
\Delta^{(n)}=\Delta-\frac{R}{n}, \,\, P^{(n)}_k=P_k-\frac{R}{n}\ .
\end{equation*}
The transverse traceless tensor part is gauge independent and therefore is not affected
by the gauge choice.
The vector part receives contributions from $\xi$, $\bar c^T_\mu$, $c^{T\mu}$, $b_\mu$, $J_V$:
\begin{eqnarray*}
&&
\frac{1}{2}\textrm{Tr}'_{(1)} \frac {\frac{d}{dt}(
\Gamma^{(2)}_{\xi\xi}(P_k)-\Gamma^{(2)}_{\xi\xi}(\Delta))}{\Gamma^{(2)}_{\xi\xi}(P_k)}
-\textrm{Tr}_{(1)}\frac{ \frac{d}{dt}\left[\bl(\alpha-\beta
P_k\br)P_k^{(4)}-\bl(\alpha-\beta \Delta\br)\Delta^{(4)}\right]}
{\bl(\alpha-\beta P_k\br) P_k^{(4)}}\nonumber\\
&& +\frac{1}{2}\textrm{Tr}_{(1)}\frac {\partial_t\left[
-\beta\bl(P_k-\Delta\br)\right] }{ \bl(\alpha-\beta P_k\br)} -
\frac{1}{2}\textrm{Tr}'_{(1)} \frac {\partial_t
\bl(P_k-\Delta\br)}{P_k^{(4)}}\ .
\end{eqnarray*}
In the first term, looking at (\ref{xiquadratic}) one sees that in the limit $\beta\to\infty$
only the gauge fixing term matters. In this limit the first term becomes simply
\begin{equation*}
\frac{1}{2}\textrm{Tr}'_{(1)}\frac
{\partial_t\left(P_k(P_k^{(4)})^2\right)}{P_k\bl(P_k^{(4)}\br)^2}=
\frac{1}{2}\textrm{Tr}'_{(1)}\frac{\partial_t
R_k}{P_k}+\textrm{Tr}'_{(1)}\frac{\partial_t R_k}{P_k^{(4)}}
\end{equation*}
Treating the other three terms in the same way, several simplifications occur.
However, one has to pay attention to the fact that some of the traces are primed
and some are not. Therefore, in the simplifications the contributions of some isolated
modes are left out. Using the optimized cutoff and specializing to $d=4$, the final result
for the vector terms is
\begin{equation}
-\frac{1}{2}\textrm{Tr}'_{(1)}\frac{\partial_t R_k}{P_k^{(4)}}
-5\frac{\partial_t R_k(\frac{R}{4})}{P_k(\frac{R}{4})}
-10\frac{\partial_t R_k(\frac{R}{4})}{P_k^{(4)}(\frac{R}{4})}\ ,
\end{equation}
where the argument of the last two terms, $\frac{R}{4}$, is the first eigenvalue of
the Laplacian on transverse vectors (see table \ref{taba2}).

Let us now come to the scalar part.
It receives contributions from $h$, $\sigma$, $\bar c$, $c$, $b$, $J_S$, $J_c$ and $J_b$.
The contribution of $h$ and $\sigma$ is given by the second line in eq. (\ref{faerge}).
One sees from (\ref{scalarpart}) that when we set $\rho=0$,
only $\Gamma^{(2)}_{\sigma\sigma}$ and $R_{\sigma\sigma}$ contain $\beta$.
Therefore in the limit $\beta\to\infty$ these terms become
\begin{equation*}
\frac{1}{2}\textrm{Tr}''_{(0)}\, \frac{ \Gamma^{(2)}_{\sigma\sigma}
\partial_t R_{hh} -2 \Gamma^{(2)}_{h\sigma}\partial_t R_{h\sigma}
+\Gamma^{(2)}_{hh} \partial_t R_{\sigma\sigma}}
{\Gamma^{(2)}_{\sigma\sigma}\Gamma^{(2)}_{hh}
-\left(\Gamma^{(2)}_{h\sigma} \right)^2} =
\frac{1}{2}\textrm{Tr}''_{(0)} \frac{\partial_t
R_{hh}}{\Gamma^{(2)}_{hh}(P_k)} +\frac{1}{2}\textrm{Tr}''_{(0)}
\frac{\partial_t
R_{\sigma\sigma}}{\Gamma^{(2)}_{\sigma\sigma}(P_k)}\ .
\end{equation*}
The second term is equal to
\begin{equation*}
\frac{1}{2}\textrm{Tr}''_{(0)}\frac{\partial_t
R_{\sigma\sigma}}{\Gamma^{(2)}_{\sigma\sigma}(P_k)} =
\frac{1}{2}\textrm{Tr}''_{(0)}\left( \frac{\partial_t
R_k}{P_k^{(4)}}+2 \frac{\partial_t R_k}{P_k^{(3)}}+\frac{\partial_t
R_k}{P_k}\right)\ .
\end{equation*}
The longitudinal ghost fields $\bar c$ and $c$ give a contribution
\begin{equation*}
-\textrm{Tr}'_{(0)}\frac{\partial_t(P_k P_k^{(3)} P_k^{(4)}-\Delta
\Delta^{(3)}\Delta^{(4)})}{P_k P_k^{(3)} P_k^{(4)}}=
-\textrm{Tr}'_{(0)}\left( \frac{\partial_t
R_k}{P_k^{(4)}}+\frac{\partial_t R_k}{P_k^{(3)}}+\frac{\partial_t
R_k}{P_k}\right)\ .
\end{equation*}
The third ghost and the Jacobians together contribute
\begin{eqnarray*}
&&\frac{1}{2}\textrm{Tr}'_{(0)}\frac{\partial_t(P_k P_k^{(4)}-\Delta
\Delta^{(4)})}{P_k P_k^{(4)}}
-\frac{1}{2}\textrm{Tr}''_{(0)}\frac{\partial_t(P_k P_k^{(3)}-\Delta
\Delta^{(3)})}{P_k P_k^{(3)}}\nonumber\\
&&+\textrm{Tr}'_{(0)}\frac{\partial_t(P_k-\Delta)}{P_k}
-\frac{1}{2}\textrm{Tr}'_{(0)}\frac{\partial_t(P_k-\Delta)}{P_k}\nonumber\\
&=&\frac{1}{2}\textrm{Tr}'_{(0)}\left(\frac{\partial_t
R_k}{P_k^{(4)}}+\frac{\partial_t R_k}{P_k}\right)
-\frac{1}{2}\textrm{Tr}''_{(0)}\left(\frac{\partial_t
R_k}{P_k^{(3)}}+\frac{\partial_t R_k}{P_k}\right)
+\textrm{Tr}'_{(0)}\frac{\partial_t R_k}{P_k}
-\frac{1}{2}\textrm{Tr}'_{(0)}\frac{\partial_t R_k}{P_k}\ .
\end{eqnarray*}
Thus the scalar terms together give
\begin{equation}
\frac{1}{2}\textrm{Tr}''_{(0)}\frac{\partial_t
R_{hh}}{\Gamma^{(2)}_{hh}(P_k)}
-\frac{1}{2}\textrm{Tr}''_{(0)}\frac{\partial_t R_k}{P_k^{(3)}}
-5\frac{\partial_t R_k(\frac{R}{3})}{P_k^{(3)}(\frac{R}{3})}
-\frac{5}{2}\frac{\partial_t
R_k(\frac{R}{3})}{P_k^{(4)}(\frac{R}{3})} \ .
\end{equation}
The last two terms are evaluated on the second eigenvalue of the Laplacian on scalars, $\frac{R}{3}$.
Finally we can collect the tensor, vector and scalar contributions to obtain
\begin{eqnarray}
%mainpart
\frac{d\Gamma_k}{dt} & = &
\frac{1}{2}\textrm{Tr}_{(2)}\,\left\lbrace \frac{\partial_t{P}_k
f'+(P_k-\Delta) \partial_t{f}'}{( P_k-\frac{R}{3}) f'+ f}
\right\rbrace - \frac{1}{2}\textrm{Tr}'_{(1)}\frac{\partial_t
R_k}{P_k-\frac{R}{4}}
-\frac{1}{2}\textrm{Tr}''_{(0)}\frac{\partial_t R_k}{P_k-\frac{R}{3}}\\
&&+\frac{1}{2}\textrm{Tr}''_{(0)}\left\lbrace \frac{ \partial_t{P}_k
(f'+6 ( P_k-\frac{R}{3}) f'')+(P_k-\Delta) (\partial_t{f}'+ 3(P_k+
\Delta-\frac{2}{3} R)\partial_t{f}'')} {\frac{2}{3} f+(
P_k-\frac{2}{3} R) f'- 3f'' (P_k-\frac{R}{3})^2} \right\rbrace
+\Sigma\nonumber
\end{eqnarray}
where $\Sigma$ is the contribution of the isolated modes.
In the gauge $\rho=0$, $\alpha=0$, $\beta\to\infty$ it is
\begin{equation}
\Sigma=-5\frac{\partial_t R_k(\frac{R}{4})}{P_k(\frac{R}{4})}
-10\frac{\partial_t R_k(\frac{R}{4})}{P_k^{(4)}(\frac{R}{4})}
-5\frac{\partial_t R_k(\frac{R}{3})}{P_k^{(3)}(\frac{R}{3})}
-\frac{5}{2}\frac{\partial_t R_k(\frac{R}{3})}{P_k^{(4)}(\frac{R}{3})}\ .
\end{equation}
The calculation in the gauge  $\rho=0$, $\alpha\to\infty$, $\beta=0$
proceeds in a similar way. The final result is the same except for
the isolated modes, which give
\begin{equation}
\Sigma=-10\frac{\partial_t R_k(\frac{R}{4})}{P_k^{(4)}(\frac{R}{4})}
-10\frac{\partial_t R_k(\frac{R}{3})P_k^{(6)}(\frac{R}{3})}{P_k(\frac{R}{3})P_k^{(3)}(\frac{R}{3})}
+5\frac{\partial_t R_k(\frac{R}{3})}{P_k(\frac{R}{3})}\ .
\end{equation}

Using the formulae in the appendix for the trace evaluation, writing $\tilde R=k^{-2}R$
and $\tilde f=k^{-4}f$, $\tilde f'=k^{-2}f'$, $\tilde f''=f''$, with the optimized cutoff this equation becomes
\begin{eqnarray}
\label{george}
\frac{d\Gamma_k}{dt} &=&
\frac{384\pi^2}{30240\tilde R^2}\Biggl\lbrace
-\frac{1008(511 \tilde R^2-360 \tilde R-1080)}{\tilde R-3}
-\frac{2016 (607 \tilde R^2-360 \tilde R-2160)}{\tilde R-4}\nonumber \\
&&+
20\frac{ (311 \tilde R^3-126 \tilde R^2-22680 \tilde R+45360) 
%k^{-2}
%
(\partial_t\tilde f'+2\tilde f' -2\tilde R\tilde f'')
%\partial_t{ f}'
%
-252 (\tilde R^2+360 \tilde R-1080) \tilde f'}
{3 \tilde f-(\tilde R-3)\tilde f'}\nonumber \\
&&+
\left[1008(29 \tilde R^2+360 \tilde R+1080) \tilde f'+4 (185 \tilde R^3+3654 \tilde R^2+22680 \tilde R+45360)
%
%k^{-2}\partial_t{ f}'
(\partial_t\tilde f'+2\tilde f' -2\tilde R\tilde f'')
\right.
\nonumber\\
&&\left.
-2016(29 \tilde R^3+273
\tilde R^2-3240)\tilde f''-9(181 \tilde R^4+3248 \tilde R^3+15288 \tilde R^2-90720)
%
%\partial_t{f}''
(\partial_t \tilde f''-2\tilde R\tilde f''')
\right]\Bigg/
\nonumber \\
&&
\left(\tilde f'' (\tilde R-3)^2+2 \tilde f+(3-2 \tilde R) \tilde f'\right)
\Biggr\rbrace+\Sigma\ .
\end{eqnarray}
where
\begin{equation}
\label{sigma}
\Sigma=-\frac{10 (\tilde R^2-20 \tilde R+54) \tilde R^2}{\tilde R^2-7 \tilde R+12}\quad{\rm or}\qquad
\Sigma=\frac{10 (11 \tilde R-36)}{\tilde R^2-7 \tilde R+12}
\end{equation}
in the $\beta$-- and $\alpha$--gauge respectively.
The (dimensionful) beta functions can be extracted from this function by taking derivatives:
\begin{equation}
\frac {dg_i}{dt}=\frac{1}{i!}\frac{\partial^i}{\partial
R^i}\frac{1}{V}\frac{d \Gamma_k}{dt}\Biggr|_{R=0}\ ,
\end{equation}
where $V=\int d^4x\sqrt{g}$.
This we have done using algebraic manipulation software.

As already mentioned in section V, different authors have different prescriptions
for the treatment of zero modes of the ghost operator.
In \cite{CPR} we have taken the point of view that since in the gauge
$\rho=0$, $\alpha=0$, $\beta\to\infty$ there is a one-to-one correspondence
between the modes of the unphysical components $\xi_\mu$ and $\sigma$
and those of the ghost field, and since the ghost contribution is
supposed to cancel the contribution of the gauge degrees of freedom of the field,
the cancellation occurs mode by mode, so that the trace over the vector part of the ghost
has to have a prime and the trace over the scalar part of the ghost has to
have a double prime.
A similar argument applies to the third ghost.
Finally, when one makes this choice for the ghosts, then also the Jacobian
determinants $J_c$ and $J_b$ have to have a double prime.
So, altogether, all vector traces would have a prime and all scalar
traces would have a double prime. This amounts to putting $\Sigma=0$.
In the next sections we shall begin by giving the results with this definition of the traces.
Later we shall also describe the effect of having $\Sigma$ as in equation (\ref{sigma}).

\subsection{Results}

\begin{figure}
%[t]\center
{
\resizebox{1 \columnwidth}{!}
{\includegraphics{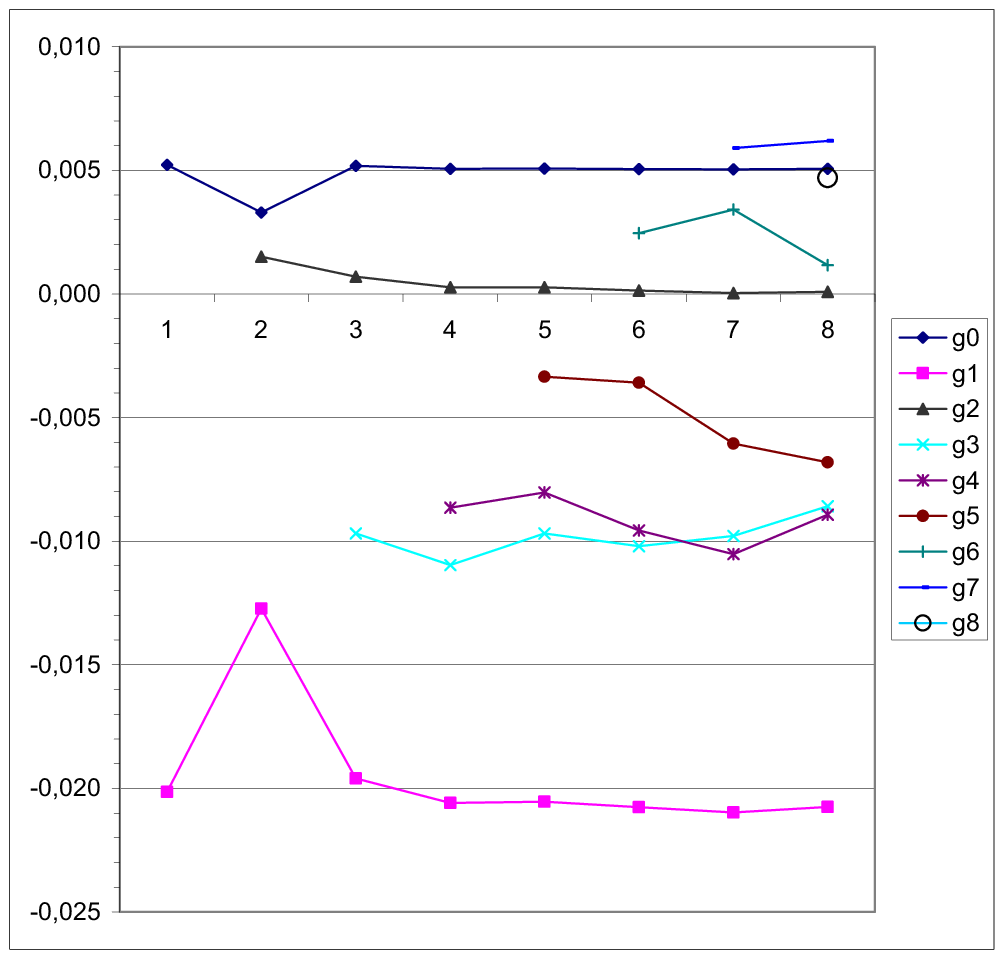}\includegraphics{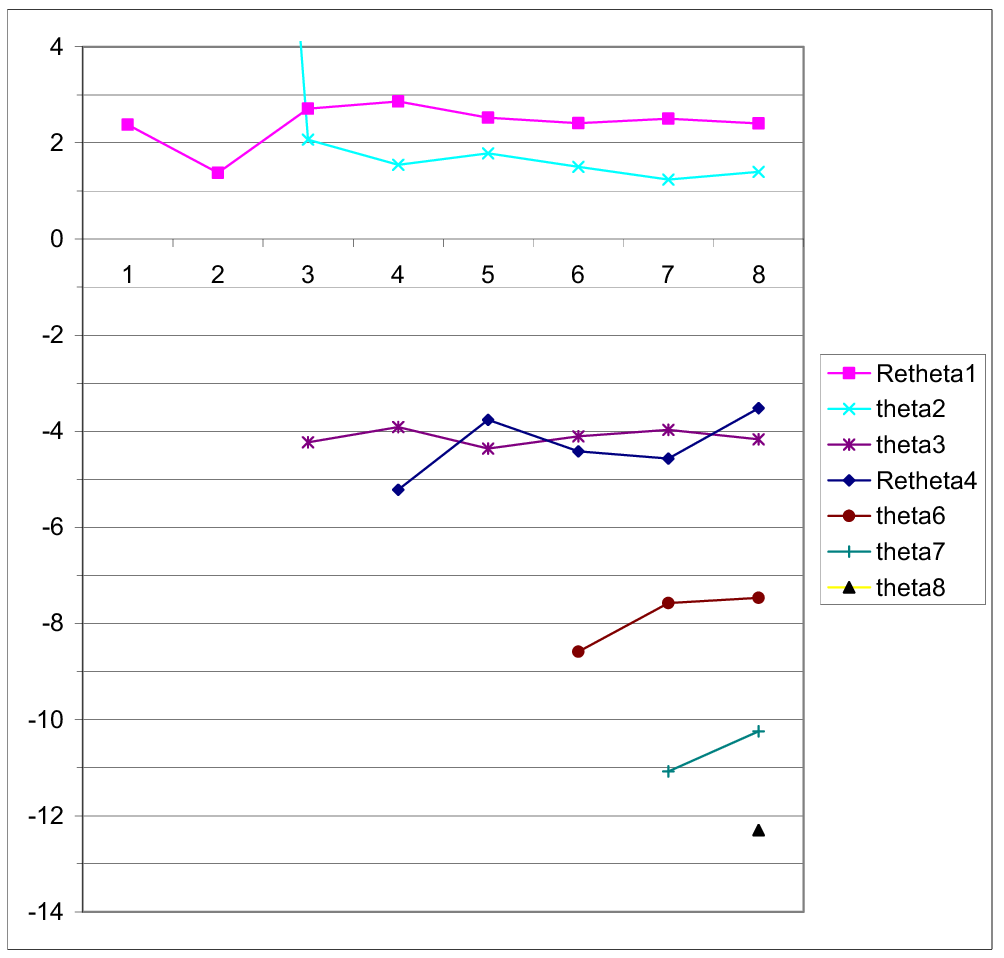}}
\caption
{
\label{fig:sss}
The position of the fixed point (left panel) and the critical
exponents (right panel) as functions of $n$, the order of the truncation.
}}
\end{figure}

We can now state our results. Table \ref{tab:mytable1} gives the
position of the nontrivial FP and table \ref{tab:mytable2} gives the
critical exponents, for truncations ranging from $n=1$ (the
Einstein--Hilbert truncation) to $n=8$. The same information is
shown graphically in figure (\ref{fig:sss}).

\begin{table}
\begin{center}
\begin{tabular}{|c|l|l|l|r|r|r|r|r|r|r|r|r|}
\hline
 $n$ & $\tilde\Lambda_*$&$\tilde G_*$ & $\tilde\Lambda_* \tilde G_*$& \multispan9 \hfil $10^3\times$ \hfil
\vline
\\
\hline
  & & & &$\tilde g_{0*}$ & $\tilde g_{1*}$ & $\tilde g_{2*}$ & $\tilde g_{3*}$
& $\tilde g_{4*}$ & $\tilde g_{5*}$ & $\tilde g_{6*}$& $\tilde g_{7*}$& $\tilde g_{8*}$\\
\hline
1&0.1297 &0.9878 &0.1282 & 5.226& -20.140& & & & & & &\\
2&0.1294 &1.5633 &0.2022 & 3.292& -12.726& 1.514& & & & & &\\
3&0.1323 &1.0152 &0.1343 & 5.184& -19.596& 0.702& -9.682& & & & &\\
4&0.1229 &0.9664 &0.1188 & 5.059& -20.585& 0.270& -10.967& -8.646& & & &\\
5&0.1235 &0.9686 &0.1196 & 5.071& -20.538& 0.269& -9.687& -8.034& -3.349& & &\\
6&0.1216 &0.9583 &0.1166 & 5.051& -20.760& 0.141& -10.198& -9.567& -3.590& 2.460 & &\\
7&0.1202 &0.9488 &0.1141 & 5.042& -20.969& -0.034& -9.784& -10.521 & -6.048& 3.421 & 5.905&\\
8&0.1221 &0.9589 &0.1171 & 5.066& -20.748& 0.088& -8.581& -8.926& -6.808 & 1.165 & 6.196 &4.695\\
\hline
\end{tabular}
\end{center}
\caption{Position of the FP for increasing order $n$ of the truncation.
To avoid writing too many decimals, the values of $\tilde g_{i*}$ have
been multiplied by 1000.}
\label{tab:mytable1}
\end{table}

\begin{table}
\begin{center}
\begin{tabular}{|c|c|c|c|c|c|c|c|c|c|}
\hline
 $n$ & $Re\vartheta_1$ & $Im\vartheta_1$ & $\vartheta_2$ & $\vartheta_3$
& $Re\vartheta_4$ & $Im\vartheta_4$ &  $\vartheta_6$ & $\vartheta_7$& $\vartheta_8$\\
\hline
1& 2.382& 2.168& & & & & & & \\
2& 1.376& 2.325& 26.862& & & & & &\\
3& 2.711& 2.275& 2.068& -4.231& & & & &\\
4& 2.864& 2.446& 1.546& -3.911& -5.216& & & & \\
5& 2.527& 2.688& 1.783& -4.359& -3.761 & -4.880 & & &\\
6& 2.414& 2.418& 1.500& -4.106& -4.418 & -5.975 & -8.583 & &\\
7& 2.507& 2.435& 1.239& -3.967& -4.568 & -4.931 & -7.572 & -11.076 &\\
8& 2.407& 2.545& 1.398& -4.167& -3.519 & -5.153 & -7.464 & -10.242 & -12.298\\
\hline
\end{tabular}
\end{center}
\caption{Critical exponents
for increasing order $n$ of the truncation.
The first two critical exponents $\vartheta_0$ and $\vartheta_1$
are a complex conjugate pair.
The critical exponent $\vartheta_4$ is real in the truncation $n=4$ but for $n\geq 5$ it becomes
complex and we have set $\vartheta_5=\vartheta_4^*$.}
\label{tab:mytable2}
\end{table}

Some comments are in order. First of all, we see that a FP
with the desired properties exists for all truncations considered.
When a new coupling is added, new unphysical FP's tend to appear;
this is due to the approximation of $f$ by polynomials.
A similar phenomenon is known to occur in scalar theory in the local potential approximation \cite{Morris:1994ki,Litim:2002cf}.
However, among the FP's it has always been possible to find one for which
the lower couplings and critical exponents have values that are close to those
of the previous truncation. That FP is then identified as the nontrivial FP
in the new truncation.

Looking at the columns of Tables \ref{tab:mytable1} and
\ref{tab:mytable2} we see that in general the properties of the FP
are remarkably stable under improvement of the truncation. In
particular the projection of the flow in the $\tilde\Lambda$-$\tilde
G$ plane agrees well with the case $n=1$. This confirms the claims
made in \cite{Lauscher} about the robustness of the Einstein--Hilbert
truncation.

The greatest deviations seem to occur in the row $n=2$, and in the columns $g_2$ and $\vartheta_2$.
The value of $g_{2*}$ decreases steadily with the truncation.
The critical exponent $\vartheta_2$ appears for the first time in the truncation $n=2$
with a very large value, but it decreases quickly and seems to converge around $1.5$.
This behavior may be related to the fact that $g_2$ is classically a marginal variable.

The beta function of $g_2$ due to the Einstein--Hilbert action (in the spirit of section V)
was considered first in \cite{granda}; the full truncation $n=2$ has been studied in
\cite{LauscherR2}.
When comparing our results for the case $n=2$ with those of \cite{LauscherR2}, one has to
keep in mind that they generally depend on the shape of the cutoff function.
A significant quantity with very weak dependence on the cutoff function
is the dimensionless product $\Lambda G$.
The value $0.12 \div 0.14$  given in \cite{LauscherR2} for $\Lambda G$
is very close to the value we find in all truncations except $n=2$.
Our value for $\tilde g_{2*}$ in the $n=2$ truncation has the same sign
but is between one half and one third of their value, depending on the cutoff function.
This is another manifestation of the relatively unstable behavior of this variable.
The value given in \cite{LauscherR2} for the critical exponent $\vartheta'$ varies in the range
$2.2 \div 3.2$
depending on the shape of the cutoff, and is in good agreement with our results,
again with the exception of the $n=2$ truncation.
Finally, in \cite{LauscherR2} the critical exponent $\vartheta_2$ has stably
large values of the order of 25 with the compact support cutoffs,
but varies between 28 and 8 with the exponential cutoffs.
The values at the high end agree well with our result in the $n=2$ truncation.
The shape dependence that is observed with exponential cutoffs
can be taken as a warning of the truncation--dependence of this quantity.

Tables \ref{tab:mytable3} and \ref{tab:mytable4} give the
position of the FP and the critical exponents in the truncation
$n=8$, using the definition of the traces with less primes (\ie\
using for $\Sigma$ the values given in equation (\ref{sigma})).
While the numerical results, especially for some of the higher couplings,
do change sensibly, the overall qualitative picture is not affected.
In this connection we also mention that in \cite{Machado:2007ea} the same
calculation has been independently repeated for $n\leq 6$.
The slight numerical differences between their results and those reported here
is due entirely to the fact that they define the traces over the Jacobians
$J_c$ and $J_b$ with a single prime; it has been checked that when the
same definition is used, the results agree perfectly.

\begin{table}
\begin{center}
\begin{tabular}{|c|l|l|l|r|r|r|r|r|r|r|r|r|}
\hline
 gauge & $\tilde\Lambda_*$&$\tilde G_*$ & $\Lambda_* G_*$& \multispan9 \hfil $10^3\times$ \hfil \vline
\\
\hline
  & & & &$\tilde g_{0*}$ & $\tilde g_{1*}$ & $\tilde g_{2*}$ & $\tilde g_{3*}$
& $\tilde g_{4*}$ & $\tilde g_{5*}$ & $\tilde g_{6*}$& $\tilde g_{7*}$& $\tilde g_{8*}$\\
\hline
$\alpha$&0.1239 &0.9674 &0.1199 & 5.096& -20.564& 0.153& -6.726& -5.722& -2.981 & 1.980 & 3.305 & 1.631\\
$\beta$&0.1242 &0.9682 &0.1202 & 5.103& -20.548& 0.138& -6.133& -4.621& -1.407 & 2.240 & 2.207 & 0.610\\
\hline
\end{tabular}
\end{center}
\caption{Position of the FP for $n=8$ taking into account the
contribution of the isolated modes given in (\ref{sigma}). To avoid
writing too many decimals, the values of $\tilde g_{i*}$ have been
multiplied by 1000.} \label{tab:mytable3}
\end{table}

\begin{table}
\begin{center}
\begin{tabular}{|c|r|r|r|r|r|r|r|r|r|}
\hline
 gauge & $Re\vartheta_1$ & $Im\vartheta_1$ & $\vartheta_2$ & $\vartheta_3$
& $Re\vartheta_4$ & $Im\vartheta_4$ &  $\vartheta_6$ & $\vartheta_7$& $\vartheta_8$\\
\hline
$\alpha$& 2.123& 2.796& 1.589& -4.212& -1.107 & 5.558 & -7.321 & -9.923 & -12.223\\
$\beta$& 2.049& 2.511& 1.438& -3.928& -0.102 & 7.320 & -7.239 & -9.664 & -12.381\\
\hline
\end{tabular}
\end{center}
\caption{Critical exponents in the $\alpha$-- and $\beta$--gauge
taking into account the contribution of the isolated modes given in
equation (\ref{sigma}). } \label{tab:mytable4}
\end{table}

\subsection{UV critical surface}

Possibly the most important result of this calculation is that in all truncations
only three operators are relevant.
One can conclude that in this class of truncations
the UV critical surface is three--dimensional. Its tangent space at the FP
is spanned by the three eigenvectors corresponding to the eigenvalues with
negative real part. In the parametrization (\ref{fofr}),
it is the three--dimensional subspace in ${\bf R}^9$ defined by the equation:
\begin{eqnarray}\label{surface}
\tilde g_3&=& 0.0006 + 0.0682\,\tilde g_0 + 0.4635\,\tilde g_1 + 0.8950\,\tilde g_2\nonumber\\
\tilde g_4&=&-0.0092 - 0.8365\,\tilde g_0 - 0.2089\,\tilde g_1 + 1.6208\,\tilde g_2\nonumber\\
\tilde g_5&=&-0.0157 - 1.2349\,\tilde g_0 - 0.7254\,\tilde g_1 + 1.0175\,\tilde g_2\nonumber\\
\tilde g_6&=&-0.0127 - 0.6226\,\tilde g_0 - 0.8240\,\tilde g_1 - 0.6468\,\tilde g_2\nonumber\\
\tilde g_7&=&-0.0008 + 0.8139\,\tilde g_0 - 0.1484\,\tilde g_1 - 2.0181\,\tilde g_2\nonumber\\
\tilde g_8&=& 0.0091 + 1.2543\,\tilde g_0 + 0.5085\,\tilde g_1 - 1.9012\,\tilde g_2
\end{eqnarray}
Unfortunately, we cannot yet conclude from this calculation
that the operators $\calo^{(n)}_i$ with $n\geq6$ would be irrelevant
if one considered a more general truncation:
the beta functions that we computed here are really mixtures of the
beta functions for various combinations of powers of Riemann or Ricci tensors,
which, in de Sitter space, are all indistinguishable.

However, there is a clear trend for the eigenvalues to grow with the power of $R$.
In fact, in the best available truncation, the real parts of the critical exponents
differ from their classical values $d_i$ by at most 2.1, and there is no tendency
for this difference to grow for higher powers of $R$.
This is what one expects to find in an asymptotically safe theory \cite{Weinberg}.

With a finite dimensional critical surface,
one can make definite predictions in quantum gravity.
The real world must correspond to one of the trajectories that emanate from the FP,
in the direction of a relevant perturbation.
Such trajectories lie entirely in the critical surface.
Thus, at some sufficiently large but finite value of $k$
one can choose arbitrarily three couplings,
for example $\tilde g_0$, $\tilde g_1$, $\tilde g_2$
and the remaining six are then determined by (\ref{surface}).
These couplings could then be used to compute the probabilities of physical processes,
and the relations (\ref{surface}), in principle, could be tested by experiments.
The linear approximation is valid only at very high energies, but it should be possible
to numerically solve the flow equations and study the critical surface
further away from the FP.

\begin{figure}
%[t]\center
{\resizebox{1\columnwidth}{!}
{\includegraphics{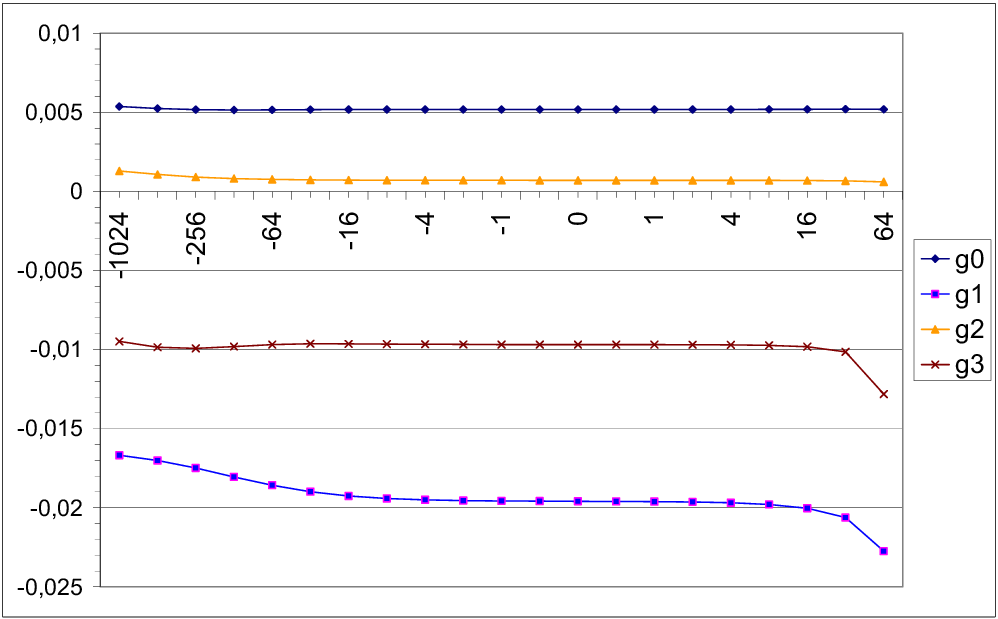}\includegraphics{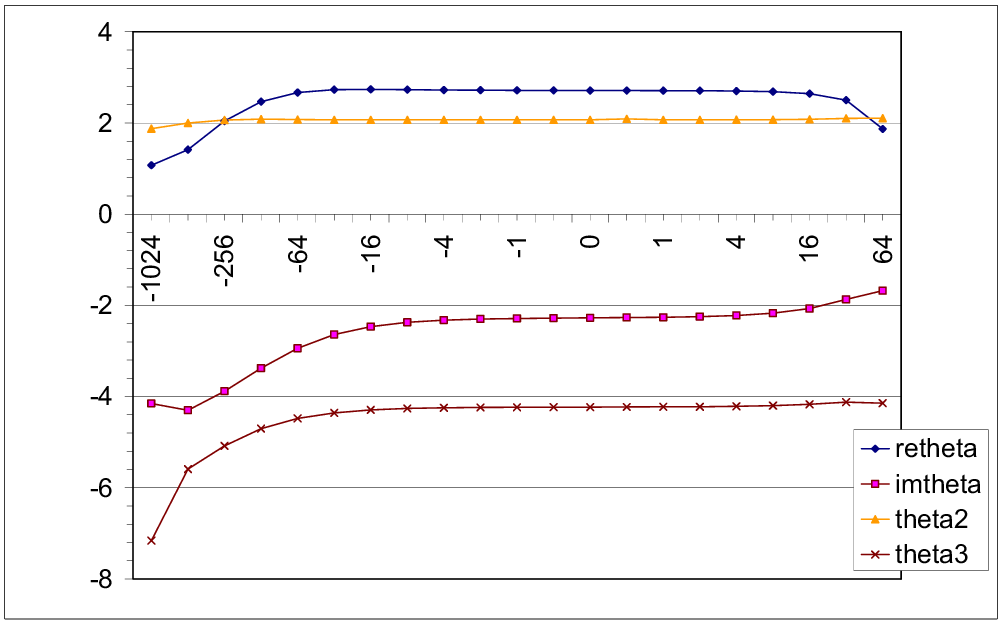}}
\caption{\label{fig:10}The position of the nontrivial fixed point (left
panel) and the critical exponents (right panel) in the $R^3$ truncation,
in the gauge $\rho=0$ and gauge parameter $a$ variable in the
range $-1024<a<64$. Note that the scales are logarithmic for $a>0$
and $a<0$ but not at $a=0$.
}}
\end{figure}

\begin{figure}
%[t]\center
{\resizebox{1\columnwidth}{!}
{\includegraphics{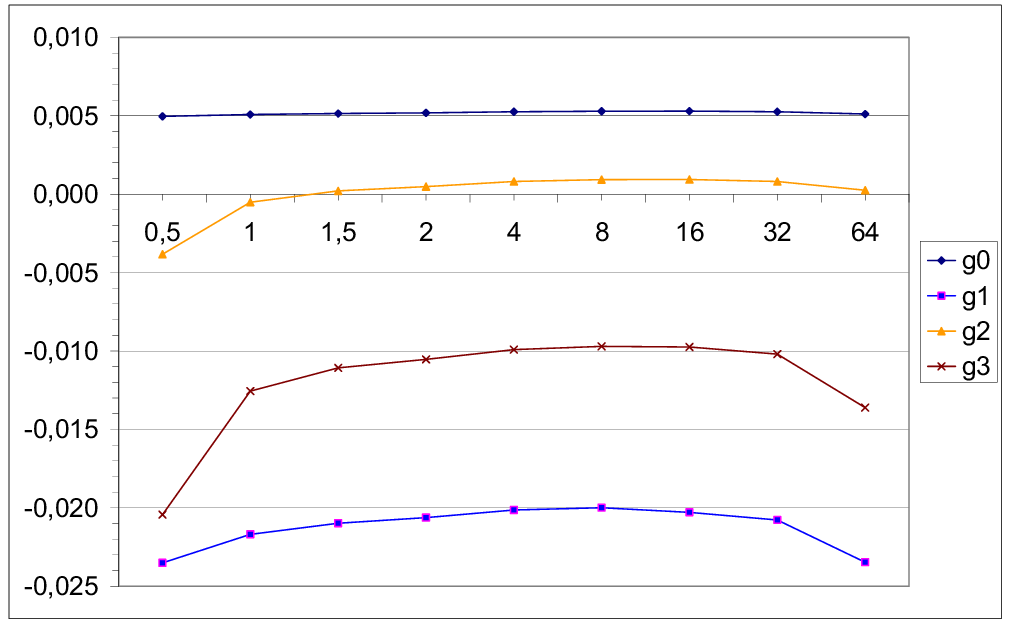}\includegraphics{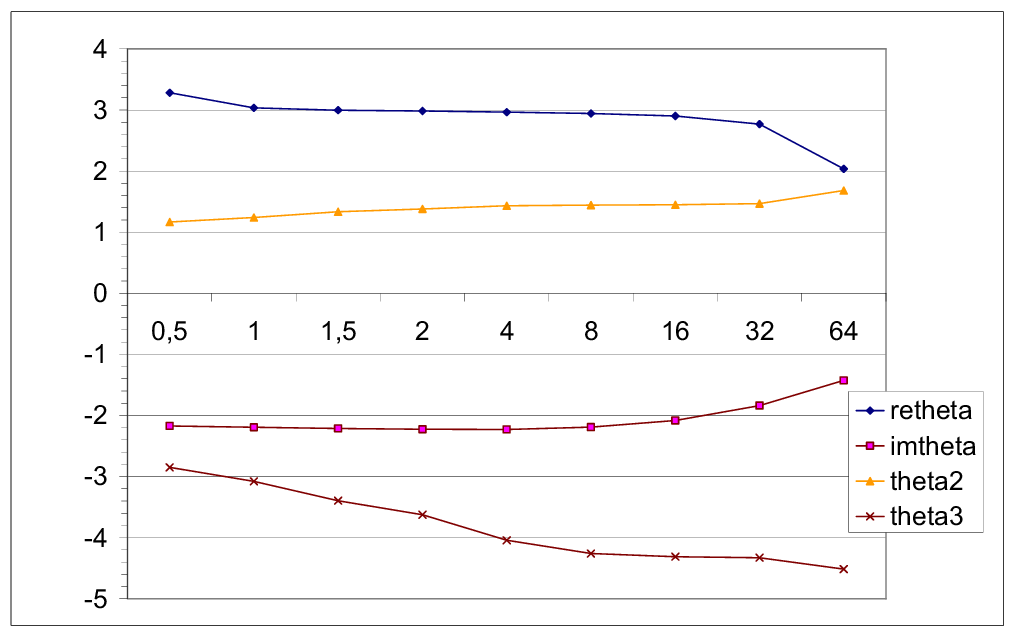}}
\caption{\label{fig:11}The position of the nontrivial fixed point (left
panel) and the critical exponents (right panel) in the $R^3$ truncation,
in the gauge $\rho=1$ and gauge parameter $a$ variable in the
range $0.5<a<64$.
}}
\end{figure}

\begin{figure}
%[t]\center
{\resizebox{1\columnwidth}{!}
{\includegraphics{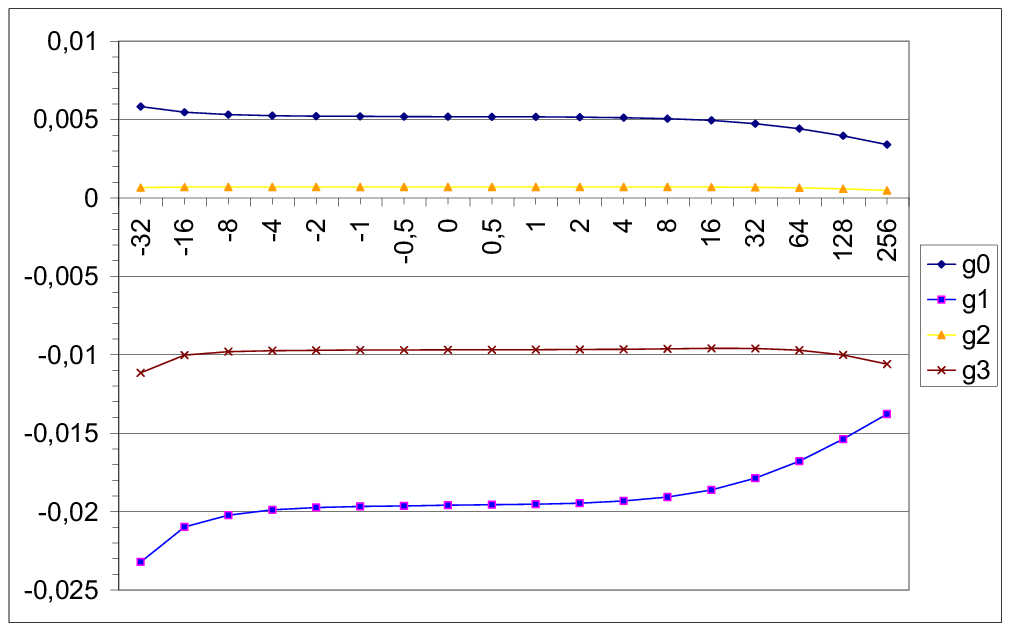}\includegraphics{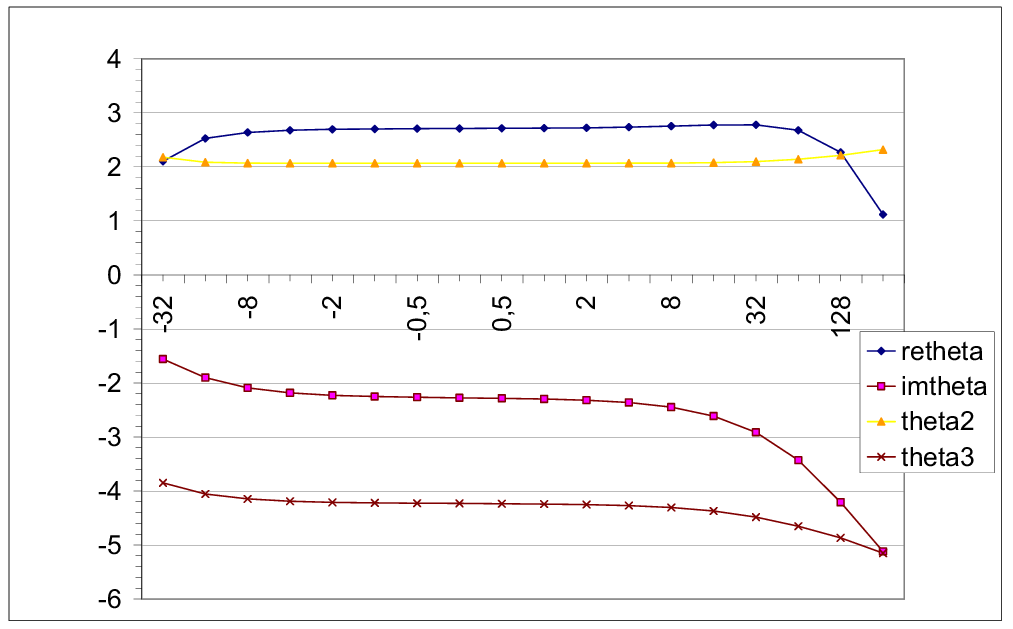}}
\caption{\label{fig:12}The position of the nontrivial fixed point (left
panel) and the critical exponents (right panel) in the $R^3$ truncation,
in the gauge $\rho=0$ and gauge parameter $b$ variable in the
range $-32<b<256$.  Note that the scales are logarithmic for $b>0$
and $b<0$ but not at $b=0$.
}}
\end{figure}

\begin{figure}
%[t]\center
{\resizebox{1\columnwidth}{!}
{\includegraphics{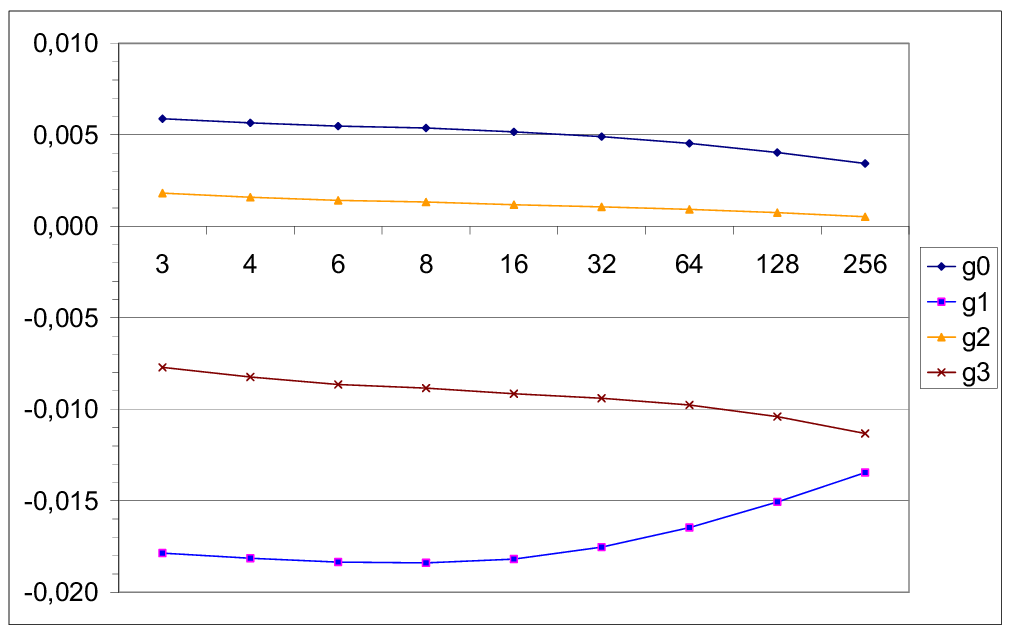}\includegraphics{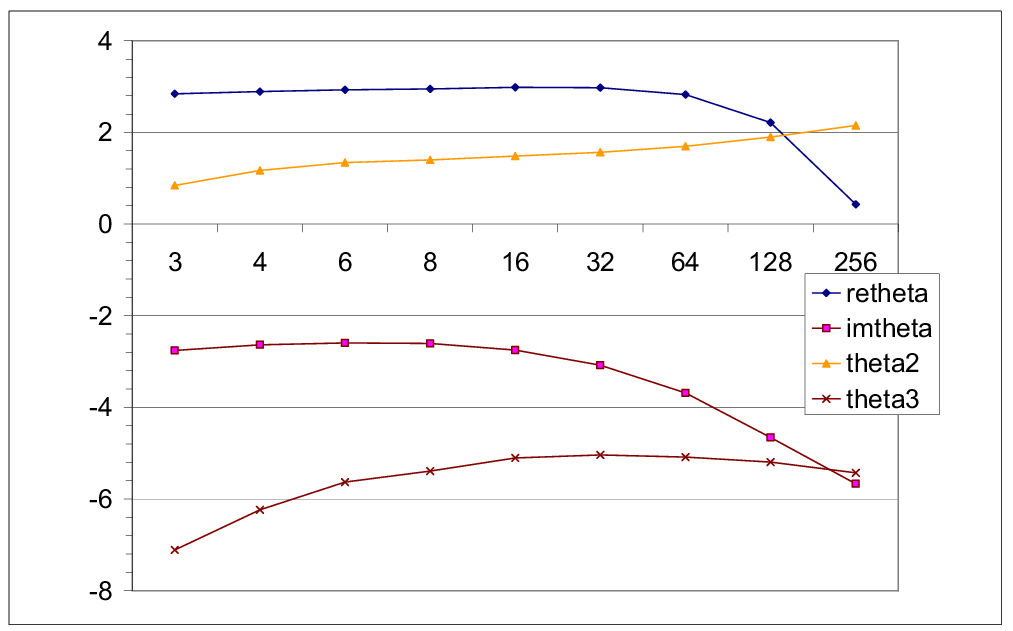}}
\caption{\label{fig:13}The position of the nontrivial fixed point (left
panel) and the critical exponents (right panel) in the $R^3$ truncation,
in the gauge $\rho=1$ and gauge parameter $b$ variable in the
range $3<b<256$.
}}
\end{figure}

\subsection{Scheme dependence in $R^3$--gravity}

The gauge parameters cannot be chosen totally independently.
For example, if one takes $\rho=0$, $\alpha$ and $\beta$
can be taken to infinity or zero, the limits exist and do commute.
If one takes instead $\rho\neq 0$, either $\alpha$ or $\beta$ has to
stay finite.

It would be very cumbersome to check the stability of the results under changes
in the gauge fixing parameters in high truncations.
However, it is possible to do so for $n=3$.
As we shall see, this is enough to verify that $g_3$ is always irrelevant
and therefore is a strong indication that the dimension of the critical surface
is stable under such variations.

Figures \ref{fig:10} and \ref{fig:12} show the position of the FP
and the critical exponents in the gauges $\rho=0$, $\beta=0$,
$\alpha$ variable and $\rho=0$, $\alpha=0$, $\beta$ variable,
respectively. It appears that one can take the limits
$\alpha\to\pm\infty$ and $\beta\to\pm\infty$ without problems and
that the results are continuous.

We have also checked the results in the gauge $\rho=1$, which was
discussed extensively in the case of the Einstein--Hilbert
truncation. Figures \ref{fig:11} and \ref{fig:13} show the position
of the FP and the critical exponents in the gauges $\rho=1$,
$\beta=0$, $\alpha$ variable and $\rho=0$, $\alpha=0$, $\beta$
variable, respectively. It appears that when $\rho=1$ one cannot
take the limits $\alpha\to\infty$ and $\beta\to\infty$, so we have
limited ourselves to positive values of $\alpha$ and $\beta$.

%%%%%%%%%%%%%%%%%%%%%%%%%%%%%%%%%%%%%%%%%%%%%%%%%%%%%%%%%%%%%%%%%%%%%%%%%%%%%%%%%%%%%%

\vfil
\eject

\section{Conclusions}

In this paper we have reviewed and extended recent work on the Wilsonian RG
approach to quantum gravity.
The central hypothesis of this approach is the existence of a nontrivial FP
for gravity, having finitely many UV--attractive directions.
Accordingly, most of the work has gone towards proving the existence of such a FP.
Let us summarize the evidence that has been obtained by applying Wilsonian RG methods to gravity
(for general reviews see also \cite{NiedermaierReuter,Niedermaierrev,revperc}).

In order to start from the simplest setting, we have begun by considering the
contribution of minimally coupled matter fields to the beta functions of the gravitational couplings,
which can be simply obtained from the heat kernel expansion for
Laplace--type operators and give beta functions that are just constants
(see equations (\ref{largenbetas},\ref{nlambdag})).
Such beta functions produce a FP {\it for all gravitational couplings}.
This result is important for two reasons:
the first reason is that in the limit when the number of matter fields
is very large, this is the dominant contribution to the beta functions.
Insofar as the number of matter fields in the real world is large,
and matter couplings can be assumed to be asymptotically free,
this may be a reasonably good approximation for some purposes.
The second reason is that the contribution of gravitons to the gravitational beta
functions, neglecting the RG improvements and considering only the leading term
in the expansion in the cosmological constant, is essentially of the same form
(compare equations (\ref{nlambdag}), (\ref{pertehflow}) and (\ref{piero})).
One may therefore see this as a bare skeleton that
can be dressed by taking into account increasingly more subtle effects.

The simplest way of approximating the ERGE consists in truncating the form of the
effective average action $\Gamma_k$, \ie\ retaining only certain operators and neglecting all others.
It is known from experience with scalar field theory that the most reliable
procedure is not to truncate on the power of the field but rather to
truncate in the number of derivatives.
In the case of gravity the lowest order of the derivative expansion is
the so--called Einstein--Hilbert truncation, where one
retains only the cosmological constant and Newton's constant.
In this case one can treat the ERGE without any further approximation.
The results are therefore ``nonperturbative'', in the sense that
couplings are not required to be small.
In section IV we have discussed in detail several ways of implementing the cutoff
procedure of the ERGE and we have shown that the results are robust
against such changes.
This adds to earlier studies concerning the dependence of the results
under changes of gauge and changes of the cutoff profile and increase
our confidence that the FP is not an artifact of the truncation.
The results have been obtained in arbitrary dimensions $d$.
Since in this approach the dimension is not used as a regulator,
one can follow the position of the FP as a function of $d$
without encountering singularities,
and compare with the results of the epsilon expansion.
We have seen that, numerically, the epsilon expansion gives
a rather poor approximation at $d=4$.

We have then shown how to recover the perturbative divergences from the ERGE.
In particular, we have seen that the one loop divergences obtained by
't Hooft and Veltman \cite{Veltman} can be reproduced starting from the Einstein--Hilbert
truncation of the ERGE, and that they are independent of the cutoff procedure and
of the profile of the cutoff functions.
We have also reproduced the known (scheme--dependent) one loop divergences in the presence
of a cosmological constant.

A more accurate treatment of these divergences requires that
the terms quadratic in curvature be retained from the start,
\ie\ that they are included in the truncations.
Unfortunately, due to technical complications, it has been impossible so far to
treat the curvature--squared truncation (which would constitute the second
order in the derivative expansion) in the same way as the Einstein--Hilbert truncation.
The most complete available treatment, which was described in section VI,
requires that further approximations be made: essentially, one
is just keeping the lowest terms in the perturbative expansion.
Still, the results obtained from the ERGE differ from those that had been calculated before
using more conventional methods.
The beta functions of the dimensionless couplings
$\lambda$, $\xi$, $\rho$ (defined in (\ref{actionansatz}))
coincide with those that had been computed previously.
In this approximation
these couplings are asymptotically free, tending logarithmically to zero from a
well--defined direction in $(\lambda,\xi,\rho)$--space.
However, the beta functions of
$\tilde\Lambda$ and $\tilde G$ contain, in addition to the terms
that were known before, also some new terms that generate
a nontrivial FP.
When the other couplings are set to their FP values,
the flow in the $(\tilde\Lambda,\tilde G)$--plane has the same form
as the perturbative Einstein--Hilbert flow  (\ref{pertehflow}).

In the perturbative approach to Einstein's theory, the one loop
divergences are at most quadratic in the curvature. But the Euler
term is a total derivative and the remaining terms vanish on shell,
so, if we neglect the cosmological constant, all counterterms can be
eliminated by a field redefinition, up to terms of higher order.
One could suspect that in pure gravity the existence of the FP in
truncations involving at most terms quadratic in curvature is in
some way related to the absence of corresponding genuine divergences
in perturbation theory. If this was the case, one might expect that
the FP ceases to exist as soon as one includes in the truncation
terms that correspond to nonrenormalizable divergences in the
perturbative treatment of Einstein's theory.
From the point of view of the ERGE, this looks quite unlikely, for
various reasons.
First of all, there exist examples of
theories that are perturbatively nonrenormalizable while
nonperturbatively renormalizable at a nontrivial FP \cite{calan,gawedzki}.
Second, when one truncates the action $\Gamma_k$, in principle all terms that are
retained could be equally important. The argument about eliminating
terms that contain the Ricci tensor only applies when the higher order
terms are considered as infinitesimal perturbations of the Hilbert
action with zero cosmological constant,
for then at leading order the on shell condition is simply $R_{\mu\nu}=0$.
However, if the higher order terms are not
infinitesimal, the on shell condition is much more complicated and
there is no indication that they can still be eliminated by field
redefinitions. In fact for certain classes of terms it is known
that they can only be eliminated at
the price of introducing a number of scalar fields with new interactions
\cite{Tomboulis2}.
Finally, the calculations reported here already provide evidence to the contrary.
In the presence of a cosmological term
neither $R_{\mu\nu}R^{\mu\nu}$ nor $R^2$ vanish on shell, so,
according to the perturbative reasoning,
none of the FP's discussed in sections VI and VII should exist.
If for some reason one is willing to neglect the cosmological constant,
in pure gravity the first genuine perturbative divergence is cubic
in the Riemann tensor, but in the presence of matter fields already
the one loop logarithmic divergences, which are quadratic in curvature,
do not vanish on shell.
Does this imply that in the presence of matter the FP ceases to exist?
We have seen explicitly in section VI that at least at one loop this is
not the case.
Admittedly, this is only a partial result, and an ``exact'' calculation
would be necessary to definitely settle this point,
but it is very strong indication that the Wilsonian approach
can handle terms that would be troublesome in perturbation theory.
For these reasons we also believe that nothing special will
happen when the Goroff--Sagnotti cubic term (which was used to prove
that pure gravity is perturbatively nonrenormalizable) will be included in the truncation.

While the systematic derivative expansion
cannot for the time being be pushed beyond the fourth order,
one can still consider different truncations that go beyond the Einstein--Hilbert one.
So far, the truncation with the greatest number of free parameters
that can be dealt with exactly is so--called $f(R)$ gravity, where
$f$ is a polynomial in the scalar curvature.
The calculation of the beta functions of this theory was briefly reported
in \cite{CPR} for polynomials up to sixth order
and has been described in greater detail in section VII
for polynomials of order up to eight.
The most important results of these calculations are the
relative stability of the results under the increase
in the number of terms in the truncation,
and the finite dimensionality of the critical surface.
It appears that the critical exponents do not deviate
very strongly from the classical dimensions, as expected,
so that the terms with six or more derivatives are irrelevant.

Further results involving infinitely many couplings were given in
\cite{Niedermaier} in the two Killing vector reduction of gravity
and more recently in \cite{reuterweyer2} in the conformal reduction,
where transverse degrees of freedom of the metric are ignored.
There is some work showing that the FP exists also in the presence
of some types of matter \cite{Perini1}.
Some independent evidence in favor of a gravitational FP is also found in numerical
simulations, both in the causal dynamical triangulation approach
\cite{Ambjorn} and in Regge calculus \cite{Williams}. This concludes
our overview of the currently available evidence for a gravitational FP.
Let us now make a few comments and discuss future prospects.

The dependence of certain results on the choice of the cutoff scheme
(which we called ``scheme dependence'') is sometimes the source of worries.
For example, could the FP for Newton's constant not disappear
if we choose the cutoff function in a perverse way?
This scheme dependence is the counterpart
in the Wilsonian approach of the regularization and renormalization scheme
ambiguities that are encountered in perturbation theory.
As we have discussed in section V, such scheme dependence is to be expected
for all results that concern dimensionful couplings.
We have seen that within certain approximations
(which we called the ``perturbative Einstein--Hilbert flow'')
all the terms with six or more derivatives can indeed be made to
vanish by a choice of cutoff.
However, as we have argued in section V, the general properties of the
cutoff functions are such that
one cannot similarly set to zero Newton's constant
and the cosmological constant.
This is the main reason for the robustness of the nontrivial FP.

Another point that is sometimes a source of misunderstandings
is the use of the term ``exact'' in relation to the ERGE,
and the ``nonperturbative'' character of these calculations.
The beta functions that we have calculated in this paper are
rational functions of the couplings $\tilde g_i$.
The appearance of the couplings in the denominators suggests
that they could be regarded as resummations of infinitely many perturbative terms.
Thus, the beta functions might still be considered ``perturbative'',
in the sense that they are {\it analytic in the coupling constants}.
We think that the ERGE is actually capable of obtaining also results
that are nonanalytic in the coupling constant,
but insofar as the FP is present already in the lowest
order of perturbation theory (see equation (\ref{betabjerrum}))
such a degree of sophistication would be unnecessary.
On the other hand, the beta functions obtained from the ERGE can be said
to be ``nonperturbative'' in the sense
that their validity is not limited to small couplings.
If we compare the one loop beta function (\ref{betabjerrum})
to the beta function (\ref{oneloopforg}) obtained from the ERGE,
they are seen to have the same form.
However, the former calculation by hypothesis is valid only for $\tilde G\ll 1$;
while the latter was obtained from a completely different procedure
and is not similarly limited.
(The difference in the coefficient is not important because the
result is scheme--dependent anyway.)
This suggests that the perturbative result is at least qualitatively
valid also for relatively large values of the coupling.
Notice that the RG improvement modifies the one loop result and
produces singularities of the beta functions.
The nontrivial FP always occurs on the side of the singularity that can be
continuously connected to the Gau\ss ian FP.

The appearance of higher derivative terms in the action at the FP
raises the old issue of unitarity.
The Wilsonian point of view puts that problem in a slightly
different perspective. A tree--level analysis of the action
(\ref{ehaction}) plus (\ref{actionansatz}) shows that it generally contains
besides a massless spin two graviton other particles with a mass of
the order of Planck's mass and a negative residue at the pole
(ghosts) \cite{Stelle}. Some authors have suggested that,
due to RG effects, these particles may not propagate \cite{Julve,Salam}.
From a general Wilsonian point of view, the presence of a propagator pole
at a given mass can only be reliably established by considering the
effective action at a value of $k$ comparable to that mass.
The effective action in the FP regime is probably quite different from
the effective action at the scale of the putative ghost mass, so any
conclusion about the mass spectrum based on the FP--effective action
is probably of little value. This fact is clearly exemplified by QCD.
The tree level analysis of the QCD action, which is known to be a
good description of strong interactions at high energies (\ie\ near the
Gau\ss ian FP), would suggest a spectrum of
particles (quarks and gluons) none of which is observed in nature.
Since the QCD coupling becomes stronger as the energy decreases,
the description of strong interactions in terms of quarks and gluons
becomes extremely complicated long before one reaches the scale of
the quark masses, and it is believed that if one could actually
do such calculations, quarks and gluons would be found not to propagate.
It is conceivable that some similar phenomenon may
occur in gravity, so the FP action (which is the analog of the QCD
action) should not be expected to be a good guide to the particle
spectrum of the theory. The confinement of quarks and gluons is one
of the outstanding problems in particle physics and it is
unfortunately possible that the analogous problem in gravity may
prove equally hard. This is related to the more general question
about the low energy action corresponding to the FP action
but this lies outside the scope of the present paper.
We refer the reader to \cite{Reutercosmology} for some discussions of this point.

Another remarkable aspect of these calculations is that the one loop
flow in the $\tilde\Lambda$--$\tilde G$ plane is essentially the same
(aside from nonuniversal numerical coefficients)
in the Einstein--Hilbert truncation (section IVG) and in the curvature
squared truncation (section VI).
In the latter, $\lambda$ and $\xi$ tend logarithmically to zero
and the corresponding terms in the action diverge.
Thus the dynamics becomes dominated by the four derivative terms,
while in the Einstein--Hilbert truncation it is dominated
by the two derivative term.
It may be somewhat surprising that the structure of the flow
should be so similar in spite of such differences in the dynamics.
This can be at least partly understood by the following argument.
In gravity at low energies (hence in the perturbative regime)
one can consider all couplings to be
scale--independent, and therefore the relative importance of the
terms in the action can be determined simply by counting the number
of derivatives of the metric. For example, at momentum scales
$p^2\ll Z$ (recall that $Z$ is the square of the Planck mass), using
standard arguments of effective field theories, the terms in the
action (\ref{actionansatz}) with four derivatives are suppressed
relative to the term with two derivatives by a factor $p^2/Z$. This
is not the case in the FP regime. If we consider phenomena occurring
at an energy scale $p$, then also the couplings should be evaluated
at $k\approx p$. But then, if there is a nontrivial FP, $Z$ runs
exactly as $p^2$ and therefore both terms are of order $p^4$.
This argument can be generalized to all terms in the derivative expansion
(\ref{Gammak}): at a FP, the running of each coefficient $g^{(n)}_i$,
which is given by the canonical dimension, exactly matches the number
of derivatives in the operator $\calo^{(n)}_i$,
so that all terms are of order $p^4$.
This may at least in part explain the robustness of the results.

When many terms are taken into account in the truncation,
it is hard to have an intuitive feeling for the mechanism that gives rise to the FP.
For example, the beta functions which are obtained by taking derivatives of (\ref{george})
with respect to curvature are exceedingly complicated.
In fact, they are manipulated by the software and one does not even see them.
This is why we have strived in the first few sections of this paper
to emphasize the simplest approximations.
They give a clear and intuitive picture suggesting the emergence of a FP
{\it to all orders in the derivative expansion}.
We would therefore like to conclude by overturning a common belief:
the existence of a nontrivial FP does not require a delicate cancellation of terms.
The FP appears essentially due to the dimensionful nature
of the coupling constants, and it can be seen already in the
perturbative Einstein--Hilbert flow
(\ie\ in the approximation where one considers just the contribution of gravitons
or matter fields with kinetic operators of the form
$-\nabla^2+\mathbf{E}$, where $\mathbf{E}$ is linear in curvature).
More advanced approximations dress up this simple result with
RG improvements and with the contribution of additional couplings.
The argument in the preceding paragraph suggests that the new couplings
will not qualitatively change the results.
And indeed, so far it seems that generically such dressing does not spoil the FP.
So, to conclude on an optimistic note, one could say that it would
actually require a special conspiracy by the new terms to undo the perturbative FP.
\bigskip
\centerline{\bf Acknowledgements} We would like to thank D. Litim and
M. Reuter for many informative discussions,
P. Machado and F. Saueressig for detailed comparisons of our calculations
and G. Narain for a careful reading.
R.P. also wishes to thank S. Weinberg for correspondence and encouragement,
and R. Loll for hospitality at the University of Utrecht, where this work was
completed.

\vfil
\eject

\appendix

\section{Trace technology}

The r.h.s. of the ERGE is the trace of a function of a differential operator.
To illustrate the methods employed to evaluate such traces,
we begin by considering the covariant Laplacian in a metric $g$, $-\nabla^2$.
If the fields carry a representation of a gauge group $G$ and are coupled
to gauge fields for $G$, the covariant derivative $\nabla$ contains also these fields.
We will denote $\Delta=-\nabla^2\mathbf{1}+\mathbf{E}$
a second order differential operator.
$\mathbf{E}$ is a linear map acting on the spacetime and internal
indices of the fields.
In our applications to de Sitter space it will have the form
$\mathbf{E}=q R\, {\bf 1}$ where $\mathbf{1}$ is the identity in the space of the fields
and $q$ is a real number.

The trace of a function $W$ of the operator $\Delta$ can be written as
\begin{equation}
\label{a1}
{\rm Tr}W(\Delta)=\sum_i W(\lambda_i)
\end{equation}
where $\lambda_i$ are the eigenvalues of $\Delta$.
Introducing the Laplace anti-transform $\tilde W(s)$
\begin{equation}
\label{a2}
W(z)=\int_{0}^{\infty}ds\,e^{-zs} \tilde W(s)
\end{equation}
we can rewrite (\ref{a1}) as
\begin{equation}
\label{a3}
{\rm Tr}W(\Delta)=\int_0^\infty ds {\rm Tr}K(s)\tilde W(s)
\end{equation}
where ${\rm Tr}K(s)=\sum_i e^{-s\lambda_i}$ is the trace of the heat kernel of $\Delta$.
We assume that there are no negative and zero eigenvalues; if present, these will
have to be dealt with separately.
The trace of the heat kernel of $\Delta$ has the well-known asymptotic expansion for $s\to 0$:
\eq
\label{heatkernel}
{\rm Tr} \bl(e^{-s\Delta}\br)=
\frac{1}{\bl(4\pi\br)^{\frac{d}{2}}}
\left[B_0\bl(\Delta\br)s^{-\frac{d}{2}}+B_2\bl(\Delta\br)s^{-\frac{d}{2}+1}
+\ldots+B_d\bl(\Delta\br)+B_{d+2}\bl(\Delta\br)s+...\right]
\feq
where $B_n=\int{\rm d}^d\, x\sqrt{g}{\rm tr}\mathbf{b}_n$ and $\mathbf{b}_n$ are linear combinations
of curvature tensors and their covariant derivatives containing $2n$ derivatives of the metric.

Assuming that $[\Delta,\mathbf{E}]=0$, the heat kernel coefficients of $\Delta$
are related to those of $-\nabla^2$ by
\eq
\label{shifted}
{\rm Tr} e^{-s(-\nabla^2+\mathbf{E})} =
\frac{1}{\bl(4\pi\br)^{\frac{d}{2}}}
\sum_{k,\ell=0}^{\infty}\frac{\bl(-1\br)^\ell}{\ell!}
\int{\rm d}^d\,x\sqrt{g}\,{\rm tr}\, \mathbf{b}_k(\Delta)\mathbf{E}^\ell s^{k+\ell-2}.
\feq

The first six coefficients have the following form \cite{Gilkey}:
\begin{eqnarray}
\mathbf{b}_0&=&\bf 1\\
\mathbf{b}_2&=& \frac{R}{6}{\bf 1}-\mathbf{E}\\
\label{b4}
\mathbf{b}_4&=& \frac{1}{180}
\bl(R^{\mu\nu\alpha\beta}R_{\mu\nu\alpha\beta}-R^{\mu\nu}R_{\mu\nu}+\frac{5}{2}R^2+6\nabla^2 R\br) {\bf 1}\nonumber\\
&&+\frac{1}{12}\mathbf{\Omega}_{\mu\nu}\mathbf{\Omega}^{\mu\nu}-\frac{1}{6}R {\bf E}+\frac{1}{2}{\bf E}^2-\frac{1}{6}\nabla^2{\bf E}
\\
\mathbf{b}_6&=& \frac{1}{6}\Biggl\{\frac{1}{180}R{\bf 1}
\bl(R^{\mu\nu\alpha\beta}R_{\mu\nu\alpha\beta}-R^{\mu\nu}R_{\mu\nu}+\frac{5}{6}R^2+\frac{7}{2}\nabla^2
R\br)\nonumber\\
&&+\frac{R}{2}{\bf E}^2+{\bf E}^3
+\frac{1}{30}\mathbf{E}\bl(R^{\mu\nu\alpha\beta}R_{\mu\nu\alpha\beta}-R^{\mu\nu}R_{\mu\nu}+\frac{5}{2}R^2+6\nabla^2
R\br)\nonumber
\\&&+\frac{R}{12}\mathbf{\Omega}_{\mu\nu}\mathbf{\Omega}^{\mu\nu}
+\frac{1}{2}\mathbf{E}\mathbf{\Omega}_{\mu\nu}\mathbf{\Omega}^{\mu\nu}
+\frac{1}{2}\mathbf{E}\nabla^2 \mathbf{E}
-\frac{1}{2}\mathbf{J}_{\mu}\mathbf{J}^{\mu}\nonumber\\
&&+\frac{1}{30}\bl(
2\mathbf{\Omega}^{\mu}_{\;\;\nu}\mathbf{\Omega}^{\nu}_{\;\;\alpha}\mathbf{\Omega}^{\alpha}_{\;\;\mu}
-2R^{\mu}_{\;\;\nu}\mathbf{\Omega}_{\mu\alpha}\mathbf{\Omega}^{\alpha\nu}
+R^{\mu\nu\alpha\beta}\mathbf{\Omega}_{\mu\nu}\mathbf{\Omega}_{\alpha\beta}\br)\nonumber\\
&&+{\bf 1}\left[-\frac{1}{630}R\nabla^2 R
+\frac{1}{140}R_{\mu\nu}\nabla^2 R^{\mu\nu}
+\frac{1}{7560}\bl(
-64R^{\mu}_{\;\;\nu}R^{\nu}_{\;\;\alpha}R^{\alpha}_{\;\;\mu}
+ 48 R^{\mu\nu}R_{\alpha\beta}R^{\alpha\;\;\beta}_{\;\;\mu\;\;\nu}\right.\right.\nonumber\\
&&+\left.\left.
6 R_{\mu\nu}R^{\mu}_{\;\;\rho\alpha\beta}R^{\nu\rho\alpha\beta}
+ 17R_{\mu\nu}^{\;\;\;\;\alpha\beta}R_{\alpha\beta}^{\;\;\;\;\rho\sigma}R_{\rho\sigma}^{\;\;\;\;\mu\nu}
-28R^{\mu\;\;\nu}_{\;\;\alpha\;\;\beta}R^{\alpha\;\;\beta}_{\;\;\rho\;\;\sigma}R^{\rho\;\;\sigma}_{\;\;\mu\;\;\nu}
\br)
\right]\Biggr\}\ ,
\end{eqnarray}
where $\mathbf{\Omega}^{\mu\nu}=\left[\nabla^{\mu},\nabla^{\nu}\right]$ is the curvature
of the connection acting on a set of fields in a particular representation
of the Lorentz and internal gauge group and $\mathbf{J}_{\mu}=\nabla_{\alpha}\mathbf{\Omega}^{\alpha}_{\;\;\mu}$.
We neglect total derivative terms.
The coefficient $\mathbf{b}_8$, which is also used in this work, is much too long
to write here, and can be found in \cite{Avramidi}.
These coefficients are for unconstrained fields. The ones for fields satisfying
differential constraints such as $h_{\mu\nu}^T$ and $\xi_{\mu}$ in the field
decompositions (\ref{decomposition}) are given in the following appendix.

Let us return to equation (\ref{a3}).
If we are interested in the local behavior of the theory (\ie\ the behavior
at length scales much smaller than the typical curvature radius)
we can use the asymptotic expansion (\ref{heatkernel}) and then evaluate
each integral separately.
Then we get
\begin{align}
\label{HKasymp}
{\rm Tr}W(\Delta)=\frac{1}{\bl(4\pi\br)^{\frac{d}{2}}}
\bigl[&Q_{\frac{d}{2}}(W)B_0(\Delta)+Q_{\frac{d}{2}-1}(W)B_2(\Delta)+\ldots\nonumber\\
&+Q_0(W)B_{d}(\Delta)+Q_{-1}(W)B_{d+2}(\Delta)
+\ldots\bigr]\ ,
\end{align}
where
\begin{equation}
\label{q1}
Q_n(W)=\int_0^\infty ds s^{-n}\tilde W(s)\ .
\end{equation}
In the case of four dimensional field theories,
it is enough to consider integer values of $n$.
However, in odd dimensions half-integer values of $n$ are needed
and we are also interested in the analytic continuation
of results to arbitrary real dimensions.
We will therefore need expressions for (\ref{q1})
that hold for any real $n$.

If we denote $W^{(i)}$ the $i$-th derivative of $W$, we have from (\ref{a2})
\begin{equation}
%\label{}
W^{(i)}(z)=(-1)^{i}\int_{0}^{\infty}ds\, s^{i}e^{-zs}\tilde W(s)\ .
\end{equation}
This formula can be extended to the case when $i$ is a real number
to define a notion of ``noninteger derivative''.
From this it follows that for any real $i$
\begin{equation}
\label{beauty}
Q_{n}(W^{(i)})  =  (-1)^{i}Q_{n-i}(W)\ .
\end{equation}
For $n$ a positive integer one can use the definition of the
Gamma function to rewrite (\ref{q1}) as a Mellin transform:
\begin{equation}
\label{Qnpos}
Q_{n}(W)=\frac{1}{\Gamma(n)}\int_{0}^{\infty}dz\, z^{n-1}W(z)
\end{equation}
while for $m$ a positive integer or $m=0$
\begin{equation}
\label{Qnneg}
Q_{-m}(W) = (-1)^{m}W^{(m)}(0)\ .
\end{equation}
More generally, for $n$ a positive real number we can define $Q_n(W)$
by equation (\ref{Qnpos}), while for $n$ real and negative we can
choose a positive integer $k$ such that $n+k>0$;
then we can write the general formula
\begin{equation}
%\label{}
Q_{n}(W)=\frac{(-1)^{k}}{\Gamma(n+k)}\int_{0}^{\infty}dz\, z^{n+k-1}W^{(k)}(z)\ .
\end{equation}
This reduces to the two cases mentioned above when $n$ is integer.
In the case when $n$ is a negative half integer $n=-\frac{2m+1}{2}$ we will set $k=m+1$ so that we have
\begin{equation}
%\label{}
Q_{-\frac{2m+1}{2}}(W)=\frac{(-1)^{m+1}}{\sqrt{\pi}}\int_{0}^{\infty}dz\,
z^{-1/2}f^{(m+1)}(z)
\end{equation}

Let us now consider some particular integrals that are needed in this paper.
As discussed in section \ref{Cutoffschemes}, there are two natural choices of cutoff function:
type I cutoff is a function $R_k(-\nabla^2)$ such that the modified inverse propagator
is $P_k(-\nabla^2)=-\nabla^2+R_k(-\nabla^2)$; type II cutoff is the same function
but its argument is now the entire inverse propagator: $R_k(\Delta)$,
such that the modified inverse propagator is
$P_k(\Delta)=\Delta+R_k(\Delta)$.

We now restrict ourselves to the case when $\mathbf{E}=q\mathbf{1}$,
so that we can write $\Delta=-\nabla^2+q\mathbf{1}$. The evaluation of the
r.h.s. of the ERGE reduces to knowledge of the heat kernel
coefficients and calculation of integrals of the form
$Q_n\left(\frac{\partial_t R_k}{(P_k+q)^\ell}\right)$. It is
convenient to measure everything in units of $k^2$. Let us define
the dimensionless variable $y$ by $z=k^2 y$; then $R_k(z)=k^2 r(y)$
for some dimensionless function $r$, $P_k(z)=k^2(y+r(y))$ and
$\partial_t R_k(z)=2 k^2(r(y)-y r'(y))$.

In general the coefficients $Q_n(W)$ will depend on the details of
the cutoff function. However, if $q=0$ and $\ell=n+1$ they turn out
to be independent of the shape of the function.
Note that they are all dimensionless.
For $n>0$, as long as $r(0)\not= 0$:
\begin{equation}
\label{univpos} Q_n\left(\frac{\partial_t R_k}{P_k^{n+1}}\right)
=\frac{2}{\Gamma(n)}\int_0^\infty dy \frac{d}{dy}
\left[\frac{1}{n}\frac{y^n}{(y+r)^n}\right] =\frac{2}{n!}\ .
\end{equation}
Similarly, if $r(0)\not= 0$ and $r'(0)$ is finite,
\begin{equation}
\label{univzero} Q_0\left(\frac{\partial_t R_k}{P_k}\right)=2\ .
\end{equation}
Finally, for $n=-m<0$
\eq Q_n\left(\frac{\partial_t
R_k}{P_k^{1-m}}\right)\Big|_{y=0}=\bl(-1\br)^m\left(\frac{\partial_t
R_k}{P_k^{1-m}}\right)^{\bl(m\br)}\Big|_{y=0}
=\sum_{n=0}^m\binom{m}{n} \bl(r-y \, r'\br)^{\bl(n\br)}
\bl(y+r\br)^{\bl(m-1\br)}\Big|_{y=0}=0 \feq as $\bl(r-y \,
r'\br)^{\bl(n\br)}=r^{\bl(n\br)}-y\,r^{\bl(n+1\br)}-r^{\bl(n\br)}=-y\,r^{\bl(n+1\br)}$
which vanishes at $y=0$.
This concludes the proof that $Q_n\left(\frac{\partial_t
R_k}{P_k^{n+1}}\right)$ are scheme--independent.

Regarding the other
coefficients $Q_n\left(\frac{\partial_t R_k}{(P_k+q)^\ell}\right)$
whenever explicit evaluations are necessary, we will use the
so-called ``optimized cutoff function'' \cite{Litim}
\begin{equation}\label{optimizedcutoff}
 R_{k}(z)=(k^{2}-z)\theta(k^{2}-z)
\end{equation}
With this cutoff $\partial_tR_k=2k^2\theta(k^2-z)$.
Since the integrals are all cut off at $z=k^2$ by the theta function in
the numerator, we can simply use $P_k(z)=k^2$ in the integrals.
For $n>0$ we have
\begin{equation}
\label{qoptpos} Q_n\left(\frac{\partial_t R_k}{(P_k+q)^\ell}\right)=
\frac{2}{n!}\frac{1}{(1+\tilde q)^\ell}k^{2(n-\ell+1)}
\end{equation}
where $\tilde q=q k^{-2}$.
For $n=0$ we have
\begin{equation}
\label{qoptzero} Q_0\left(\frac{\partial_t
R_k}{(P_k+q)^\ell}\right)=\frac{\partial_t
R_k}{(P_k+q)^\ell}\Biggr|_{z=0}= \frac{2}{(1+\tilde
q)^\ell}k^{2(-\ell+1)}\ .
\end{equation}
Finally, owing to the fact that the function $\frac{\partial_t R_k(z)}{(P_k(z)+q)^\ell}$
is constant in an open neighborhood of $z=0$, we have
\begin{equation}
\label{qoptneg} Q_n\left(\frac{\partial_t
R_k}{(P_k+q)^\ell}\right)=0\ \ {\rm for}\ n<0\ .
\end{equation}
This has the remarkable consequence that with the optimized cutoff the trace in the ERGE
consists of finitely many terms.

For noninteger $n$ let us calculate
\eq
Q_{-\frac{2n+1}{2}}\left(\frac{\partial_{t}R_{k}}{P_{k}}\right)=
\frac{(-1)^{n+1}}{\sqrt{\pi}}\int_{0}^{\infty}dz\, z^{-1/2}
\frac{d^{n+1}}{dx^{n+1}}\frac{\partial_{t}R_{k}(z)}{P_{k}(z)}
\feq
where $P_{k}(z)=z+(k^{2}-z)\theta(k^{2}-z)$. We change the variable
to $x=z/k^{2}$ so we have
\eq
Q_{-\frac{2n+1}{2}}\left(\frac{\partial_{t}R_{k}}{P_{k}}\right)=
\frac{(-1)^{n+1}k^{-(2n+1)}}{\sqrt{\pi}}\int_{0}^{\infty}dx\, x^{-1/2}\frac{d^{n+1}}{dx^{n+1}}\frac{2x\theta(1-x)}{x+(1-x)\theta(1-x)}
\feq
We find
\begin{eqnarray}
\int_{0}^{\infty}dx\, x^{-1/2}\frac{d}{dx}f(x) & = & 2 \nonumber\\
\int_{0}^{\infty}dx\, x^{-1/2}\frac{d^{2}}{dx^{2}}f(x) & = & -5
\end{eqnarray}
so that
\begin{eqnarray}
Q_{-1/2}\left(\frac{\partial_{t}R_{k}}{P_{k}}\right) & = & -\frac{2}{\sqrt{\pi}k}\\
Q_{-3/2}\left(\frac{\partial_{t}R_{k}}{P_{k}}\right) & = & -\frac{5}{\sqrt{\pi}k^{3}}
\nonumber
\end{eqnarray}

We also need some $Q$-functionals of $\frac{R_k}{(P_k+q)^\ell}$.
For $n>0$ we have
\begin{equation}
\label{qoptposR}
Q_n\left(\frac{R_k}{(P_k+q)^\ell}\right)=
\frac{1}{(n+1)!}\frac{1}{(1+\tilde q)^\ell}k^{2(n-\ell+1)}\ .
\end{equation}
The function $\frac{R_k(z)}{(P_k(z)+q)^\ell}$
is equal to $\frac{k^2-z}{(k^2+q)^\ell}$ in an open neighborhood of $z=0$; therefore
\begin{equation}
\label{qoptzeroR}
Q_0\left(\frac{R_k}{(P_k+q)^\ell}\right)=\frac{R_k}{(P_k+q)^\ell}\Biggr|_{z=0}=
\frac{1}{(1+\tilde q)^\ell}k^{2(-\ell+1)}
\end{equation}
\begin{equation}
\label{qoptnegR}
Q_{-1}\left(\frac{R_k}{(P_k+q)^\ell}\right)=\frac{1}{(1+\tilde q)^\ell}k^{-2\ell}\ ,\ \ \
Q_n\left(\frac{R_k}{(P_k+q)^\ell}\right)=0\ \ {\rm for}\ n<-1\ .
\end{equation}
Finally, for the type III cutoff one also needs the following
\begin{equation}
\label{qoptone}
Q_n\left(\frac{1}{(P_k+q)^\ell}\right)=
\frac{1}{n!}\frac{k^{2(n-\ell)}}{(1+\tilde q)^\ell}\ \ {\rm for}\ n\geq 0\ ;\ \
Q_n\left(\frac{1}{(P_k+q)^\ell}\right)=0\ \ {\rm for}\ n<0\ .
\end{equation}

In conclusion let us address a general problem concerning the choice of
the operator $\calo$, whose eigenfunctions are taken as a basis in the functional space.
In some calculations the r.h.s. of the ERGE takes the form
$\frac{1}{2}{\rm Tr}\frac{\partial_t R_k(\Delta+q\mathbf{1})}{P_k(\Delta+q\mathbf{1})}$ where
$\Delta$ is an operator and $q$ is a constant.
Equation (\ref{HKasymp}) tells us how to compute the trace of this function, regarded
as a function of the operator $\Delta+q\mathbf{1}$.
In the derivation of this result it was implicitly assumed that $\calo=\Delta+q\mathbf{1}$.
However, the trace must be independent of the choice of basis in the functional space.
It is instructing to see this explicitly, namely to evaluate the trace
regarding $\frac{1}{2}{\rm Tr}\frac{\partial_t R_k(\Delta+q\mathbf{1})}{P_k(\Delta+q\mathbf{1})}$ as a
function of $\Delta$.
Given any function $W(z)$ we can define $\bar W(z)=W(z+q)$; in general, expanding in $q$ we then have
\begin{eqnarray}
\label{qfunctions}
Q_{n}(\bar W) & = & \frac{1}{\Gamma(n)}\int_0^\infty dz\, z^{n-1} W(z+q)\nonumber\\
 & = &
\frac{1}{\Gamma(n)}\int_0^\infty dz\, z^{n-1} (W(z)+q W'(z)+\frac{1}{2!}q^2 W''(z)
+\frac{1}{3!}q^3 W'''(z)\ldots)\nonumber\\
 & = & Q_{n}(W)+q Q_{n}(W')+\frac{1}{2!}q^2 Q_n(W'')+\frac{1}{3!}q^3 Q_n(W''')
+\ldots\nonumber\\
 & = & Q_{n}(W)-q Q_{n-1}(W)+\frac{1}{2!}q^2 Q_{n-2}(W)
-\frac{1}{3!}q^3 Q_{n-3}(W)\ldots
\end{eqnarray}
where in the last step we have used equation (\ref{beauty}).
Using (\ref{HKasymp}) for the function $\bar W$ we then have
\begin{eqnarray}
\label{tracedelta}
{\rm Tr}\bar W[\Delta] & = &
\frac{1}{\bl(4\pi\br)^{\frac{d}{2}}}
\left[Q_{\frac{d}{2}}(\bar W)B_0(\Delta)
+Q_{\frac{d}{2}-1}(\bar W)B_2(\Delta)+\ldots
+Q_0(\bar W)B_{2d}(\Delta)+\ldots\right]\nonumber\\
 & = &
\frac{1}{\bl(4\pi\br)^{\frac{d}{2}}}
\Biggl[
\left(Q_{\frac{d}{2}}(W)-q Q_{\frac{d}{2}-1}(W)+\frac{1}{2!}q^2 Q_{\frac{d}{2}-2}(W)-\frac{1}{3!}q^3 Q_{\frac{d}{2}-3}(W)+\ldots\right)B_0(\Delta)\nonumber\\
 &  & +\left(Q_{\frac{d}{2}-1}(W)-q Q_{\frac{d}{2}-2}(W)+\frac{1}{2!}q^2 Q_{\frac{d}{2}-3}(W)-\frac{1}{3!}q^3 Q_{\frac{d}{2}-4}(W)+\ldots\right)B_2(\Delta)
\nonumber\\
 &  & +\qquad\qquad\ldots\nonumber \\
 &  & +\left(Q_{0}(W)-q Q_{-1}(W)+\frac{1}{2!}q^2 Q_{-2}(W)-\frac{1}{3!}q^3 Q_{-3}(W)+\ldots\right)B_{2d}(\Delta)\nonumber\\
 &  & + \qquad\qquad\ldots\Biggr]
\end{eqnarray}
We can now collect the terms that have the same $Q$-functions.
They correspond to the anti-diagonal lines in (\ref{tracedelta}).
Using equation (\ref{shifted}) one recognizes that the coefficient of
$Q_{\frac{d}{2}-k}$ is $B_{2k}(\Delta+q\mathbf{1})$. Therefore
\begin{align}
{\rm Tr}\bar W[\Delta]  &=& \frac{1}{\bl(4\pi\br)^{\frac{d}{2}}}
\Bigl[Q_{\frac{d}{2}}(\bar W)B_0(\Delta+q\mathbf{1})
+Q_{\frac{d}{2}+1}(\bar W)B_2(\Delta+q\mathbf{1})\nonumber\\
&&+\ldots+Q_0(\bar W)B_{2d}(\Delta+q\mathbf{1})+\ldots\Bigr]
\end{align}
which coincides term by term with the expansion of ${\rm Tr} W[\Delta+q]$
using the basis of eigenfunctions of the operator $\calo=\Delta+q\mathbf{1}$.
This provides an explicit check, at least in this particular example, that the trace
of this function is independent of the basis in the functional space.

\section{Spectral geometry of differentially constrained fields}

In this appendix we work on a sphere.
Consider the decomposition of a vector field $A_{\mu}$ into its
transverse and longitudinal parts:
$$
A_{\mu}\rightarrow A^T_{\mu}+\nabla_{\mu}\Phi
$$
The spectrum of $-\nabla^2$ on vectors is the disjoint union
of the spectrum on transverse and longitudinal vectors.
The latter can be related to the spectrum of $-\nabla^2-\frac{R}{d}$ on scalars
using the formula
\eq
-\nabla^2\nabla_{\mu}\Phi=-\nabla_{\mu}\bl(\nabla^2+\frac{R}{d}\br)\Phi.
\feq
Therefore one can write for the heat kernel
\eq
\label{vt}
{\rm Tr}\,{\rm e}^{-s\bl(-\nabla^2\br)}\mid_{A_{\mu}}=
{\rm Tr}\,{\rm e}^{-s\bl(-\nabla^2\br)}\mid_{A^T_{\mu}}
+{\rm Tr}\,{\rm e}^{-s\bl(-\nabla^2-\frac{R}{d}\br)}\mid_{\Phi}
-{\rm e}^{\bl(s\frac{R}{d}\br)}.\feq
The last term has to be subtracted because a constant scalar is an eigenfunction
of $-\nabla^2-\frac{R}{d}$ with negative eigenvalue, but does not
correspond to an eigenfunction of $-\nabla^2$ on vectors.
The spectrum of $-\nabla^2$ on scalars and transverse vectors
is obtained from the representation theory of $SO(d+1)$
and is reported in table \ref{taba2}.

A similar argument works for symmetric tensors, when using the decomposition
(\ref{decomposition}).
One can use equation
\eq
-\nabla^2\left( \nabla_{\mu}\xi_{\nu}+\nabla_{\nu}\xi_{\mu}\right) =
\nabla_{\mu}\left( -\nabla^2-\frac{d+1}{d\left( d-1\right) }R\right)\xi_{\nu}+
\nabla_{\nu}\left( -\nabla^2-\frac{d+1}{d\left( d-1\right) }R\right)\xi_{\mu}
\feq
and equation
\eq
-\nabla^2\left(\nabla_{\mu}\nabla_{\nu}-\frac{1}{d}g_{\mu\nu}\nabla^2 \right) \sigma=
\left(\nabla_{\mu}\nabla_{\nu}-\frac{1}{d}g_{\mu\nu}\nabla^2 \right) \left(-\nabla^2-\frac{2}{d-1}R\right)\sigma
\feq
to relate the spectrum of various operators on vectors and scalars
to the spectrum of $-\nabla^2$ on tensors.
One has to observe that the $d(d+1)/2$ Killing vectors are eigenvectors
of $-\nabla^2-\frac{d+1}{d\left( d-1\right) }R$ on vectors but give a vanishing tensor $h_{\mu\nu}$,
so they do not contribute to the spectrum of $-\nabla^2$ on tensors.
Likewise, a constant scalar
and the $d+1$ scalars proportional to the Cartesian coordinates of the
embedding $\mathbf{R}^n$,
which correspond to two the lowest eigenvalues of $-\nabla^2-\frac{2}{d-1}R$,
also do not contribute to the spectrum of tensors.
So one has for the heat kernel on tensors
\begin{eqnarray}
\label{ttt}
{\rm Tr}\,{\rm e}^{\bl(-s\bl(-\nabla^2\br)\br)}\Big |_{h_{\mu\nu}}&=&
{\rm Tr}\,{\rm e}^{\bl(-s\bl(-\nabla^2\br)\br)}\Big |_{h^T_{\mu\nu}}
+{\rm Tr}\,{\rm e}^{\bl(-s\bl(-\nabla^2-\frac{\bl(d+1\br)R}{d\bl(d-1\br)}\br)\br)}\Big|_{\xi}
+{\rm Tr}\,{\rm e}^{\bl(-s\bl(-\nabla^2\br)\br)}\Big |_h
\\
&&
\!\!\!\!\!\!\!\!\!\!\!\!\!\!\!\!\!\!\!\!\!
+{\rm Tr}\,{\rm e}^{\bl(-s\bl(-\nabla^2-\frac{2}{d-1}R\br)\br)}\mid_{\sigma}\nonumber
-{\rm e}^{\bl(\frac{2}{d-1}sR\br)}-\bl(d+1\br)\,{\rm e}^{\bl(\frac{1}{d-1}sR\br)}
-\frac{d\bl(d+1\br)}{2}\,{\rm e}^{\bl(\frac{2}{d\bl(d-1\br)}sR\br)}\ .
\end{eqnarray}

The last exponentials can be expanded in Taylor series as $\sum_{m=0}^{\infty}c_m R^m$
and these terms can be viewed as modifications of the heat kernel coefficients
of $-\nabla^2$ acting on the differentially constrained fields.
To see where these modifications enter, recall that the volume of the sphere is
\eq
\label{volumesphere}
V_{\rm dS}=\bl(4\pi\br)^{\frac{d}{2}}\bl(\frac{d\bl(d-1\br)}{R}\br)^{\frac{d}{2}}\frac{\Gamma\bl(\frac{d}{2}\br)}{\Gamma\bl(d\br)}
\feq
so that
\eq
\int {\rm d}^d x \sqrt{g}\,{\rm tr}\,\mathbf{b}_n\propto R^{\frac{n-d}{2}}\ .
\feq
This means a coefficent $c_m$ from the Taylor series will contribute to a heat kernel
coefficient for which $2m=n-d$.
So there are contributions to $\mathbf{b}_n$ only for $n\ge d$.

We have discussed how the negative and zero modes from constrained
scalar and vector fields affect the heat kernel coefficients of the
decomposed vector and tensor fields. These modes have to be excluded
also from the traces over the constrained fields; this is denoted
by one or two primes, depending on the number of excluded modes.
This can be done by calculating the trace and subtracting the contributions
to the operator trace from the excluded modes.
Thus the trace with $m$ primes is
\eq
{\rm Tr}^{'\ldots '}\left[W(-\nabla^2)\right]={\rm
Tr}\left[W(-\nabla^2)\right]-\sum_{l=1}^m
D_l\bl(d,s\br)W\bl(\lambda_l\bl(d,s\br)\br)
\feq
where $\lambda_l(d,s)$ are the eigenvalues, $D_l(d,s)$ their multiplicities,
both depending on the dimension $d$ and on the spin of the field, $s$.
The eigenvalues and multiplicities for the $m$-th mode of the Laplacian
on the sphere are given in table \ref{taba2}.

\begin{table}
\begin{center}
\begin{tabular}{|c|c|c|}\hline
Spin s  & Eigenvalue $\lambda_l(d,s)$ & Multiplicity $D_l(d,s)$\\\hline
0 & $\frac{l(l+d-1)}{d(d-1)}R$; $l=0,1\ldots$& $\frac{(2l+d-1)(l+d-2)!}{l!(d-1)!}$\\\hline
1 & $\frac{l(l+d-1)-1}{d(d-1)}R$; $l=1,2\ldots$& $\frac{l(l+d-1)(2l+d-1)(l+d-3)!}{(d-2)!(l+1)!}$\\\hline
2 & $\frac{l(l+d-1)-2}{d(d-1)}R$; $l=2,3\ldots$& $\frac{(d+1)(d-2)(l+d)(l-1)(2l+d-1)(l+d-3)!}{2(d-1)!(l+1)!}$\\\hline
\end{tabular}\end{center}
\caption{Eigenvalues and their multiplicities of the Laplacian on the d-sphere}
\label{taba2}
\end{table}

The expressions that we will need are those for the cases where one mode is excluded from the transverse vector trace
($s=1$, $m=1$), or one or two modes from the scalar trace ($s=0$, $m=1,2$), each one in two and four dimensions.
The results obtained by calculating the corresponding multiplicity and eigenvalue from table \ref{taba2}
are given in table \ref{taba21}.
To see what is the relevant contribution to one of the heat-kernel coefficients,
one can expand the obtained expression in $R$. For the case $s=0$, $d=4$, $m=2$ one has for example
\begin{eqnarray}
&&\sum_{l=1}^2 D_l\bl(4,0\br)W\bl(\lambda_l\bl(4,0\br)\br)= W\bl(0\br)+5W\bl(\frac{R}{3}\br)\nonumber\\
&&=\frac{R^2}{4\bl(4\pi\br)^2} \int {\rm d}x\sqrt{g}
\bl(W\bl(0\br)+\frac{5R}{18}W'\bl(0\br)+\frac{5}{108}R^2W''\bl(0\br)+
\frac{5}{36\cdot 27}R^3W'''\bl(0\br)+\ldots\br).
\end{eqnarray}
From this one sees that, in this case, the $\mathbf{b}_{2n}$ receive a correction for $n\ge 2$.
In two dimensions, that would be already the case for $n\ge 1$.
The full list of heat kernel coefficients of $-\nabla^2$
in 4d is given in table \ref{taba6}.

\begin{table}
\begin{center}
\begin{tabular}{|c|c|c|}\hline
  & s=1 & s=0\\\hline
$m=1$, $d=2$ & $3 W\left(\frac{R}{2}\right)$ & $W\left(0\right)$\\\hline
$m=1$, $d=4$ & $10 W\left(\frac{R}{4}\right)$& $W\left(0\right)$\\\hline
$m=2$, $d=2$ & & $W\left(0\right)+3 W\left(R\right)$\\\hline
$m=2$, $d=4$ & & $W\left(0\right)+5 W\left(\frac{R}{3}\right)$\\\hline
\end{tabular}\end{center}
\caption{$\sum_{l=1}^m
D_l\bl(d,s\br)W\bl(\lambda_l\bl(d,s\br)\br)$ for $s=0,1$, $d=2,4$, $m=1,2$}
\label{taba21}
\end{table}

\begin{table}
\begin{center}
\begin{tabular}[c]{|c|c|c|c|c|c|c|}\hline
\rule[-4mm]{0mm}{10mm}& S& V& VT& T& TT&TTT\\\hline
\rule[-4mm]{0mm}{10mm}\raisebox{-0.8ex}[0.8ex]{\hphantom{ak}
$\mathrm{tr}\mathbf{b}_0$\hphantom{ak}}&1&4&3&10&9&5\\\hline
\rule[-4mm]{0mm}{10mm}$\mathrm{tr}\mathbf{b}_2$&$\frac{R}{6}$&$\frac{2R}{3}$&$\frac{R}{4}$&$\frac{5R}{3}$&$\frac{3 R}{2}$&$-\frac{5R}{6}$
\\\hline
\rule[-4mm]{0mm}{10mm}$\mathrm{tr}\mathbf{b}_4$&$\frac{29R^2}{2160}$&$\frac{43R^2}{1080}$&$-\frac{7R^2}{1440}$&$\frac{11R^2}{216}$
&$\frac{81 R^2}{2160}$&$-\frac{R^2}{432}$\\
\hline \rule
[-4mm]{0mm}{10mm} $\mathrm{tr}\mathbf{b}_6$
&\hphantom{adk}$\frac{37R^3}{54432}$\hphantom{adk}
&\hphantom{ak}$-\frac{R^3}{17010}$ \hphantom{ak}
&\hphantom{ak}$-\frac{541R^3}{362880}$\hphantom{ak}
&\hphantom{ak}$-\frac{1343R^3}{136080}$\hphantom{ak}
&\hphantom{ak}$\frac{-319 R^3}{30240}$\hphantom{ak}
&\hphantom{akd}$\frac{311 R^3}{54432}$\hphantom{akd}\\
\hline
\rule[-4mm]{0mm}{10mm} $\mathrm{tr}\mathbf{b}_8$
&\hphantom{adk}$\frac{149R^4}{6531840}$\hphantom{adk}
&\hphantom{ak}$-\frac{2039R^4}{13063680}$ \hphantom{ak}
&\hphantom{ak}$-\frac{157R^4}{2488320}$\hphantom{ak}
&\hphantom{ak}$-\frac{2999R^4}{3265920}$\hphantom{ak}
&\hphantom{ak}$\frac{683 R^4}{725760}$\hphantom{ak}
&\hphantom{akd}$\frac{109 R^4}{1306368}$\hphantom{akd}\\ \hline
\end{tabular}\end{center}
\caption{Heat kernel coefficients for $S^4$. The columns for the transverse
vector (VT) and transverse traceless tensor (TTT) are obtained from
equations (\ref{vt}) and (\ref{ttt}) in $d=4$. Note that the excluded modes
contribute to $\mathrm{tr}\mathbf{b}_n$ only for $n\geq4$.}
\label{taba6}
\end{table}

\section{Proper time ERGE}

Let us start from the ERGE for gravity in the Einstein--Hilbert truncation
with a type III cutoff, written in equation (\ref{typethree}).
Define the functions:
\begin{equation}
A_{k}(z)=\frac{\partial_t R_{k}(z)}{z+R_{k}(z)}\qquad\quad
B_{k}(z)=\frac{R_{k}(z)}{z+P_{k}(z)}\qquad\quad
C_{k}(z)=\frac{\partial_{z}R_{k}(z)}{z+R_{k}(z)}\ .
\end{equation}
The term in equation (\ref{typethree}) containing $C$ is nontrivial.
To rewrite it in a managable form we take the Laplace transform:
\begin{equation}
C_{k}\left(z\right)=\int_{0}^{\infty}ds\,\tilde{C}_{k}(s)\, e^{-sz}\ .
\end{equation}
Since the operator $\partial_{t}(\Delta_2-2\Lambda)$ commutes
with $\Delta_2-2\Lambda$, we can write
\begin{eqnarray}
C_{k}\left(\Delta_2-2\Lambda\right)\partial_{t}(\Delta_2-2\Lambda) & = & \int_{0}^{\infty}ds\,\tilde{C}_{k}(s)\,\partial_{t}(\Delta_2-2\Lambda)\, e^{-s(\Delta_2-2\Lambda)}\nonumber \\
 & = & -\int_{0}^{\infty}\frac{ds}{s}\,\tilde{C}_{k}(s)\,\partial_{t}e^{-s(\Delta_2-2\Lambda)}\,.\label{4}
\end{eqnarray}
Laplace transforming also $A_k$ and $B_k$,
the first term in equation (\ref{typethree}) becomes
\begin{equation}
\frac{1}{2}\int_{0}^{\infty}ds\left[
\tilde{A}_{k}(s)
+\tilde{B}_{k}(s)\,\eta
-\frac{1}{s}\,\tilde{C}_{k}(s)\,\partial_{t}\right]\textrm{Tr}\,e^{-s(\Delta_2-2\Lambda)}\ .
\end{equation}
This is the functional RG equation in ``proper time'' form
\cite{Litim:2001ky,Zappala,Bonanno:2004sy}. Note that the first term
corresponds precisely to the one loop approximation. The trace of
the heat kernel can be expanded
\begin{eqnarray*}
\textrm{Tr}\, e^{-s(\Delta_2-2\Lambda)}
 & = & e^{-s(-2\Lambda)}\frac{1}{(4\pi)^{d/2}}\int dx\sqrt{g}\,\textrm{tr}\left[\mathbf{1}s^{-\frac{d}{2}}
+\left(\mathbf{1}\frac{R}{6}-\mathbf{W}\right)s^{-\frac{d}{2}+1}+O(R^{2})\right]\\
 & = & e^{-s(-2\Lambda)}\frac{1}{(4\pi)^{d/2}}\int dx\sqrt{g}\left[\frac{d(d+1)}{2}s^{-\frac{d}{2}}
+\frac{d(7-5d)}{12}R\, s^{-\frac{d}{2}+1}+O(R^{2})\right]\ ,
\end{eqnarray*}
whereas for the ghosts
\begin{eqnarray*}
\textrm{Tr}\, e^{-s(\delta_{\nu}^{\mu}\Delta-R_{\nu}^{\mu})}
& = & \frac{1}{(4\pi)^{d/2}}\int dx\sqrt{g}\,\textrm{tr}\left[\delta_{\nu}^{\mu}s^{-\frac{d}{2}}+\left(\delta_{\nu}^{\mu}\frac{R}{6}+R_{\nu}^{\mu}\right)s^{-\frac{d}{2}+1}+O(R^{2})\right]\\
 & = & \frac{1}{(4\pi)^{d/2}}\int dx\sqrt{g}\left[ds^{-\frac{d}{2}}+\frac{d+6}{6}R\, s^{-\frac{d}{2}+1}+O(R^{2})\right]\,.\end{eqnarray*}
The ERGE then takes the form:
\begin{eqnarray*}
\partial_{t}\Gamma_{k} & = & \frac{1}{(4\pi)^{d/2}}\int dx\sqrt{g}
\Bigl\lbrace
\frac{d(d+1)}{4}Q_{\frac{d}{2}}\left(\bar A_{k}+\eta \bar B_{k}-2\partial_{t}\Lambda\,\bar C_{k}\right)
-dQ_{\frac{d}{2}}\left(A_{k}\right)
\nonumber\\
 &  & +\left[\frac{d(7-5d)}{12}\,
Q_{\frac{d}{2}-1}\left(\bar A_{k}+\eta \bar B_{k}-2\partial_{t}\Lambda\,\bar C_{k}\right)
%\nonumber\\
% &  &
-\frac{d+6}{d}\,Q_{\frac{d}{2}-1}\left(A_{k}\right)\right]\,R+O(R^{2})\Bigr\rbrace\ .
\end{eqnarray*}
where $\bar W(z)=W(z-2\Lambda)$.
Using an optimized cutoff one can now reproduce equations (\ref{firstsum}) and (\ref{secondsum}).
However, in this way the sums in equation (\ref{gammatypethree}) can
be resummed for any type of cutoff shape.

\section{Cutoff of type Ib without field redefinitions}

We collect here the formulae for the beta functions of $\Lambda$ and $G$
in the Einstein--Hilbert truncation, using a cutoff of type Ib and
without redefining the fields $\xi_\mu$ and $\sigma$.
The ERGE, including the contributions of the Jacobians, is
\begin{eqnarray}
 \frac{d \Gamma_k}{dt}
&=& \frac{1}{2} \textrm{Tr}_{(2)}
\frac{\partial_t R_k+\eta R_k}{P_k-2\Lambda+\frac{d^2-3 d+4}{d(d-1)}R}\nonumber\\
&&+ \frac{1}{2} \textrm{Tr}'_{(1)} \frac{\partial_t
R_k\bl(2P_k+\frac{d-4}{d}R-2\Lambda\br) +\eta
R_k\bl(P_k+z+\frac{d-4}{d}R-2\Lambda\br)-2\partial_t\Lambda
R_k}{\bl(P_k-\frac{R}{d}\br)\bl(P_k+\frac{d-3}{d}R-2\Lambda\br)}
\nonumber\\
&&+\frac{1}{2}
 \textrm{Tr}_{(0)}  \frac{\partial_t R_k+\eta R_k}{P_k
- 2 \Lambda + \frac{d-4}{d} R} \nonumber\\
&&+\frac{1}{2} \textrm{Tr}''_{(0)}
\frac{1}{
P_k
   ( P_k-\frac{R}{d-1}) ( P_k+\frac{d-4}{d} R-2  \Lambda )}
\times\nonumber\\
&&
\Biggl\lbrace
\partial_t R_k \left(
3  P_k^2+2 P_k \left(\frac{d^2-6 d+4}{d(d-1)} R-2 \Lambda \right)
-\frac{R}{d-1} \left(\frac{d-4}{d} R-2  \Lambda \right)\right)
\nonumber\\
&&+\eta
R_k \left(3  P_k^2-P_k \left(3  R_k-2 \left(\frac{d^2-6 d+4}{d(d-1)} R-2  \Lambda \right)\right)
%\right.\nonumber\\
%&&\left.
-\left( R_k+\frac{R}{d-1}\right) \left(-
   R_k+\frac{d-4}{d} R-2  \Lambda \right)\right)
   \nonumber\\
&&
- 2\partial_t\Lambda
  R_k \left(2  P_k- R_k-\frac{R}{d-1}\right)
\Biggr\rbrace 
\nonumber\\&&
- \textrm{Tr}_{(1)} \frac{\partial_t R_k}{P_k-\frac{R}{d}}
-\textrm{Tr}'_{(0)} \frac{2\left(P_k-\frac{R}{d}\right)\partial_t
R_k}{\left(P_k-
\frac{2R}{d}\right)P_k}\nonumber\\
&&- \frac{1}{2}\textrm{Tr}'_{(1)} \frac{ \partial_t R_k}{
P_k-\frac{R}{d}}
 - \frac{1}{2}\textrm{Tr}''_{(0)}. \frac{2\left(P_k-\frac{R}{2 (d-1)}\right)
\partial_t R_k}{P_k (P_k-\frac{R}{d-1} )}
+ \textrm{Tr}'_{(0)} \frac{\partial_t R_k}{P_k}
\end{eqnarray}

which gives

\begin{eqnarray}
\frac{d \Gamma_k}{dt} &=&\frac{1}{(4\pi)^{\frac{d}{2}}}\int dx\,\sqrt{g}\Biggl\lbrace
Q_{\frac{d}{2}}\left(  \frac{d ((d-3) P_k+8 \Lambda ) \left(\partial_t R_k\right)}{4 P_k (P_k-2 \Lambda )}     \right)
\nonumber\\
&&
+\eta \,Q_{\frac{d}{2}}\left( \frac{R_k \left(\left(d^2+3 d+2\right) P_k^2-2 P_k \left((d+2) R_k+2 (d+1) \Lambda \right)+2 R_k \left(R_k+2 \Lambda \right)\right)}{4 P_k^2 \left(P_k-2
   \Lambda \right)} \right)
   \nonumber\\
   &&
-\partial_t\Lambda \,Q_{\frac{d}{2}}\left(     \frac{R_k \left((d+1) P_k-R_k\right)}{P_k^2 (P_k-2 \Lambda )}  \right)
\nonumber\\
&&
+R\left[
Q_{\frac{d}{2}}\left(   -\frac{\left(-16 \left(d^2-1\right) \Lambda  P_k+16 \left(d^2-1\right) \Lambda ^2+\left(d^4-2 d^3+3 d^2-4 d-2\right) P_k^2\right) \partial_t R_k}{4 (d-1)
   d P_k^2 \left(P_k-2 \Lambda \right){}^2}   \right)
  \right. \nonumber\\
&&\left.
+\eta \,Q_{\frac{d}{2}}\left(
\left(R_k \left(2 P_k \left(2 \left(d^2+d+1\right) \Lambda  R_k-\left(d^2-6 d+4\right) R_k^2+4 \left(d^2-d+1\right) \Lambda ^2\right)
\right.\right.\right.\right.\nonumber\\
&&\left.\left.\left.\left.
+P_k^2 \left(2 \left(d^3-4
   d^2-3 d+4\right) R_k-8 \left(d^2-d+1\right) \Lambda \right)+d \left(-d^3+2 d^2+3 d+2\right) P_k^3
\right.\right.\right.\right.   \nonumber\\
&&\left.\left.\left.\left.   -4 d \Lambda  R_k \left(R_k+2 \Lambda \right)\right)\right)
 \Big/
   4 (d-1) d P_k^3 \left(P_k-2 \Lambda \right){}^2         \right)
   \right.\nonumber\\
&&\left.
-\partial_t\Lambda \,Q_{\frac{d}{2}}\left(  -\frac{R_k \left(P_k \left(2 \left(d^2-d+1\right) \Lambda -\left(d^2-6 d+4\right) R_k\right)+\left(d^3-4 d^2-2 d+4\right) P_k^2-2 d \Lambda  R_k\right)}{(d-1)
   d P_k^3 \left(P_k-2 \Lambda \right){}^2}  \right)
  \right. \nonumber\\
 &&\left. 
+Q_{\frac{d}{2}-1}\left( \frac{\partial_t R_k \left(12 (d-1) d \delta _{2,d} \left(3 P_k+2 \Lambda \right)+8 \left(d^3-d^2-6 d+6\right) \Lambda +\left(d^4-4 d^3-9 d^2-12\right)
   P_k\right)}{24 (d-1) d P_k \left(P_k-2 \Lambda \right)}     \right)
  \right. \nonumber\\
&&\left.
\eta \,Q_{\frac{d}{2}-1}\left(\frac{R_k }{ 24 (d-1) d P_k^2 \left(P_k-2 \Lambda \right)}
\left(2 (d-1) P_k \left(R_k \left(3 d \delta _{2,d}+d^2+2 d-6\right)
\right.\right.\right.\right.   \nonumber\\
&&\left.\left.\left. \left.
+2 \Lambda  \left(3 d \delta _{2,d}+d^2+d-6\right)\right)
\right.\right.\right.   \nonumber\\
&&\left.\left.\left. 
    +P_k^2 \left(-54 (d-1)
   d \delta _{2,d}-d^4-2 d^3+13 d^2+38 d-24\right)-2 (d-1) d R_k \left(R_k+2 \Lambda \right)\right)\right)
   \right.\nonumber\\
&&\left.
-\partial_t\Lambda \,Q_{\frac{d}{2}-1}\left( \frac{R_k \left(P_k \left(3 d \delta _{2,d}+d^2+d-6\right)-d R_k\right)}{6 d P_k^2 \left(P_k-2 \Lambda \right)}       \right)
\right]
\end{eqnarray}
With the optimized cutoff one obtains the coefficients
\begin{eqnarray}
\frac{A_1}{16\pi}&=& 
\frac{8 \left(d^2+5 d+3\right) \tilde{\Lambda }+d^3+3 d^2-10 d-24}
{(4\pi)^{\frac{d}{2}}(d^2+6 d+8) (1-2 \tilde{\Lambda}) \Gamma \left(\frac{d}{2}\right)}
\nonumber\\
\frac{A_2}{16\pi}&=& 
\frac{-4 \left(d^2+11 d+30\right) \tilde{\Lambda }+d^3+13 d^2+48 d+28}
{(4\pi)^{\frac{d}{2}}(d^3+12 d^2+44 d+48) (1-2 \tilde{\Lambda }) \Gamma \left(\frac{d}{2}\right)}
\nonumber\\
\frac{A_3}{16\pi}&=&-
\frac{4 (d+5) }
{(4\pi)^{\frac{d}{2}}(d^2+6 d+8) (1-2 \tilde{\Lambda }) \Gamma \left(\frac{d}{2}\right)}
\nonumber\\
\frac{B_1}{16\pi}&=&
\Big(
 \left(8 (d+6) \left(  d^3 (d+3)^2-12d^2-88d+96\right) \tilde{\Lambda }^2
   \right.\nonumber\\
   &&\left.
   -2 (d+2) (d^6+14d^5+43d^4-118d^3-628d^2-648d+576) \tilde{\Lambda }
   \right.\nonumber\\
  &&\left.
   +d ( d^6+2d^5-49d^4-62d^3+84d^2-3432d+2688)\right)\Big)
   \nonumber\\
   &&\Big/
   12  (4\pi )^{\frac{d}{2}} (1-2 \tilde{\Lambda })^2
   d^2(d^4+11 d^3+32 d^2+4 d-48) \Gamma \left(\frac{d}{2}\right)
\nonumber\\
\frac{B_2}{16\pi}&=&
\left(
8 (d+6) \left(d^3(d+3)^2-12d^2-88d+96\right) \tilde{\Lambda }^2
\right.\nonumber\\
&&\left.
   -2 (d+2) (d^6+14d^5+43d^4-118d^3-628d^2-648d+576) \tilde{\Lambda }
\right.\nonumber\\
&&   \left.
   +d (
   d^6+2d^5-49d^4-62d^3+84d^2-3432d+2688)\right)\Big)
   \nonumber\\
   &&\Big/
   12  (4\pi )^{\frac{d}{2}} (1-2 \tilde{\Lambda })^2
   d^2(d^4+11 d^3+32 d^2+4 d-48) \Gamma \left(\frac{d}{2}\right)
\nonumber\\
\frac{B_3}{16\pi}&=&
\frac{ 
  2 \left(d^3 (d+3)^2-12d^2-88d+96\right) \tilde{\Lambda }
  -d^5-6d^4-3d^3+216d^2+244d-48}
   {3 (4\pi )^{\frac{d}{2}} (1-2 \tilde{\Lambda })^2 
   \Gamma \left(\frac{d}{2}\right)  d^2(d^3+5 d^2+2 d-8)
   }
   \nonumber
\end{eqnarray}
and the beta functions in four dimensions
\begin{eqnarray}
\beta_{\tilde\Lambda} &=&-2\tilde\Lambda +
\frac{ \tilde{G}}{24\pi}
\frac{  (1-2 \tilde{\Lambda }) (-200 \tilde{\Lambda }^3+20\tilde{\Lambda }^2-33\tilde{\Lambda }+12)
+
\frac{\tilde{G}}{240\pi}  (- 80 \tilde{\Lambda }^3+36096\tilde{\Lambda }^2-43442\tilde{\Lambda }+12079)
   }
  {
     (1-2 \tilde{\Lambda })^3
  +\frac{\tilde{G}}{720\pi} (61-657 \tilde{\Lambda }) (1-2 \tilde{\Lambda })
   -\frac{\tilde{G}^2}{5760\pi^2} (1406-1945 \tilde{\Lambda })
 }
\nonumber\\
\beta_{\tilde G} &=& 2\tilde G 
-\frac{ \tilde{G}^2}{24\pi}\frac{
(1-2 \tilde{\Lambda })
    (200 \tilde{\Lambda }^2-212\tilde{\Lambda }+105)
+\frac{\tilde{G}}{24\pi} (8 \tilde{\Lambda }^2-1524\tilde{\Lambda }+1021)    
    }
 {
     (1-2 \tilde{\Lambda })^3
  +\frac{\tilde{G}}{720\pi} (61-657 \tilde{\Lambda }) (1-2 \tilde{\Lambda })
   -\frac{\tilde{G}^2}{5760\pi^2} (1406-1945 \tilde{\Lambda })
 }
\end{eqnarray}

\end{document}